\documentclass[fleqn,usenatbib]{mnras}

\usepackage{newtxtext,newtxmath}
\usepackage{chemformula}

\usepackage[T1]{fontenc}

\usepackage{graphicx}	
\usepackage{amsmath}	
\usepackage{siunitx}
\usepackage[version=4]{mhchem}
\usepackage{orcidlink}
\usepackage{pdflscape}
\usepackage{calc}

\newcommand{\Rsun}{R$_{\odot}$}
\newcommand{\Msun}{M$_{\odot}$}
\newcommand{\Rjup}{R$_{\mathrm{J}}$}
\newcommand{\Mjup}{M$_{\mathrm{J}}$}

\newcommand{\Teff}{T$_{\mathrm{eff}}$}
\newcommand{\degree}{$^{\circ}$}

\defcitealias{PenzlinBooth2024}{Penzlin \& Booth et al.}

\title[BOWIE-ALIGN: the misaligned WASP-15b]{BOWIE-ALIGN: JWST reveals hints of planetesimal accretion and complex sulphur chemistry in the atmosphere of the misaligned hot Jupiter WASP-15b}

\author[J. Kirk et al.]{James Kirk$^{\orcidlink{0000-0002-4207-6615},1}$\thanks{E-mail: j.kirk22@imperial.ac.uk (JK)},
Eva-Maria Ahrer$^{\orcidlink{0000-0003-0973-8426},2}$,
Alastair B. Claringbold$^{\orcidlink{0000-0003-1309-5558},3,4}$,
Maria Zamyatina$^{{\orcidlink{0000-0002-9705-0535}},5}$,
Chloe Fisher$^{{\orcidlink{0000-0003-0652-2902}},6}$,
\newauthor
Mason McCormack$^{\orcidlink{0000-0002-1463-9847},7}$,
Vatsal Panwar$^{\orcidlink{0000-0002-2513-4465},3,4}$,
Diana Powell$^{\orcidlink{0000-0002-4250-0957},7}$,
Jake Taylor$^{{\orcidlink{0000-0003-4844-9838}},6}$,
Daniel P. Thorngren$^{{\orcidlink{0000-0002-5113-8558}},8}$,
\newauthor
Duncan A. Christie$^{\orcidlink{0000-0002-4997-0847},2}$,
Emma Esparza-Borges$^{\orcidlink{0000-0002-2341-3233},9,10}$,
Shang-Min Tsai$^{\orcidlink{0000-0002-8163-4608},11}$,
Lili Alderson$^{\orcidlink{0000-0001-8703-7751},12,13}$,
\newauthor
Richard A. Booth$^{{\orcidlink{https://orcid.org/0000-0002-0364-937X}},14}$,
Charlotte Fairman$^{\orcidlink{0000-0001-9665-5260}12}$,
Mercedes L\'opez-Morales$^{{\orcidlink{0000-0003-3204-8183}},15}$,
N. J. Mayne$^{{\orcidlink{0000-0001-6707-4563}},5}$,
\newauthor
Annabella Meech$^{\orcidlink{0000-0002-7500-7173}6,16}$,
Paul Molli\`{e}re$^{\orcidlink{0000-0003-4096-7067},2}$,
James E. Owen$^{\orcidlink{0000-0002-4856-7837},1}$, 
Anna B.T. Penzlin$^{{\orcidlink{0000-0002-8873-6826}},1}$,
Denis E. Sergeev$^{{\orcidlink{0000-0001-8832-5288}},5}$,
\newauthor
Daniel Valentine$^{\orcidlink{0000-0002-2643-6836},12}$,
Hannah R. Wakeford$^{{\orcidlink{0000-0003-4328-3867}},12}$,
and Peter J.\ Wheatley$^{\orcidlink{0000-0003-1452-2240},3,4}$
\\
$^{1}$Department of Physics, Imperial College London, Prince Consort Road, SW7 2AZ, London, UK\\
$^{2}$Max-Planck-Institut f\"{u}r Astronomie, K\"{o}nigstuhl 17, 69117 Heidelberg, Germany \\
$^{3}$Centre for Exoplanets and Habitability, University of Warwick, Gibbet Hill Road, Coventry CV4 7AL, UK\\
$^{4}$Department of Physics, University of Warwick, Gibbet Hill Road, Coventry CV4 7AL, UK\\
$^{5}$Department of Physics and Astronomy, Faculty of Environment, Science and Economy, University of Exeter, Exeter EX4 4QL, UK\\
$^{6}$Department of Physics, University of Oxford, Denys Wilkinson Building, Keble Road, Oxford, OX1 3RH, United Kingdom\\
$^{7}$Department of Astronomy \& Astrophysics, University of Chicago, Chicago, IL 60637, USA\\
$^{8}$Department of Physics and Astronomy, Johns Hopkins University, Baltimore, MD 21218, USA\\
$^{9}$Instituto de Astrof\'isica de Canarias, E-38200 La Laguna, Tenerife, Spain\\
$^{10}$Departamento de Astrof\'isica, Universidad de La Laguna, E-38206 La Laguna, Tenerife, Spain\\
$^{11}$Department of Earth and Planetary Sciences, University of California, Riverside, CA, USA \\
$^{12}$School of Physics, University of Bristol, HH Wills Physics Laboratory, Tyndall Avenue, Bristol, BS8 1TL, UK\\
$^{13}$Department of Astronomy, Cornell University, 122 Sciences Drive, Ithaca, NY 14853, USA\\
$^{14}$School of Physics and Astronomy, University of Leeds, Leeds, LS2 9JT\\
$^{15}$Space Telescope Science Institute, 3700 San Martin Drive,  Baltimore, MD 21218, USA\\
$^{16}$Center for Astrophysics ${\rm \mid}$ Harvard {\rm \&} Smithsonian, 60 Garden St, Cambridge, MA 02138, USA\\}
\date{Accepted 2025 January 30. Received 2025 January 14; in original form 2024 October 1}

\pubyear{\the\year{}}

\begin{document}
\label{firstpage}
\pagerange{\pageref{firstpage}--\pageref{lastpage}}
\maketitle

\begin{abstract}

We present a transmission spectrum of the misaligned hot Jupiter WASP-15b from 2.8--5.2 microns observed with JWST's NIRSpec/G395H grating. Our high signal to noise data, which has negligible red noise, reveals significant absorption by H$_2$O ($4.2\sigma$) and CO$_2$ ($8.9\sigma$). From independent data reduction and atmospheric retrieval approaches, we infer that WASP-15b's atmospheric metallicity is super-solar ($\gtrsim 15\times$ solar) and its carbon-to-oxygen ratio is consistent with solar, that together imply planetesimal accretion. Our general circulation model simulations for WASP-15b suggest that the carbon-to-oxygen we measure at the limb is likely representative of the entire photosphere due to the mostly uniform spatial distribution of \ch{H2O}, \ch{CO2} and CO. We additionally see evidence for absorption by SO$_2$ and absorption at 4.9\,\micron, for which the current leading candidate is OCS, albeit with several caveats. If confirmed, this would be the first detection of OCS in an exoplanet atmosphere and point towards complex photochemistry of sulphur-bearing species in the upper atmosphere. These are the first observations from the BOWIE-ALIGN survey which is using JWST's NIRSpec/G395H instrument to compare the atmospheric compositions of aligned/low-obliquity and misaligned/high-obliquity hot Jupiters around F stars above the Kraft break. The goal of our survey is to determine whether the atmospheric composition differs across two populations of planets that have likely undergone different migration histories (disc versus disc-free) as evidenced by their obliquities (aligned versus misaligned).

\end{abstract}




\section{Introduction}\label{sec:intro}

One of the main goals behind many exoplanet atmosphere observational programmes is to learn about exoplanet formation and evolution. A primary focus of the field has been to use a planet's carbon-to-oxygen ratio (C/O) to infer where a planet formed relative to ice lines in a protoplanetary disc, largely motivated by \cite{Oberg2011}. However, there are many competing physical processes which make inferences from C/O challenging. These include the evolving ice lines within a disc \citep[e.g.][]{Morbidelli2016,Owen2020}, the drift of volatile-carrying solids in the disc \citep[e.g.,][]{Booth2017,Schneider2021}, the relative importance of solid vs gaseous accretion in setting a planet's atmospheric composition \citep[e.g.][]{Espinoza2017}, and the diversity found from observations of protoplanetary discs \citep[e.g.,][]{Law2021}. Furthermore, it is likely that observations of exoplanetary atmospheres probe a limited range of atmospheric pressures \citep{Dobbs-Dixon2017}, which might not be representative of the bulk planet's atmospheric composition due to processes such as local atmospheric mixing \citep[e.g.,][]{Zamyatina24_quenchingdriven}, cloud formation \citep[e.g.,][]{Helling2016}, or a planet's interior evolution \citep{Muller2024}. Therefore, it remains to be observationally demonstrated that C/O, and atmospheric composition in general, are reliable tracers of planet formation. 

As \citetalias{PenzlinBooth2024} \citeyear{PenzlinBooth2024} showed, the high dimensionality and unconstrained nature of key planet formation and disc parameters make it challenging to predict the atmospheric composition of an exoplanet from planet formation models. However, they demonstrated that comparing populations of planets with different migration histories could constrain planet formation models. Specifically, they showed that the C/O and metallicity of exoplanets that migrate through a disc should diverge from exoplanets that undergo disc-free migration, due to the fact that disc-migrated planets accrete inner disc material, a result that builds upon earlier work of \cite{Madhusudhan2014} and \cite{Booth2017}. However, the amplitude and sign of this divergence is dependent on whether silicates from the inner disc release their oxygen into the planetary atmosphere upon accretion and the uncertain form of the dominant carbon carriers in discs. 

These theoretical considerations motivate our observational survey which seeks to compare the compositions of four aligned/low-obliquity hot Jupiters that likely migrated through their protoplanetary discs versus four misaligned/high-obliquity hot Jupiters that likely underwent disc-free migration via high-eccentricity migration \citep[e.g.,][]{Rasio1996,Wu2003,Ford2008,2016Munoz}. In our sample, we only include hot Jupiters orbiting F stars above the Kraft break (effective temperatures $\gtrsim 6100$\,K) where tidal realignment is inefficient due to the stars' radiative envelopes \citep[e.g.,][]{Albrecht2012}. This reduces the likelihood of our aligned sample being polluted with initially misaligned planets that have had their obliquities damped. \cite{Kirk2024_align} gives more detail regarding our survey (`BOWIE-ALIGN', JWST program ID: GO 3838, PIs: Kirk \& Ahrer). By comparing aligned versus misaligned hot Jupiters, our goals are to constrain planet formation models (e.g., \citetalias{PenzlinBooth2024} \citeyear{PenzlinBooth2024}) and to robustly test the reliability of C/O and metallicity as tracers of planet formation. 

In this work, we present the first observations from our programme, those of the hot Jupiter WASP-15b. WASP-15b, discovered by \cite{West2009}, has a mass of $0.542^{+0.054}_{-0.053}$\,\Mjup\,, a radius of $1.428 \pm 0.077$\,\Rjup\ \citep{Bonomo2017}, and an equilibrium temperature of $1676 \pm 29$\,K \citep{Southworth2013}. Importantly for our programme, it has a precisely measured sky-projected obliquity ($\lambda = -139.6^{+4.3}_{-5.2}$, \citealt{Triaud2010}) and orbits an F7 dwarf star above the Kraft break ($1.18 \pm 0.12$\,\Msun, $1.477 \pm 0.072$\,\Rsun, \citealt{Bonomo2017}; \Teff\ $= 6372 \pm 13$\,K, \citealt{GaiaDR3}). Our study is the first published transmission spectrum of WASP-15b.

We describe our observations in Section \ref{sec:observations} and our data reduction in Section \ref{sec:data_reduction} that results in the planet's transmission spectrum (Section \ref{sec:trans_spec}). In Section \ref{sec:interior_model} we present constraints on the planet's atmospheric metallicity derived from interior structure models. We interpret the transmission spectrum using 1D atmospheric models in Section \ref{sec:1D_models}, a 3D general circulation model (GCM) in Section \ref{sec:GCM}, and photochemical models in Section \ref{sec:photochem}. We discuss our results in Section \ref{sec:discussion} and conclude in Section \ref{sec:conclusions}.

\section{Observations}
\label{sec:observations}

We observed one transit of WASP-15b on January 26, 2024 with the JWST/NIRSpec instrument \citep{Jakobsen2022} in Bright Object Time Series mode. We used the G395H grating, F290LP filter, 2048 subarray and the NRSRAPID readout pattern. This setup covers opacity from H$_2$O, CO$_2$, CO and SO$_2$ \citep{Alderson2023} between wavelengths of 2.8--5.2\,\micron\ at an average spectral resolution of $R = 2700$. We chose to use 44 groups per integration, which was informed by WASP-15's K magnitude of 9.7, and we acquired 685 integrations over 7.72 hours, which included a baseline of 2.26 hours pre-ingress and 1.65 hours post-egress. Due to the brightness of WASP-15, we used a nearby, fainter star for target acquisition (2MASS\,J13554509-3209041), observed with the SUB32 array and CLEAR filter.

\section{Data reduction}
\label{sec:data_reduction}

We performed three independent reductions of the data, one using the \texttt{Tiberius} pipeline \citep{Kirk2017,Kirk2021} and the other two using the \texttt{Eureka!} pipeline \citep{Bell2022Eureka}. This approach was motivated by the work of the Early Release Science programme that demonstrated the benefits of independent data reductions \citep{JWST2023,Alderson2023,Ahrer2023,Feinstein2023,Rustamkulov2023}. We give more information about our survey's data analysis strategy in \cite{Kirk2024_align}. We describe the approaches taken by the independent reductions for this work in the following subsections. 

\subsection{Tiberius reduction}
\label{sec:tiberius_reduction}

\texttt{Tiberius} \citep{Kirk2017,Kirk2021} is an open-source Python-based code that has been used in several studies of JWST data from multiple instruments \citep[e.g.,][]{Rustamkulov2023,Alderson2023,Kirk2024}.

\subsubsection{Light curve extraction}

We began by processing the raw images (\texttt{uncal.fits} files) through the standard set of stage 1 steps\footnote{\url{https://jwst-pipeline.readthedocs.io/en/latest/jwst/pipeline/calwebb_detector1.html\#calwebb-detector1}} of the \texttt{jwst} pipeline (v1.8.2). However, we did not perform the \texttt{jump} step as it has been found to increase the noise in exoplanet transit light curves \citep[e.g.,][]{Rustamkulov2023},  and performed our own 1/f correction before the \texttt{ramp\_fit} step. This involved subtracting the median value for each column on the detector after masking the 22 pixels centred on the stellar trace. The result of stage 1 was the production of \texttt{gainscalestep.fits} files with flux units of DN/s. We additionally performed the \texttt{assign\_wcs} and \texttt{extract\_2d} steps of the \texttt{jwst} pipeline to obtain the wavelength solution. 

Next, we performed our own cosmic ray / bad pixel identification and removal. First, we calculated the running median for every pixel's time series with a sliding box of three pixels and subtracted this running median from every pixel's time series. We then used the residuals to identify $5\sigma$ outliers and replaced these outliers with the running median's value. We chose a sliding box of three pixels to locate sharply varying features and to allow the pixel replacement to be informed by the neighbouring values in the time series. We also generated our own bad pixel mask at this stage. To do this, we combined the pixels flagged during stage 1 as bad, saturated, dead or hot with pixels identified as $5\sigma$ outliers in a median-combined science integration from the first segment of data (260 integrations).

For each outlier-clipped integration, we then located the stellar trace by fitting a Gaussian to each row in the cross-dispersion direction, followed by a smoothing with a fourth order polynomial. We then performed standard aperture photometry with an aperture full width of 8 pixels, after subtracting the background flux which was calculated as the median of the 10 pixels after masking 22 pixels centred on the trace. This aperture width was selected because of the lower noise in the resulting white light curves as compared to aperture widths of 6 and 10 pixels. We extracted the stellar flux between pixel rows 608 and 2044 (zero-indexed) for NRS1 and between rows 3 and 2043 for NRS2. With the stellar spectra in hand, we proceeded to create our light curves. For the white light curves, we integrated between wavelengths of 2.75 and 3.72\,\micron\ for the NRS1 detector and between 3.82 and 5.18\,\micron\ for the NRS2 detector. For our spectroscopic light curves, we adopted three different binning schemes: $R=100$ ($\sim61$ pixels wide), $R=400$ ($\sim15$ pixels wide) and 1 pixel resolution. Here we present results from the $R=100$ and $R=400$ analyses, with the high resolution (1-pixel) analysis presented in a follow-up study (Esparza-Borges et al., in prep.).

\subsubsection{Light curve fitting}

We fit the NRS1 and NRS2 white light curves independently to obtain our own set of system parameters, using a \texttt{batman} \citep{batman} analytic transit light curve model multiplied by a linear-in-time polynomial. The free parameters in our white light curve fits were: the time of mid-transit ($T_0$), the planet's inclination ($i$), the planet's semi-major axis relative to the stellar radius ($a/R_*$), the relative planet-to-star radii ($R_P/R_*$) and the two coefficients of the linear polynomial. We held the orbital period fixed to 3.7520998 days \citep{Patel2022} and the planet's eccentricity to 0 \citep{West2009}. We used a quadratic limb darkening law and fixed the coefficients to the values computed by \texttt{ExoTiC-LD} \citep{Grant2024_exoticld} using the \texttt{Stagger} grid of 3D stellar atmosphere models \citep{Magic2015} and the stellar parameters from \cite{Bonomo2017} ($\log g = 4.17$\,cgs, [Fe/H] $=-0.17$) and \cite{GaiaDR3} (\Teff $=6372$\,K). 

We used a Levenberg-Marquadt algorithm, implemented through \texttt{SciPy} \citep{scipy}, to explore the parameter space and determine the best-fit parameters. For each light curve, we ran two sets of fits. The first fit allowed us to rescale the photometric uncertainties to give $\chi^2_\nu = 1$ for the best-fit model. The second fit was performed with the rescaled uncertainties. The results from this second fit were used in the rest of our analysis. 

\begin{figure}
    \centering
    \includegraphics[width=1\columnwidth]{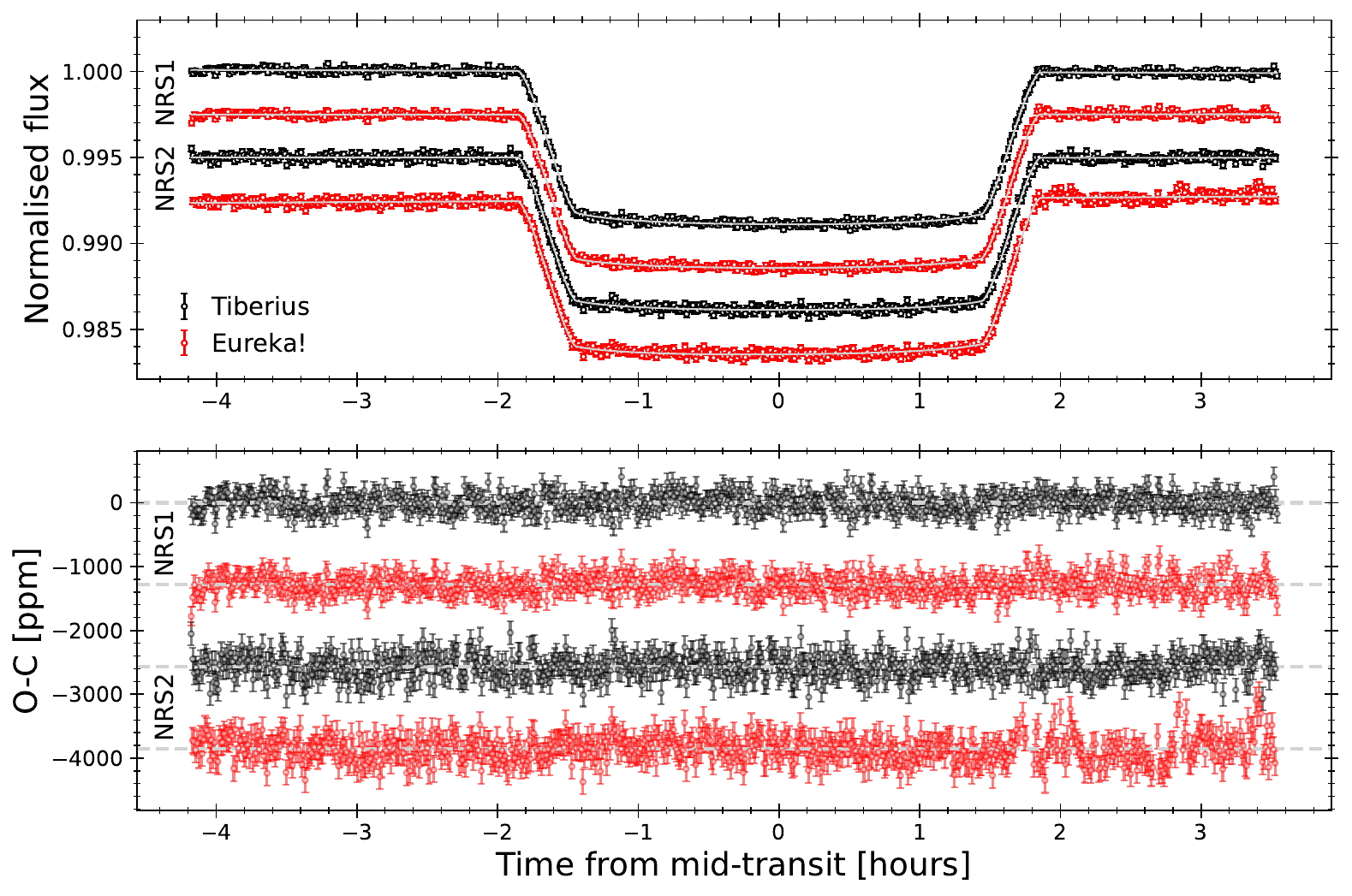}
    \caption{The white light curves and fits from two independent reductions. Top panel: the white light curves from \texttt{Tiberius} (black) and \texttt{Eureka!} (red). The thin grey lines indicate the best-fitting models. The light curves from NRS1 and NRS2 are offset from one another for visualisation. Bottom panel: the residuals from the fits in the top panel.}
    \label{fig:Tiberius_wl_fit}
\end{figure}

\begin{figure}
    \centering
    \includegraphics[width=1\columnwidth]{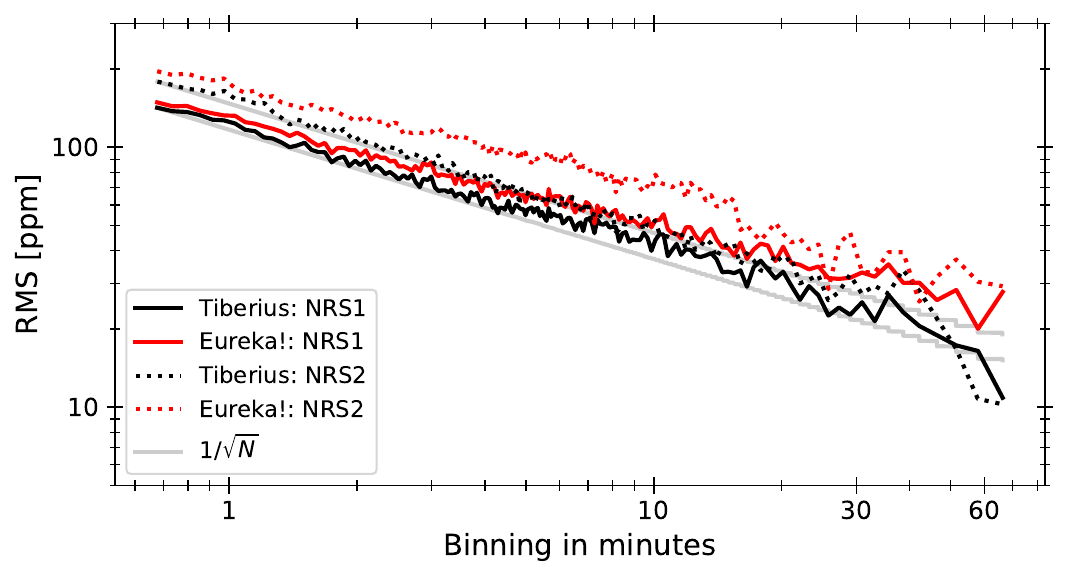}
    \caption{The Allan variance plot from the \texttt{Tiberius} (black) and \texttt{Eureka!} (red) white light curve fits. The results for NRS1 are shown by the solid lines and NRS2 by the dotted lines. These closely follow the expectations from pure white noise (grey lines) which indicates a lack of red noise in the residuals.}
    \label{fig:Allan_variance_WL}
\end{figure}

The results from our white light curve fits are shown in Figure \ref{fig:Tiberius_wl_fit} and Table \ref{tab:system_params}. The system parameters from both detectors are consistent with one another and with values from TESS \citep{Patel2022}. As shown by the Allan variance plots in Figure \ref{fig:Allan_variance_WL}, the residuals from the white light curve fits show minimal red noise. This is likely due to a combination of the relatively high groups/integration of our observations (44), the choice to fill 80\,\% of the full well and the relatively quiet nature of this F-type star. The Allan variance plots for the spectroscopic light curves are given in Appendix \ref{sec:Allan_variance_spec} and show that there is also minimal red noise in the spectroscopic light curves. 

\begin{figure}
    \centering
    \includegraphics[width=1\columnwidth]{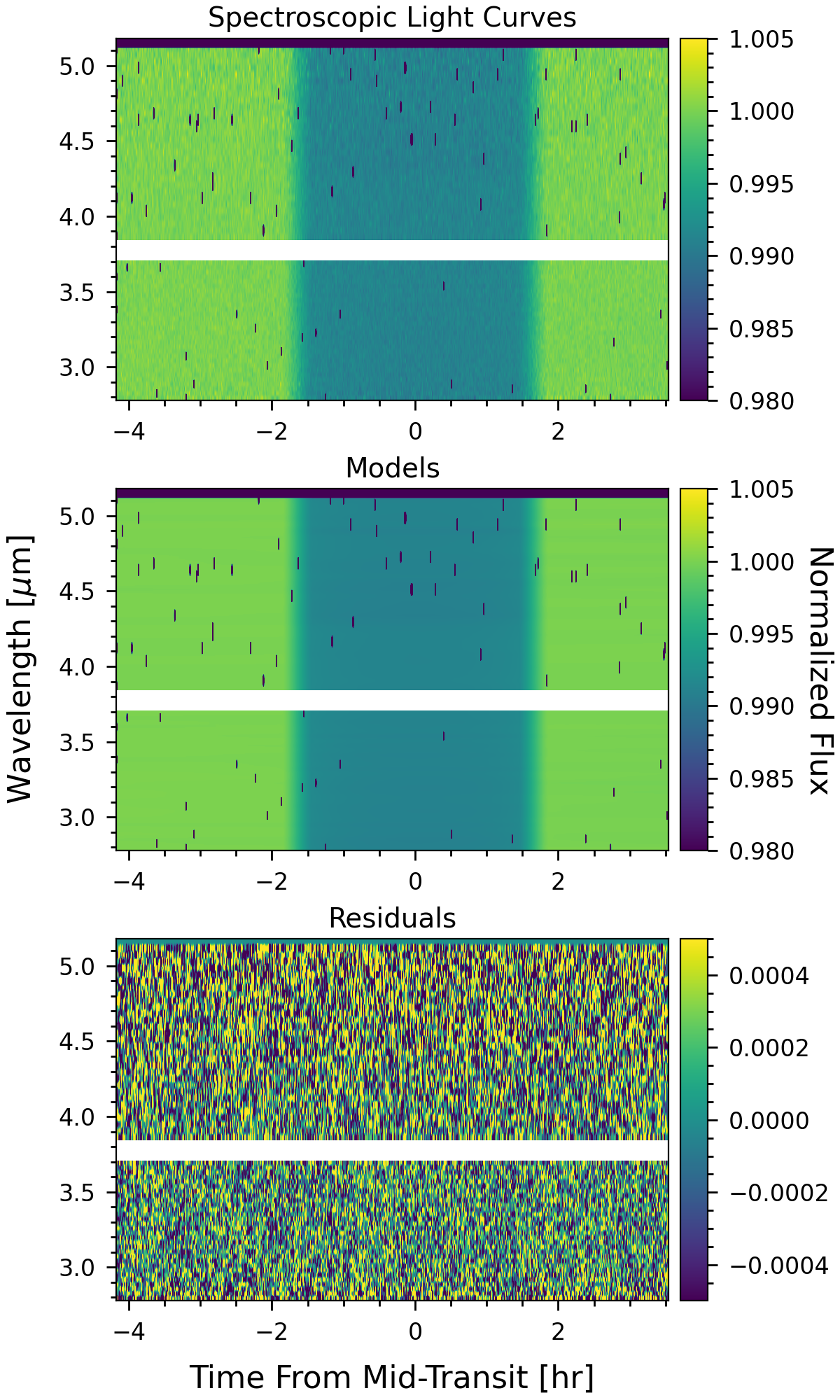}
    \caption{The $R=100$ spectroscopic light curves and fits from the \texttt{Tiberius} reduction. The horizontal white bars correspond to the wavelengths of the gap between the detectors. Top panel: the spectroscopic light curves for NRS1 and NRS2. Middle panel: the best-fitting light curve models. Bottom panel: the residuals from the light curve fits.}
    \label{fig:Tiberius_wb_fit}
\end{figure}

After the white light curve fits, we fit the spectroscopic light curves following the same procedure but with $a/R_*$, $i$ and $T_0$ fixed to the weighted mean values in Table \ref{tab:system_params}. The spectroscopic light curves, models and residuals are shown in Figure \ref{fig:Tiberius_wb_fit}. 

\begin{table*}
    \caption{The resulting system parameters from our fits to the JWST NIRSpec/G395H white light curves. We include the values from the TESS analysis of \protect\cite{Patel2022} for comparison. The TESS $T_0$ has been propagated to the JWST transit epoch and accounts for the uncertainties in $T_0$ and P from \protect\cite{Patel2022}.}
    \label{tab:system_params}
    \centering
    \begin{tabular}{c|c|c|c|c|c} \hline
         Pipeline & Instrument & $T_0$ (BJD) & $R_P/R_*$ & $a/R_*$ & $i$ (\degree) \\ \hline \hline
         \texttt{Tiberius} & NRS1 & $ 2460336.666378 \pm 0.000035 $ & $ 0.092858 \pm 0.000068 $ & $ 7.524 \pm 0.033 $ & $ 86.160 \pm 0.068 $  \\
         \texttt{Tiberius} & NRS2 & $ 2460336.666412 \pm 0.000043 $ & $ 0.092961 \pm 0.000084 $ & $ 7.535 \pm 0.042 $ & $ 86.184 \pm 0.085 $  \\
         \texttt{Tiberius} & Weighted mean & $ 2460336.666392 \pm 0.000027 $ & $ 0.092899 \pm 0.000053 $ & $ 7.528 \pm 0.026 $ & $ 86.167 \pm 0.053 $ \\ \hline
         \texttt{Eureka!} & NRS1 & $2460336.666404 \pm 0.000038$ & $0.093377 \pm 0.000089$ & $ 7.522 \pm 0.064 $ & $ 86.130 \pm 0.065 $ \\
         \texttt{Eureka!} & NRS2 & $2460336.666484 \pm 0.000053$ & $0.093759 \pm 0.000098$ & $ 7.545_{-0.037}^{+0.035} $ & $ 86.217_{-0.074}^{+0.070} $\\ 
         \texttt{Eureka!} & Weighted mean & $2460336.666441 \pm 0.000033$ & $0.093577 \pm 0.000070$ & $7.536_{-0.024}^{+0.023}$ & $86.177_{-0.050}^{+0.047}$  \\ \hline
         \cite{Patel2022} & TESS & $ 2460336.667068 \pm 0.004240 $ & $0.0938${\raisebox{0.5ex}{\tiny$\substack{+0.0009 \\ -0.0012}$}} & $ 7.59^{+0.33}_{-0.28}$ & $86.22${\raisebox{0.5ex}{\tiny$\substack{+0.44 \\ -0.63}$}}  \\ \hline
    \end{tabular}
\end{table*}

\subsection{Eureka! reduction}
\label{sec:eureka_reduction}

We utilised the open-source pipeline \texttt{Eureka!} \citep[v0.11.dev245+ge8ea1d1c.d20240701;][]{Bell2022Eureka} to reduce our data for two additional analyses. \texttt{Eureka!} has been successfully applied to JWST data sets and benchmarked against other pipelines \citep[e.g.,][]{Ahrer2023, MoranStevenson2023}. We describe the principal \texttt{Eureka!} reduction in the following section. The second \texttt{Eureka!} reduction was done independently with a different choice of reduction parameters and is presented in Appendix \ref{sec:eureka_VP}. 

\subsubsection{Light curve extraction}

We started our analysis with the raw images (\texttt{uncal.fits} files) and ran Stages\,1 and 2 of \texttt{Eureka!}, which are wrapped around the default \texttt{jwst} pipeline (v1.12.2) steps. We followed the default steps similar to the \texttt{Tiberius} but including the \texttt{jump} step with a threshold of $10$\,$\sigma$ which is larger than the \texttt{jwst} pipeline default value. We used the additional 1/f background subtraction at the group-level using the routine in \texttt{Eureka!} before the ramp fit and we opted to use a custom scale factor (using a smoothing filter calculated from the first group) for the bias correction. We performed this step because it has been found to minimise transit depth offsets between the NRS1 and NRS2 detectors in other datasets \citep{MoranStevenson2023}.

\texttt{Eureka!}'s Stage\,3 performs the spectral extraction of the data. First, we rejected outliers $>3$ times the median absolute deviation in the spatial direction and performed double-iterative masking of $>5\sigma$ outliers along the time axis. We also masked bad pixels flagged by the \texttt{jwst} pipeline's data quality (\texttt{dq\_init}) step.  This is followed by a correction for the curvature of the spectral trace and a median-subtraction of the background for each frame using the area $>8$ pixels away from the central pixel of the spectral trace (i.e., masking the trace with a full width of 17 pixels). Then we used a full width of 9 pixels for the optimal spectral extraction. 

The extracted spectra were generated and binned in Stage\,4, where we clipped $>5\sigma$ outliers from both NRS1 and NRS2 based on a rolling median of 5\,pixels. 

We followed the same binning schemes as in the \texttt{Tiberius} pipeline, i.e., we computed a broadband (white) light curve, as well as $R=100$, $R=400$ and pixel-level light curves. At this point we also generated limb-darkening coefficients using the same approach as in the \texttt{Tiberius} reduction.

\subsubsection{Light curve fitting}

In \texttt{Eureka!}'s Stage\,5 we fit our extracted light curves using a \texttt{batman} transit light curve model \citep{batman} and a linear-in-time polynomial. We used the MCMC sampling algorithm \texttt{emcee} \citep{emcee} where the starting parameter values were set to the results of an initial least-squares fit. 

Similarly to the \texttt{Tiberius} reduction, we fit the NRS1 and NRS2 white light curves independently and retrieved the system parameters $a/R_*$, inclination and mid-transit time. Like \texttt{Tiberius}, we held the period fixed to 3.7520998 days \citep{Patel2022} and assumed a circular orbit. We used the quadratic limb-darkening law, with the first coefficient, $u1$, fixed to the values obtained by \texttt{ExoTiC-LD} \citep{Grant2024_exoticld} using the same stellar parameters as used by \texttt{Tiberius} (Section \ref{sec:tiberius_reduction}) and the same 3D \texttt{Stagger} models \citep{Magic2015}.  However, unlike the \texttt{Tiberius} reduction, we fitted the second coefficient, $u2$, as a free parameter in all light curve fits. The retrieved system parameters for NRS1 and NRS2 were fixed when fitting the spectroscopic light curves for the respective detectors. These system parameters are given in Table \ref{tab:system_params}. The white light curves and best-fit models are shown in Figure \ref{fig:Tiberius_wl_fit}. The Allan variance plots from the white light curve fits are shown in Figure \ref{fig:Allan_variance_WL} and the spectroscopic light curves in Figure \ref{fig:Allan_variance_spec}. 

\subsection{The transmission spectrum of WASP-15\MakeLowercase{b}}
\label{sec:trans_spec}

\begin{figure}
    \centering
    \includegraphics[width=1\columnwidth]{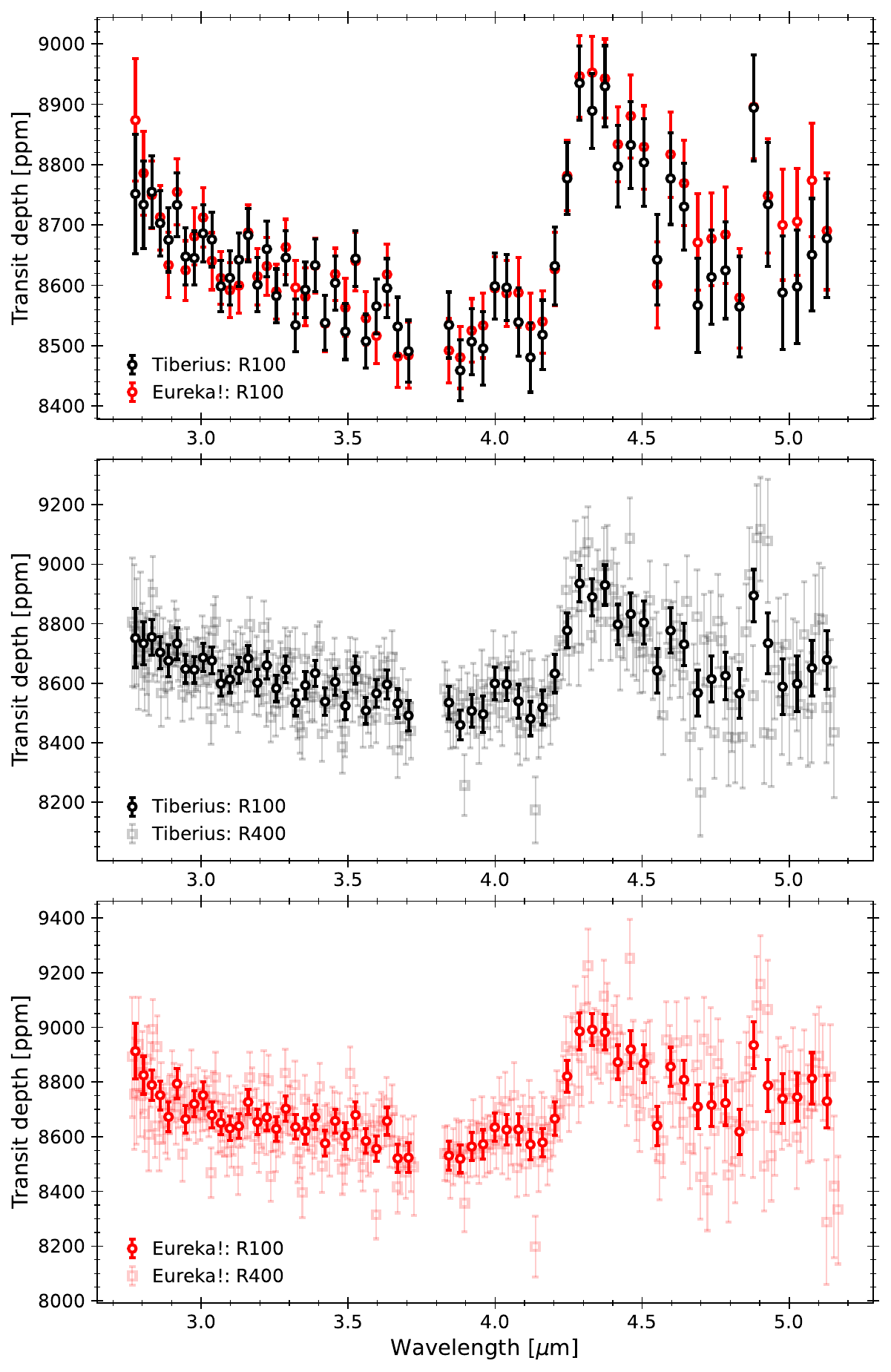}
    \caption{WASP-15b's transmission spectrum. Top panel: the comparison between the spectra obtained with \texttt{Tiberius} (black) and \texttt{Eureka!} (red) at $R=100$. The \texttt{Eureka!} spectrum has been offset by -38\,ppm to match the median transit depth of \texttt{Tiberius}. Middle panel: The spectrum at $R=100$ (black) and $R=400$ (gray squares), both obtained with \texttt{Tiberius}. Bottom panel: The spectrum at $R=100$ (red) and $R=400$ (light red squares), both obtained with \texttt{Eureka!}.}
    \label{fig:trans_spec_R100_R400_comparison}
\end{figure}

Figure \ref{fig:trans_spec_R100_R400_comparison} shows WASP-15b's transmission spectrum at $R=100$ and $R=400$ from \texttt{Tiberius} and \texttt{Eureka!}. There are two things to note from this comparison. Firstly, there is a median transit depth offset between the two pipelines' spectra of 38\,ppm, which is 70\,\% of the median $1\sigma$ transit depth uncertainty of 55\,ppm. Secondly, there is a slope difference between the NRS2 spectra. We discuss the possible causes of these differences in Section \ref{sec:discussion_differences_reduction}. Looking beyond the comparison between the pipelines, the spectra reveal significant absorption features, which we explore in Sections \protect\ref{sec:1D_models} and \ref{sec:GCM}.

\section{Constraints from interior structure models}
\label{sec:interior_model}

Before we infer WASP-15b's atmospheric metallicity from its transmission spectrum, we investigated plausible metallicities for this planet using the interior structure models of \cite{Thorngren2019}. These models solve the 1D structure equations for giant planets, with the hot Jupiter heating power set according to \citet{Thorngren2018}, and a metallicity prior set according to the mass-metallicity relation of \citet{Thorngren2016}.  In order to set an upper limit on the potential atmospheric metallicity, these models assume a fully-mixed interior -- no core to hide additional bulk metal.  

\begin{figure}
    \centering
    \includegraphics[width=1\columnwidth]{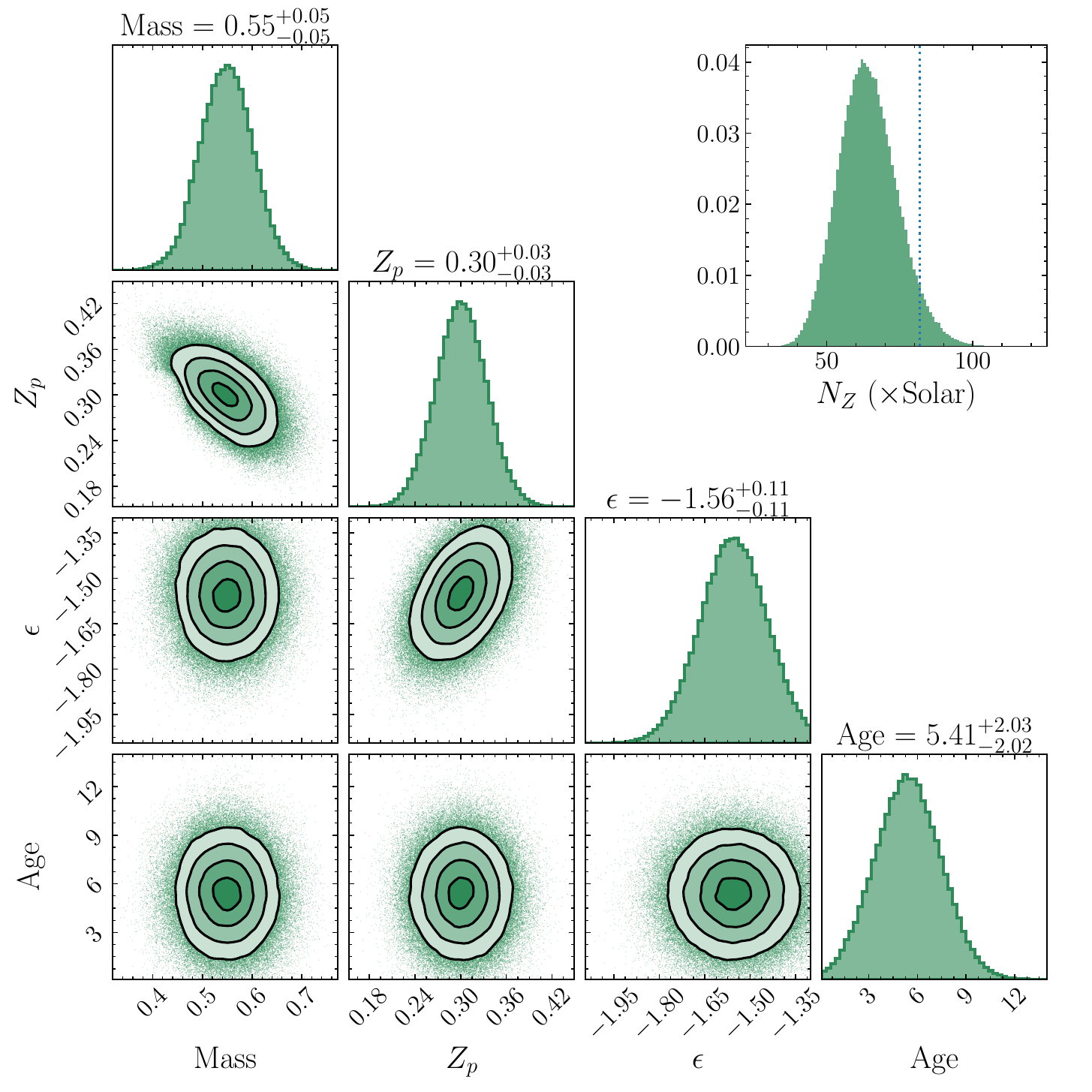}
    \caption{The posterior of the parameters of our interior structure model of WASP-15b: mass ($\mathrm{M_{J}}$), $Z_p$ (unitless mass ratio), $\epsilon$ (unitless heating power parameter), and the age (Gyr).  The interior and statistical models used are identical to that of \citet{Thorngren2019}.  We find that the planet is moderately metal-rich with $Z_p=0.30\pm0.03$, corresponding to a 95\% upper limit on the atmospheric composition of $82\times$ solar.} 
    \label{fig:interior_model}
\end{figure}

We fit the models using the same Bayesian framework as \citet{Thorngren2019}. For the planet's mass, we adopted the value from \cite{Bonomo2017} of $0.542 \pm 0.054$\,\Mjup. For the age, we used a value of $5.40 \pm 2.05$\,Gyr which we estimate as the median of \cite{Bonomo2017}'s asymmetric age distribution ($3.9^{+2.8}_{-1.3}$\,Gyr). However, at these mature ages and for hot Jupiters in particular, the choice of age has little effect on the inferred bulk metallicity. For the planet's radius and flux/irradiation, we use values derived from our JWST data ($1.335 \pm  0.065$\,\Rjup\ and 1.65\,Gerg\,s$^{-1}$\,cm$^{-2}$, respectively\footnote{The planet's radius was derived from our \texttt{Tiberius} $R_P/R_*$, and \cite{Bonomo2017}'s $R_*$. The flux was derived from Gaia DR3's $T_\mathrm{eff}$ \citep{Gaia}, our \texttt{Tiberius} $a/R_*$ and \cite{Bonomo2017}'s $R_*$.}). This method leads us to infer that the bulk metallicity is $Z_p = 0.30 \pm 0.03$ in mass ratio (Figure \ref{fig:interior_model}).  Note that this is the statistical uncertainty and does not account for theoretical uncertainties in e.g. the equations of state used.  Still, it serves as a useful upper limit when converted from mass ratio to number ratio in $\times$ solar units.  Here we take the solar ratio of Z:H to be 0.00104 \citep{Asplund2009}, though note that other authors choose 0.00208 (solar composition gas under planetary conditions); it is important for consistency to be clear which definition is used.  We find that the 95th percentile of the metallicity distribution is $82\times$ solar, corresponding to a mean molecular weight $\mu=3.4$\,amu, which we adopt as an upper limit on plausible atmospheric metallicities.

\section{Interpreting the spectrum with 1D models}
\label{sec:1D_models}

In this section we describe the inferences we make from interpreting the spectrum with 1D atmospheric forward models and retrievals. We describe the results from three independent approaches in this Section and compare these in the Discussion (Section \ref{sec:discussion_differences_retrieval}). 

\subsection{petitRADTRANS forward models}
\label{sec:forward_model_pRT}

We implement a 1D forward model grid of synthetic transmission spectra using the {\tt petitRADTRANS} package \citep[v2.7.7,][]{molliere2019petitradtrans,Nasedkin2024}. 

In both our forward models and retrievals with \texttt{petitRADTRANS} (Section \ref{sec:retrievals_pRT}), we use correlated-$k$ radiative transfer with opacity tables at $R=1000$ to calculate the transmission spectra. The opacity tables were pre-computed from spectral lines, using the following line-lists from HITEMP: \ch{H2O} and \ch{CO} \citep{Rothman2010}; and ExoMol: \ch{CO2} \citep{Yurchenko2020}, \ch{SO2} \citep{Underwood2016}, \ch{CH4} \citep{yurchenko2017hybrid}, \ch{H2S} \citep{azzam16exomol}, \ch{OCS} \citep{owens2024exomol}, \ch{HCN} \citep{barber2014exomol}, \ch{C2H2} \citep{chubb2021exomolop}, \ch{SO} \citep{brady2024exomol}, and \ch{NH3} \citep{coles2019exomol}. We assume a \ch{H2}- and He-dominated atmosphere and include opacity from \ch{H2-H2} and \ch{H2-He} collision-induced absorption as well as Rayleigh scattering from \ch{H2} and He.

\begin{figure}
    \centering
    \includegraphics[width=\columnwidth]{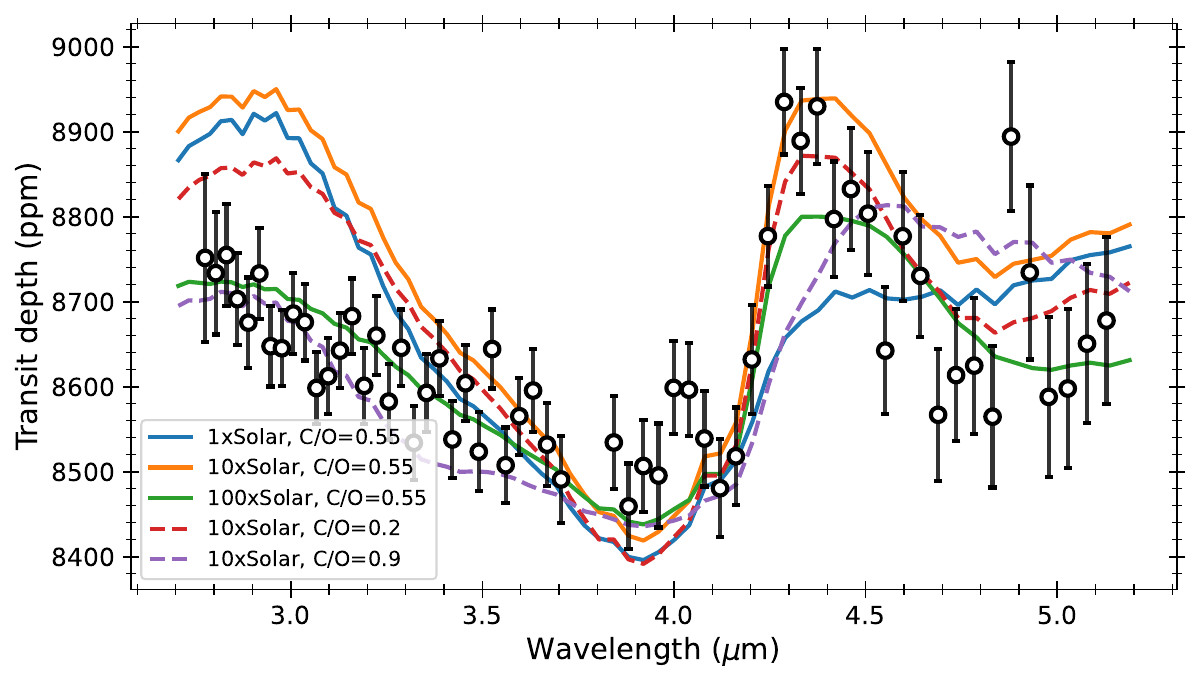}
    \includegraphics[width=\columnwidth]{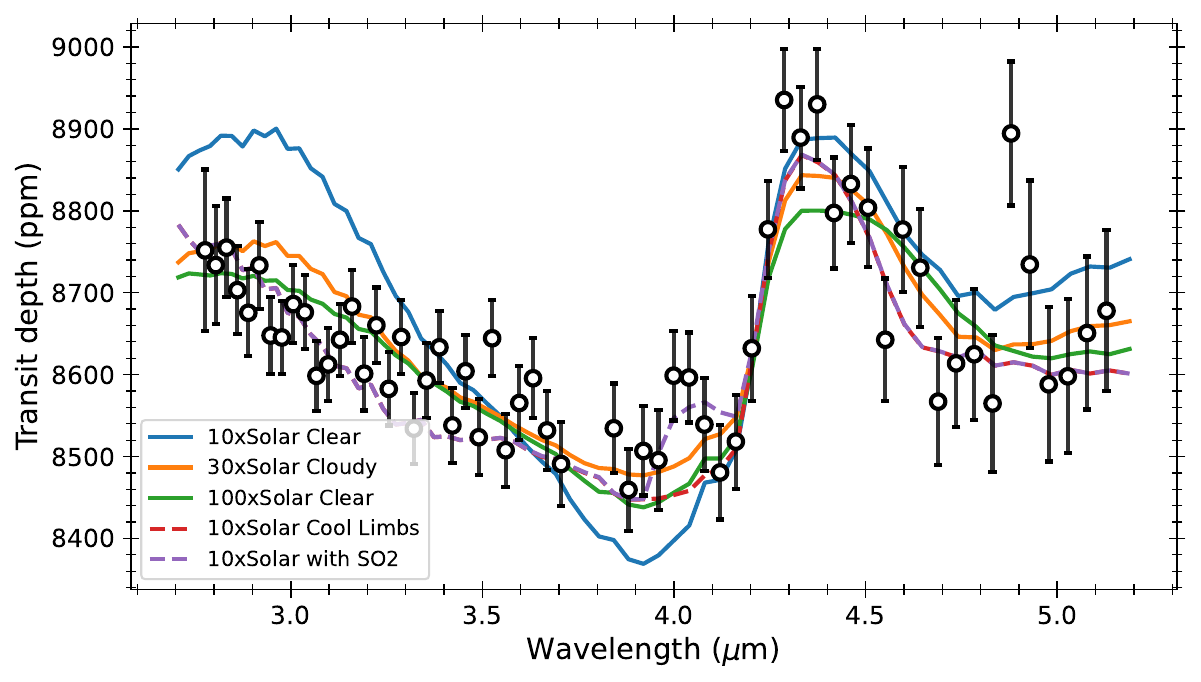}
    \caption{Top panel: Selected synthetic transmission spectra from our \texttt{petitRADTRANS} 1D equilibrium chemistry forward model grid, plotted against our \texttt{Tiberius} $R=100$ spectrum of WASP-15b. The temperature profile is fixed to an isotherm at $T_{\mathrm{eq}}=1676$\,K. We find that a $\sim$solar  C/O ratio, and a metallicity $>10\times$ solar is required to give the appropriate feature shapes, identifiable as \ch{H2O} and \ch{CO2}, but we are not able to match the feature amplitudes with this simple model. Bottom panel: Selected synthetic transmission spectra from our expanded \texttt{petitRADTRANS} 1D equilibrium chemistry forward model grid. This demonstrates that the inclusion of higher altitude clouds at 0.1 mbar (the cloudy model, orange) or fixing the isothermal temperature to a cooler value of 1000 K (the cool limbs model, green), combined with a $10-50\times$ solar metallicity and $\mathrm{C/O}=0.55$ allows us to more successfully fit the feature amplitude compared to the original grid with no clouds and equilibrium temperature limbs (top panel). We additionally include a model with 5\,ppm of \ch{SO2} (purple dashed line, bottom panel).}
    \label{fig:prt_forward_model1}
\end{figure}

We use an initial grid of equilibrium chemistry simulations, with an isothermal temperature profile and the radius, mass, and equilibrium temperature of WASP-15b reported in Section \ref{sec:intro}, varying the C/O ratio from 0.1--1.0 and the metallicity from 0.1--100$\times$ solar. The equilibrium chemical composition is interpolated from a large grid pre-calculated using \texttt{easyCHEM}, as described in \citet{molliere2017observing}. We use the parametrization described in \citet{molliere2015model} using metallicity [Fe/H] and C/O ratio. The abundance of all atoms, including the C/H ratio, is scaled by metallicity relative to solar \citep[as defined in][]{Asplund2009}, except the O/H ratio, which is further scaled by the C/O ratio relative to the C/H ratio.

Comparing this forward model grid by eye to both the \texttt{Tiberius} and \texttt{Eureka!} $R=100$ spectra revealed that a super-solar metallicity of 10 -- 100$\times$ solar \citep{Asplund2009}, and a C/O ratio from 0.2 -- 0.8 was required to give a similar spectrum, as seen in Figure \ref{fig:prt_forward_model1} (top panel). Leaving out opacity from one molecule at a time from the model revealed \ch{H2O} and \ch{CO2} as the primary opacity sources, with potential minor contributions from CO and \ch{H2S}. None of these clear atmosphere simulations were simultaneously able to predict the strength of the main \ch{H2O} and \ch{CO2} features, with $<80\times$ solar metallicity models over-predicting the strength of the \ch{H2O} feature, and $>50\times$ solar metallicity models under-predicting the strength of the \ch{CO2} feature. We therefore extended our grid with simulations with a grey cloud deck at 0.1\,mbar, a grey cloud deck at 10\,mbar, or a cooler 1000\,K isothermal atmosphere. We found that 0.1\,mbar clouds, a cooler terminator temperature, or some combination thereof create a good fit to both features, as portrayed in Figure \ref{fig:prt_forward_model1} (bottom panel).

We were unable to generate satisfactory fits to the apparent features at 4.0\,$\mu$m and 4.9\,$\mu$m using equilibrium chemistry models. Motivated by the detection of \ch{SO2} at 4.0\,\micron\ in the transmission of WASP-39b \citep{Tsai2023}, we tested this scenario for WASP-15b and found that the inclusion of a modest 5\,ppm abundance of \ch{SO2} can fit the feature (purple dashed line, Figure \ref{fig:prt_forward_model1}, bottom panel), as described in Section \ref{sec:retrievals_pRT}. We tested a wide variety of molecules to explain the feature at 4.9\,$\mu$m. We found that both OCS and \ch{O3} had absorption centred at the right wavelengths but were generally broader than the feature width. We discuss the interpretation of this feature in more detail in Section \ref{subsec:OCS}. Using the {\tt TriArc} package \citep{claringbold2023prebiosignature}, we found that, given the equilibrium chemistry forward model and precision of the observation, the abundances of \ch{SO2} and OCS would need to be $\geq 10$ and 1\,ppm to constitute 3$\sigma$ detections.


\subsection{petitRADTRANS retrievals}
\label{sec:retrievals_pRT}

For a more detailed exploration of our transmission spectrum, we performed free chemistry and equilibrium chemistry retrievals on the $R=100$ spectra from both \texttt{Tiberius} and \texttt{Eureka!}. \texttt{petitRADTRANS} uses Bayesian nested-sampling \citep{Skilling2004} implemented through \texttt{MultiNest} \citep{Feroz2008} with \texttt{PyMultiNest} \citep{buchner2014x}. We used the same correlated-$k$ opacities and isothermal temperature structure as the forward models of Section \ref{sec:forward_model_pRT}. We also included a grey cloud deck, with the cloud-top pressure as a free parameter, with a log uniform prior from 1 $\mu$bar to 100 bar. We used a Gaussian prior for the planet's gravity based on \citet{Bonomo2017}, and wide, uniform priors for the planet's radius ($0.8-1.8$ \Rjup) and limb temperature ($500-3000$ K). The stellar radius was fixed to 1.477\,$R_{\odot}$. We experimented with fitting the reference pressure at fixed planetary radius, to determine that a reference pressure of 1\,mbar, at which the radius and gravity were defined, was most appropriate. We used a wide prior on the planet's radius because of our decision to fix the reference pressure in our retrieval; as the true reference pressure is unknown, we cannot use our prior knowledge of the radius in the retrieval without potentially biasing it. This is the most common approach for atmospheric retrievals \citep[e.g.,][]{Alderson2022,taylor2023awesome,banerjee2024atmospheric}. We also ran a retrieval with a tight radius prior based on the white-light transit depth and a uniform reference pressure prior and found that it didn't affect our results. 

For our free chemistry retrievals, we permitted the log volume mixing ratio to vary from $-12$ to $-0.5$ for each molecule. For our chemical equilibrium retrievals, the elemental ratios were parameterized by C/O ratio and metallicity, as previously described. We used a uniform prior of 0.1 to 1.5 for C/O, and a log uniform prior of $-2$ to $3$ on metallicity. Given the strong impact of photochemistry on sulphur species, which is not accounted for by chemical equilibrium, we also ran a hybrid retrieval where the \ch{H2O}, \ch{CH4}, \ch{CO}, and \ch{CO2} abundances are determined by chemical equilibrium, while the photochemically-active sulphur species \ch{H2S}, \ch{SO2}, and OCS abundances are free parameters. 

\begin{figure}
     \centering
     \includegraphics[width=\columnwidth]{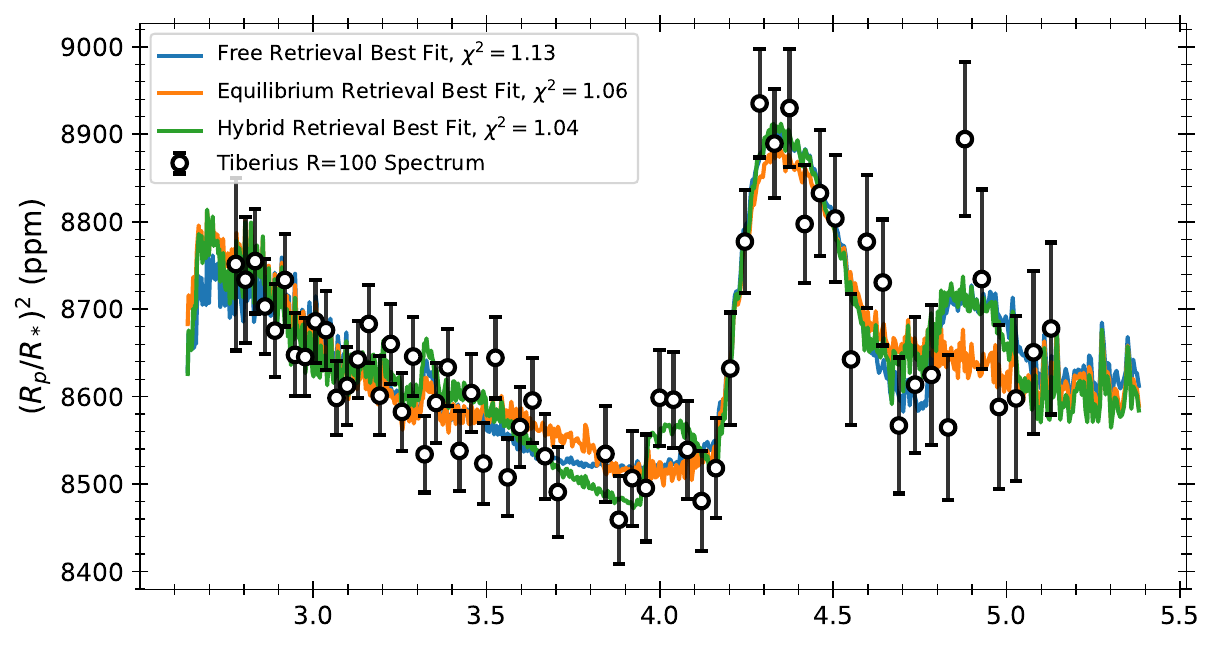}
     \includegraphics[width=\columnwidth]{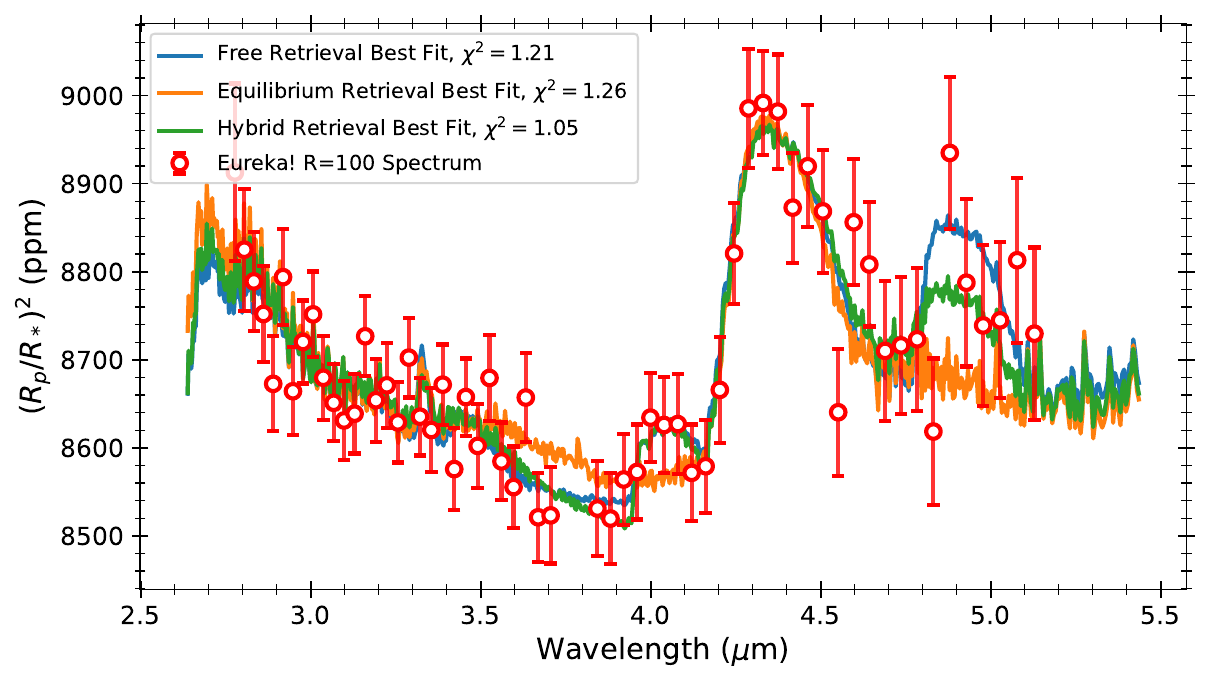}
     \caption{Best-fitting models from our \texttt{petitRADTRANS} free chemistry (blue), equilibrium chemistry (orange), and hybrid chemistry (equilibrium chemistry plus free S-bearing species, green) retrievals, fitted to the \texttt{Tiberius} $R=100$ reduction (top panel) and \texttt{Eureka!} $R=100$ reduction (bottom panel).}
     \label{fig:prt_best_fit_models}
 \end{figure}

For each pair of retrievals, we found complete consistency in the retrieved parameters between the \texttt{Tiberius} and \texttt{Eureka!} spectra. The complete retrieval results are summarised in Table \ref{tab:all_retrievals}. The best-fit models from each retrieval setup are shown in Figure \ref{fig:prt_best_fit_models}. 

Our equilibrium chemistry retrievals infer a C/O ratio of $0.48${\raisebox{0.5ex}{\tiny$\substack{+0.11 \\ -0.16}$}} for the \texttt{Tiberius} spectrum and $0.53${\raisebox{0.5ex}{\tiny$\substack{+0.09 \\ -0.16}$}} for the \texttt{Eureka!} spectrum, consistent with the solar value of 0.55 \citep[using the solar metallicity of][]{Asplund2009}. The metallicity was determined to be super-solar, at $18${\raisebox{0.5ex}{\tiny$\substack{+22 \\ -8}$}}$\times$ solar and
$22${\raisebox{0.5ex}{\tiny$\substack{+27 \\ -9}$}}$\times$ solar, respectively. We also retrieved a limb temperature of $\sim900$ K, much colder than the equilibrium temperature, and minimal constraints on the cloud-top pressure, with a 3$\sigma$ lower limit of 0.1 mbar.

Our free chemistry retrievals yielded similar results, ruling out \ch{H2O} and \ch{CO2} abundances of less than 1\,ppm. The log vertical mixing ratios were found to be $-2.83${\raisebox{0.5ex}{\tiny$\substack{+0.59 \\ -0.95}$}} for \ch{H2O} and $-4.40${\raisebox{0.5ex}{\tiny$\substack{+0.73 \\ -1.02}$}} for \ch{CO2} in the \texttt{Tiberius} spectrum, with similar results for the \texttt{Eureka!} spectrum. There were no firm constraints for other species in the \texttt{Tiberius} spectrum, but the posterior corner plot from the \texttt{Eureka!} spectrum (see Figure \ref{fig:prt_free_both_corner}) indicates that one of CO or OCS is likely present in the spectrum (the posteriors are inconsistent with neither being present to $\>3\sigma$), but they cannot be distinguished. Our free chemistry retrievals favoured a cloudy atmosphere, placing a maximum cloud-top pressure of $\sim0.1$ bar.

\begin{figure*}
    \centering
    \settototalheight{\dimen0}{\includegraphics[width=\textwidth]{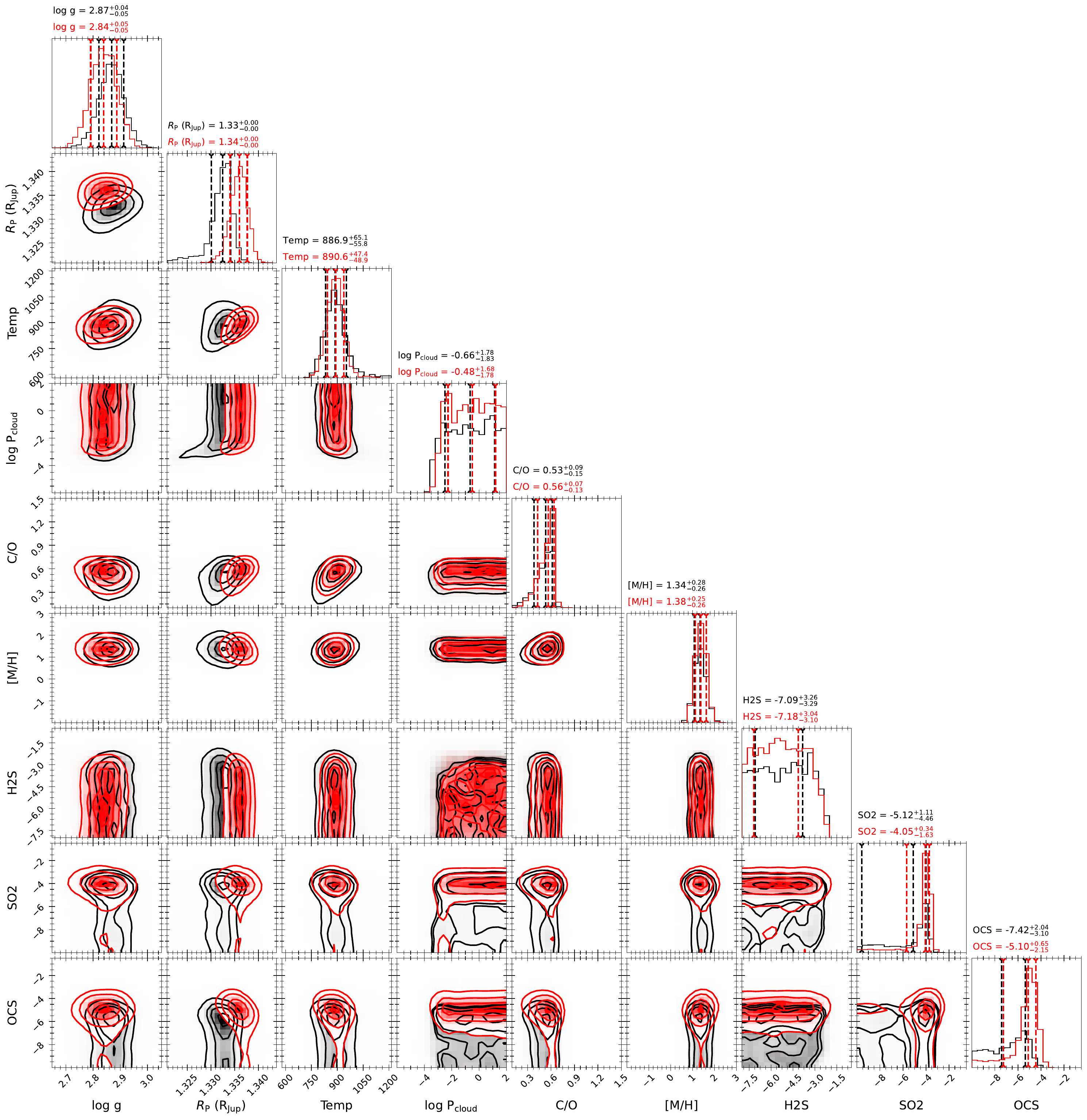}}
    \includegraphics[width=\textwidth]{figures/prt_hybrid_both_corner.pdf}%
    \llap{\raisebox{\dimen0-5cm}{%
    \includegraphics[height=5cm]{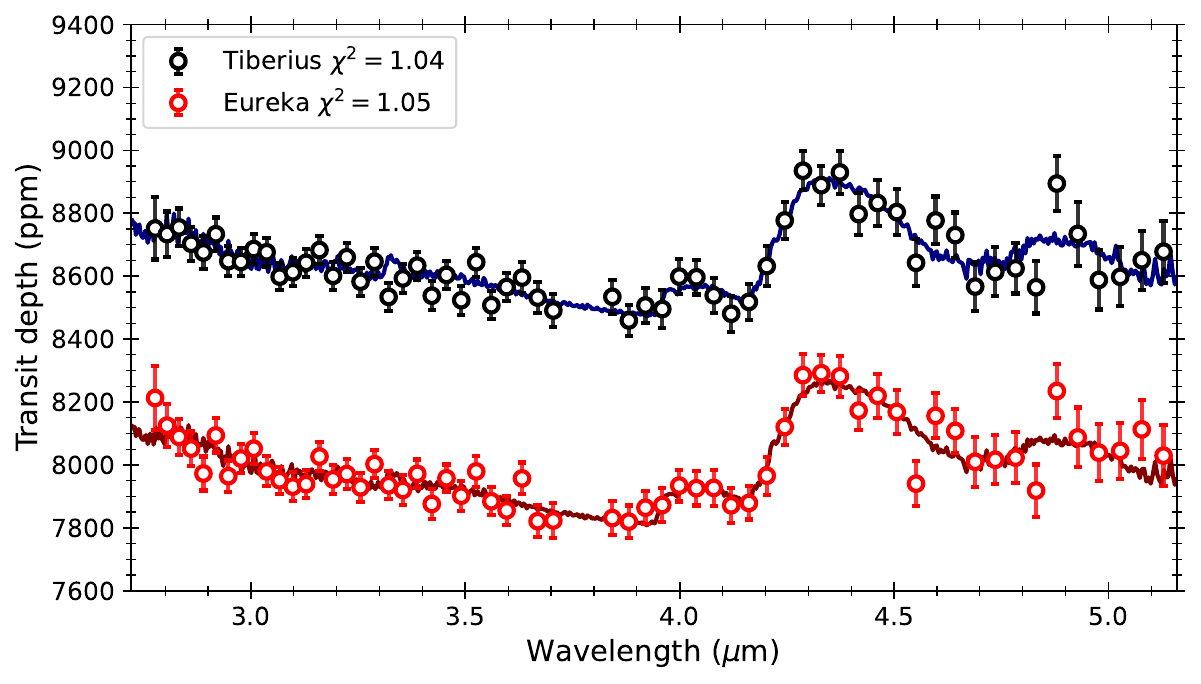}%
    }}
    \caption{Corner plot showing the posterior PDFs from the hybrid \texttt{petitRADTRANS} retrieval performed on the \texttt{Tiberius} $R=100$ spectrum (black) and \texttt{Eureka!} $R=100$ spectrum (red). In the hybrid retrievals, the abundances of \ch{CH4}, \ch{CO2}, \ch{CO}, and \ch{H2O} are fixed to their abundances at chemical equilibrium, while the abundances of S-bearing species \ch{H2S}, \ch{SO2}, and OCS are free parameters. The best fitting model and residuals are displayed in the top right.}
    \label{fig:prt_Tiberius_hybrid_corner}
\end{figure*}

Our hybrid retrievals provided the best fit to the data, with reduced $\chi_\nu^2=1.04$ for \texttt{Tiberius} and $\chi_\nu^2=1.05$ for \texttt{Eureka!}, as depicted in Figure \ref{fig:prt_best_fit_models}. By treating the sulphur species as free parameters, we could fit the features at 4.0 $\mu$m and 4.9 $\mu$m with \ch{SO2} and OCS, respectively, while also permitting \ch{H2S} to be depleted due to photochemical destruction or a low S/H ratio, impacting the spectrum either side of the detector gap. This approach resulted in more precise constraints on the C/O ratio, with a retrieved C/O ratio of $0.53${\raisebox{0.5ex}{\tiny$\substack{+0.09 \\ -0.15}$}} for the \texttt{Tiberius} spectrum and $0.56${\raisebox{0.5ex}{\tiny$\substack{+0.07 \\ -0.13}$}} for the \texttt{Eureka!} spectrum. The corner plot for the hydrid retrieval is shown in Figure \ref{fig:prt_Tiberius_hybrid_corner}.

We also ran additional retrievals for the purpose of model comparison. By running additional free chemistry retrievals omitting a single species on the \texttt{Tiberius} $R=100$ spectrum and comparing the Bayesian evidence \citep{trotta2008bayes,benneke2013distinguish}, we were able to place the detection significance of \ch{CO2} at 8.9$\sigma$, and \ch{H2O} at 4.2$\sigma$. We also repeated the hybrid chemistry retrievals, omitting in turn \ch{SO2}, \ch{OCS}, and then both. We present a comparison of retrieval evidences in Table \ref{tab:prt_evidences}. In the \texttt{Eureka!} spectrum, $\Delta\ln{Z}=+2.2$ between the equilibrium and hybrid retrievals, equivalent to a 2.6$\sigma$ preference. This is the result of small improvements of evidence, none of which are individually significant, from including each of \ch{H2S}, \ch{SO2}, and \ch{OCS} as free parameters. The \texttt{Tiberius} spectrum is agnostic between equilibrium and hybrid models, with a $\Delta\ln{Z}$ of only $+0.5$ between the best (hybrid with \ch{H2S} and OCS) and worst (hybrid with \ch{H2S}, \ch{SO2}, and OCS) models. We also included hybrid retrievals on the $R=400$ spectra, to determine if the additional resolution could better resolve the \ch{SO2} and OCS features. These results were completely consistent with the $R=100$ retrievals, with the evidence marginally improving for each S-species added with the \texttt{Eureka!} spectrum, but changing by less than 0.4 between different \texttt{Tiberius} setups.

\begin{table}
    \centering
    \caption{The Bayesian evidence ($\ln{Z}$) and degrees of freedom (DOF) from each of our \texttt{petitRADTRANS} retrievals, including equilibrium chemistry, free chemistry, and hybrid retrievals with equilibrium \ch{CH4}, \ch{H2O}, \ch{CO}, and \ch{CO2} abundances, but the abundance of specified S-bearing species as free parameters. The evidence is very agnostic about the inclusion of various S-species in the \texttt{Tiberius} spectrum, but there is a moderate ($2.6\sigma$) preference for the hybrid chemistry model in the \texttt{Eureka!} spectrum, implying some combination of \ch{H2S} depletion, \ch{OCS} enrichment, and the presence of \ch{SO2}.}
    \label{tab:prt_evidences}
    \begin{tabular}{llll}
    \hline Retrieval          & \texttt{Tiberius} $\ln{Z}$ & \texttt{Eureka!} $\ln{Z}$ & DOF\\
    \hline Equilibrium      & -40.0    & -47.5 & 6    \\
    Hybrid (\ch{H2S} only)     & -39.9    & -46.3 & 7   \\
    Hybrid (\ch{H2S}+\ch{SO2}) & -40.0    & -46.0  & 8 \\
    Hybrid (\ch{H2S}+\ch{OCS}) & -39.8    & -45.7 & 8  \\
    Hybrid (\ch{H2S}+\ch{OCS}+\ch{SO2})             & -40.3    & -45.3 & 9  \\
    Free                       & -43.0    & -48.2 & 11\\
    \hline
    \end{tabular}
\end{table}

\subsection{BeAR retrieval}
\label{sec:retrievals_bear}

As a comparison, we also perform free-chemistry retrievals with the GPU-accelerated open-source Bern Atmospheric Retrieval code (\texttt{BeAR}\footnote{\texttt{BeAR} can be found at \url{https://github.com/newstrangeworlds/bear}}). This is an updated version of the retrieval code previously known as \texttt{Helios-r2} \citep{Kitzmann2020}. \texttt{BeAR} uses the \texttt{MultiNest} library \citep{Feroz2008} to perform the retrieval using Bayesian nested-sampling \citep{Skilling2004}. It uses line-by-line opacity sampling, which we sampled at a resolution of \SI{1.0}{\per\cm} in wavenumber (equivalent to $R\sim2500$), and in this work we include the following molecules and their associated \texttt{ExoMol} line-lists: \ch{H2O} \citep{Polyansky2018}, \ch{CH4} \citep{Yurchenko2013}, \ch{CO} \citep{Li2015}, \ch{CO2} \citep{Yurchenko2020}, \ch{SO2} \citep{Underwood2016}, \ch{H2S} \citep{azzam16exomol}, and \ch{OCS} \citep{Wilzewski2016}. We also include opacity due to collision-induced absorption from \ch{H2}-\ch{H2} \citep{Abel2011} and \ch{H2}-\ch{He} \citep{Abel2012}, as well as Rayleigh scattering from \ch{H2} \citep{Vardya1962}. We assume free-chemistry, for which the molecular abundances are allowed to vary from volume mixing ratios of $10^{-12}$ to 1.0. The rest of the atmosphere is composed of \ch{H2} and \ch{He}, assuming a solar ratio of 0.17 \citep{Asplund2009}. The atmosphere is divided into 200 equal layers in log-pressure space, assuming a top pressure of $10^{-8}$ bar, and a bottom pressure of 10 bar. We assume an isothermal temperature, sampled between 500 and 2500 K, and include a grey cloud deck, for which we retrieve a cloud-top pressure sampled between $10^{-5}$ and 10 bar. The stellar radius is fixed to 1.477 $R_{\odot}$, and the planet radius and gravity are free parameters. The planet radius uses a uniform prior of 1.27--1.40 $R_{\rm J}$, and the gravity uses a Gaussian prior on $\log{g}$ with a mean of 2.828 and a standard deviation of 0.041, in cgs units. All \texttt{BeAR} retrievals in this work use 1000 live points. 

We applied this retrieval code to both the \texttt{Tiberius} and \texttt{Eureka!} reductions at $R=100$ and $R=400$. For the $R=100$ case (shown in Figure \ref{fig:BeAR_cornerplot_R100_unrestricted}), we find a bimodal distribution for a number of parameters, representing two families of models. The first, favoured by the \texttt{Eureka!} data, is composed of \ch{H2O}, \ch{CO}, and \ch{CO2}, with a moderate water volume mixing ratio of $\sim10^{-5}$. The second, favoured by the \texttt{Tiberius} data, only shows evidence for \ch{H2O} and \ch{CO2}, but now the water abundance is $\sim15$\%. This corresponds to a very high-metallicity scenario, which is unphysical according to our interior structure model (see Section \ref{sec:interior_model}). For the $R=400$ case, the retrieval favours the higher metallicity model in both reduction cases (not shown). 

\begin{figure*}
    \centering
    \settototalheight{\dimen0}{\includegraphics[width=\textwidth]{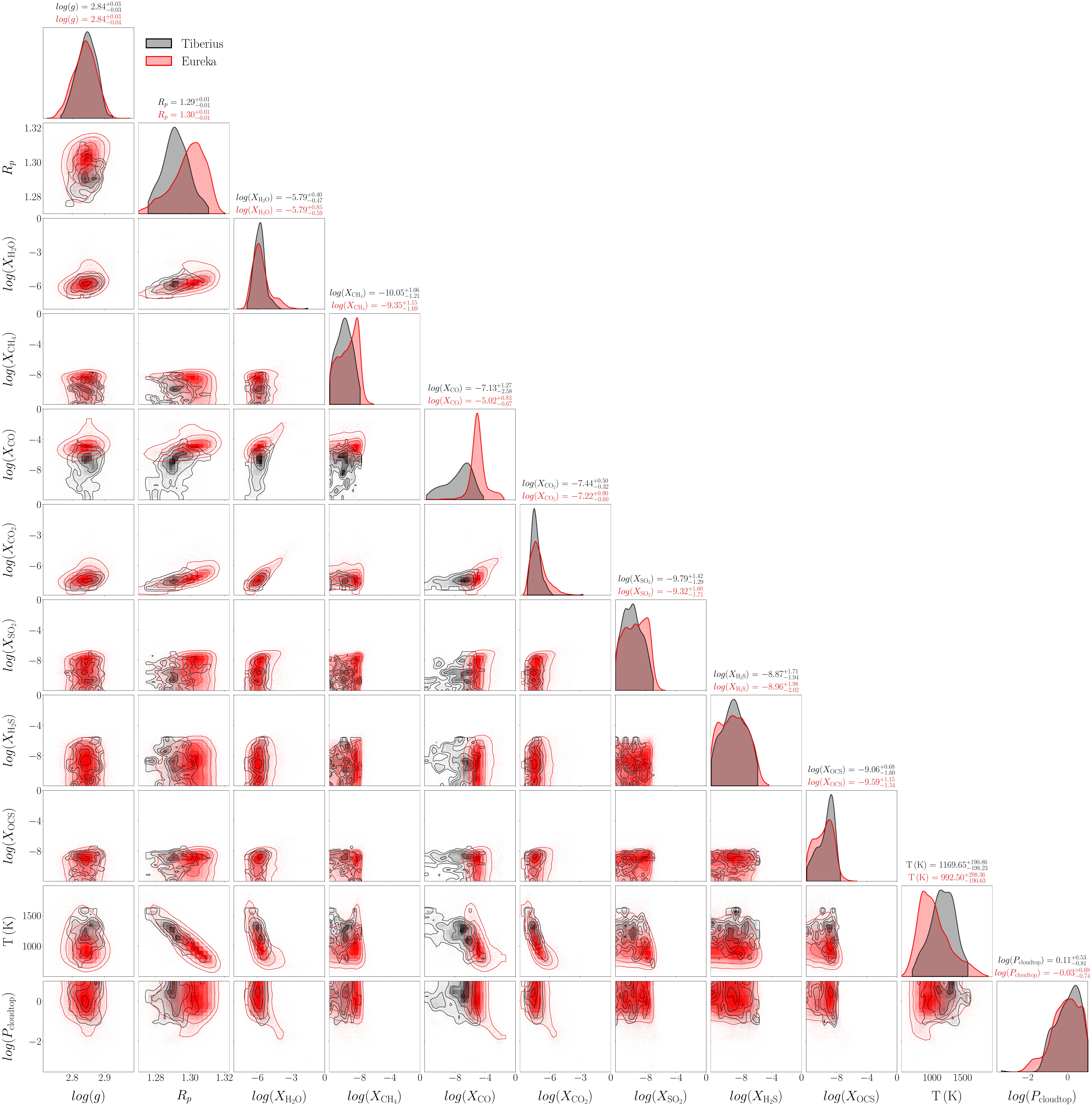}}
    \includegraphics[width=\textwidth]{figures/BeAR_retrieval_comparison_R100_lowmet.pdf}%
    \llap{\raisebox{\dimen0-6cm}{%
    \includegraphics[height=6cm]{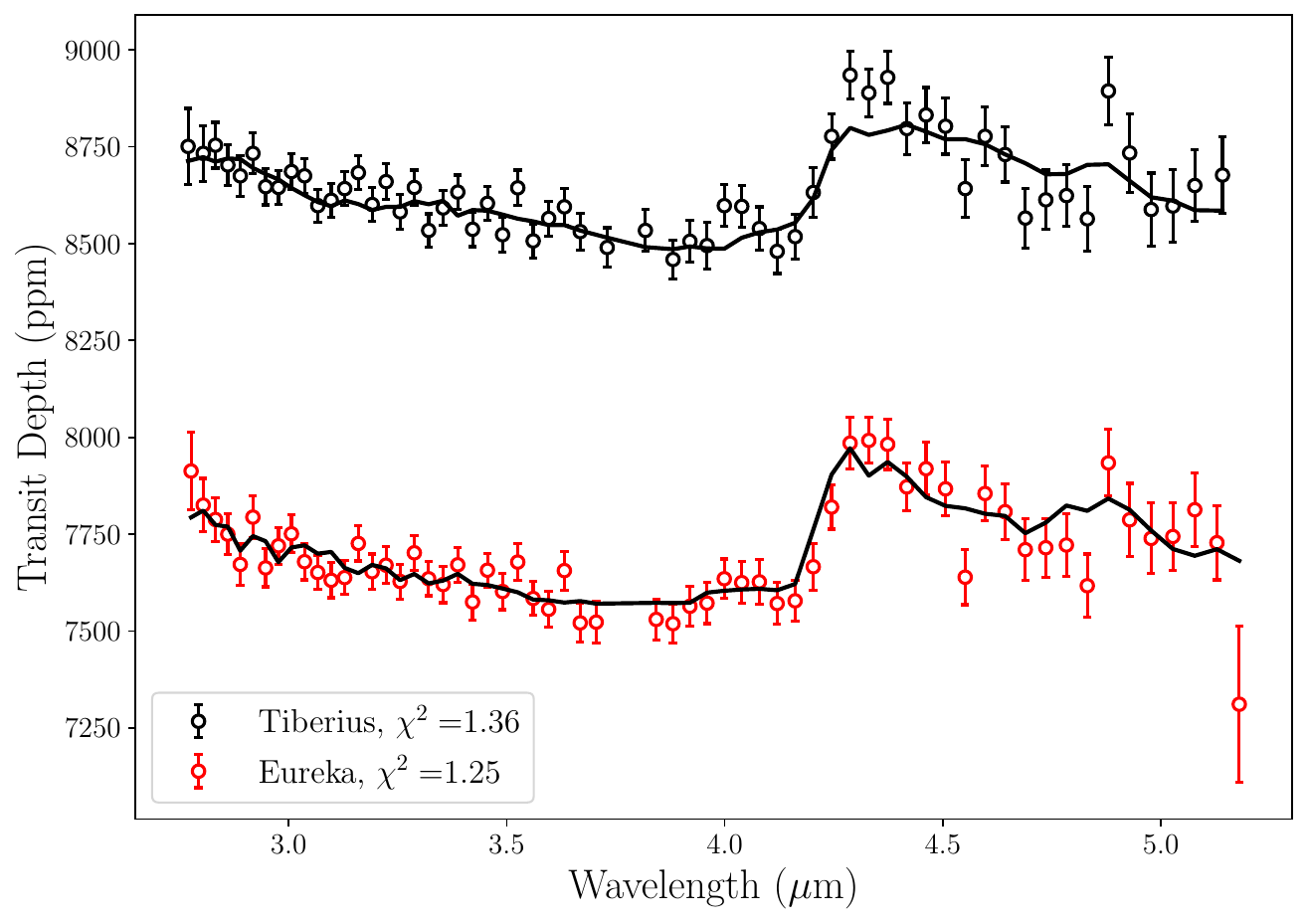}%
    }}
    \caption{Cornerplot showing the posteriors from the BeAR retrievals on the \texttt{Tiberius} (black) and \texttt{Eureka!} (red) reductions of WASP-15 b at $R=100$, restricted to the metallicities allowed by the interior structure model. The top right insert shows the best-fit models for the \texttt{Tiberius} (black) and \texttt{Eureka!} (red) reductions. The \texttt{Eureka!} spectrum is offset by 1000 ppm for visualisation purposes. The legend in the bottom left indicates the reduced $\chi^2$ values for each of the fits.}
    \label{fig:BeAR_cornerplot_R100}
\end{figure*}

Motivated by the upper limits on WASP-15 b's atmospheric metallicity and mean molecular weight derived from our interior structure model, we recomputed our \texttt{BeAR} posteriors after excluding the high metallicity modes from our posteriors. Specifically, we summed the mass mixing ratios of the molecules and limited the metal-mass ratio to 0.36 (equal to the $2\sigma$ upper limit on $Z_p$, Section \ref{sec:interior_model}). The resulting posteriors are shown in Figure \ref{fig:BeAR_cornerplot_R100}. Now we see a good agreement between the \texttt{Eureka!} and \texttt{Tiberius} retrievals, with a water volume mixing ratio of $\sim10^{-6}$. In agreement with \texttt{petitRADTRANS}, we retrieve limb temperatures of $\sim900-1100$ K. The cloud-top pressure is unconstrained in both reduction cases. The upper-right insert of Figure \ref{fig:BeAR_cornerplot_R100} shows the best-fit spectra for each of the retrievals, as well as the reduced $\chi^2$ values for these fits.

\subsection{PLATON retrieval}
\label{sec:retrievals_platon}

We also used the open source \texttt{PLATON} package \citep{Zhang2019,Zhang2020} to interpret our transmission spectra. \texttt{PLATON} assumes equilibrium chemistry in 1D and an isothermal pressure-temperature profile. In its default configuration, which we used here, \texttt{PLATON} includes opacities from 34 chemical species with equilibrium abundances calculated using \texttt{GGchem} \citep{Woitke2018} over a large grid of metallicities, C/O, temperatures, and pressures. The full list of species, along with line lists, is given in \cite{Zhang2020}. The dominant species at the wavelengths and temperatures we are concerned with here are: H$_2$O \citep{Polyansky2018}, CO$_2$ \citep{Tashkun2011} and CO \citep{Faure2013,Gordon2017}. \texttt{PLATON} also includes SO$_2$ \citep{Underwood2016} by default but not OCS which, as we discuss in Section \ref{subsec:OCS}, may be responsible for the feature at 4.9\,\micron. For our analysis, we used the line lists generated at a spectral resolution of $R=10000$ and the opacity sampling method of radiative transfer, rather than correlated-$k$. 

\begin{figure*}
    \centering
    \settototalheight{\dimen0}{\includegraphics[width=\textwidth]{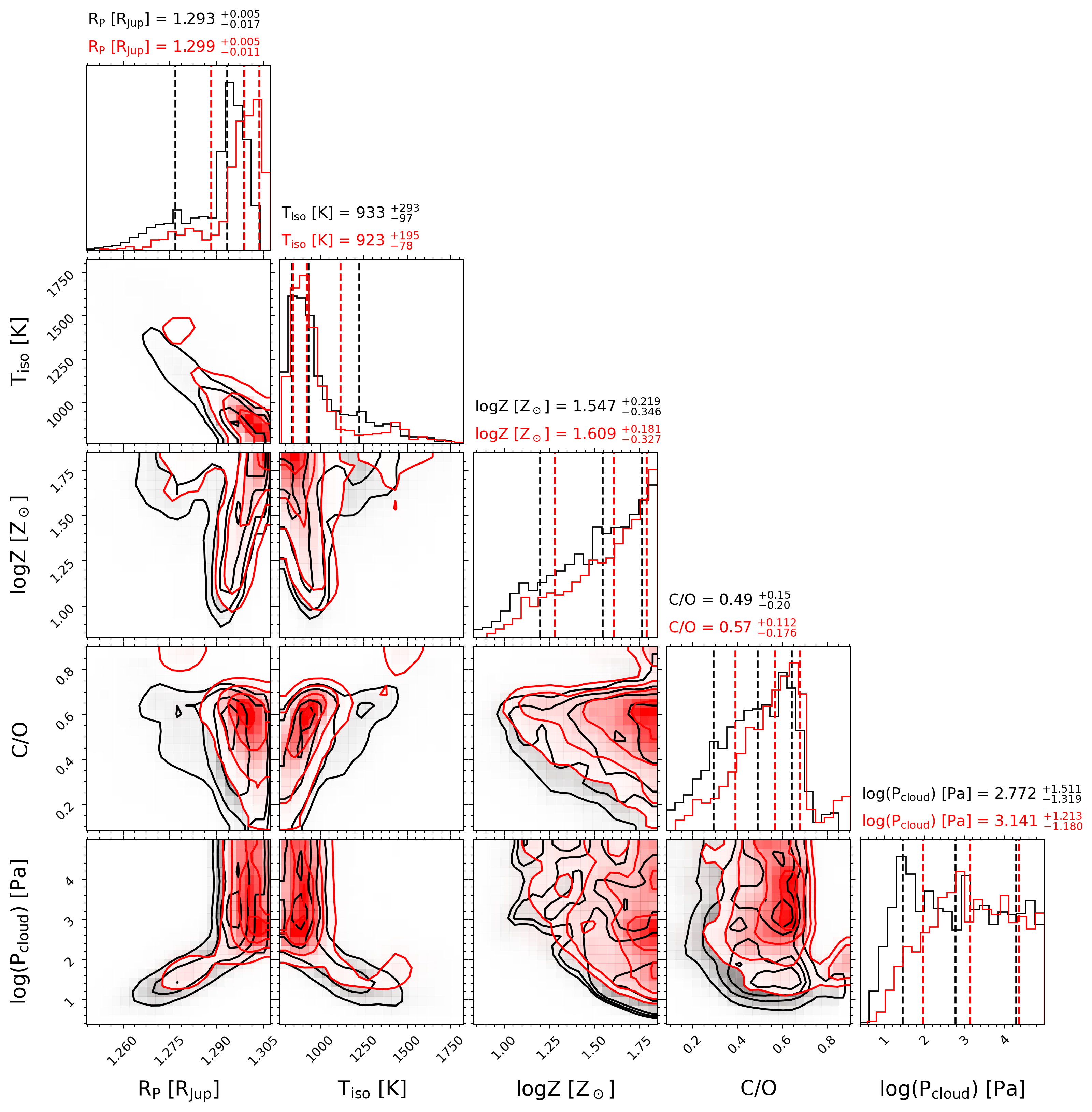}}
    \includegraphics[width=\textwidth]{figures/platon_corner_R100_Tiberius+Eureka_lowZ_solutions.png}%
    \llap{\raisebox{\dimen0-5cm}{
    \includegraphics[height=5cm]{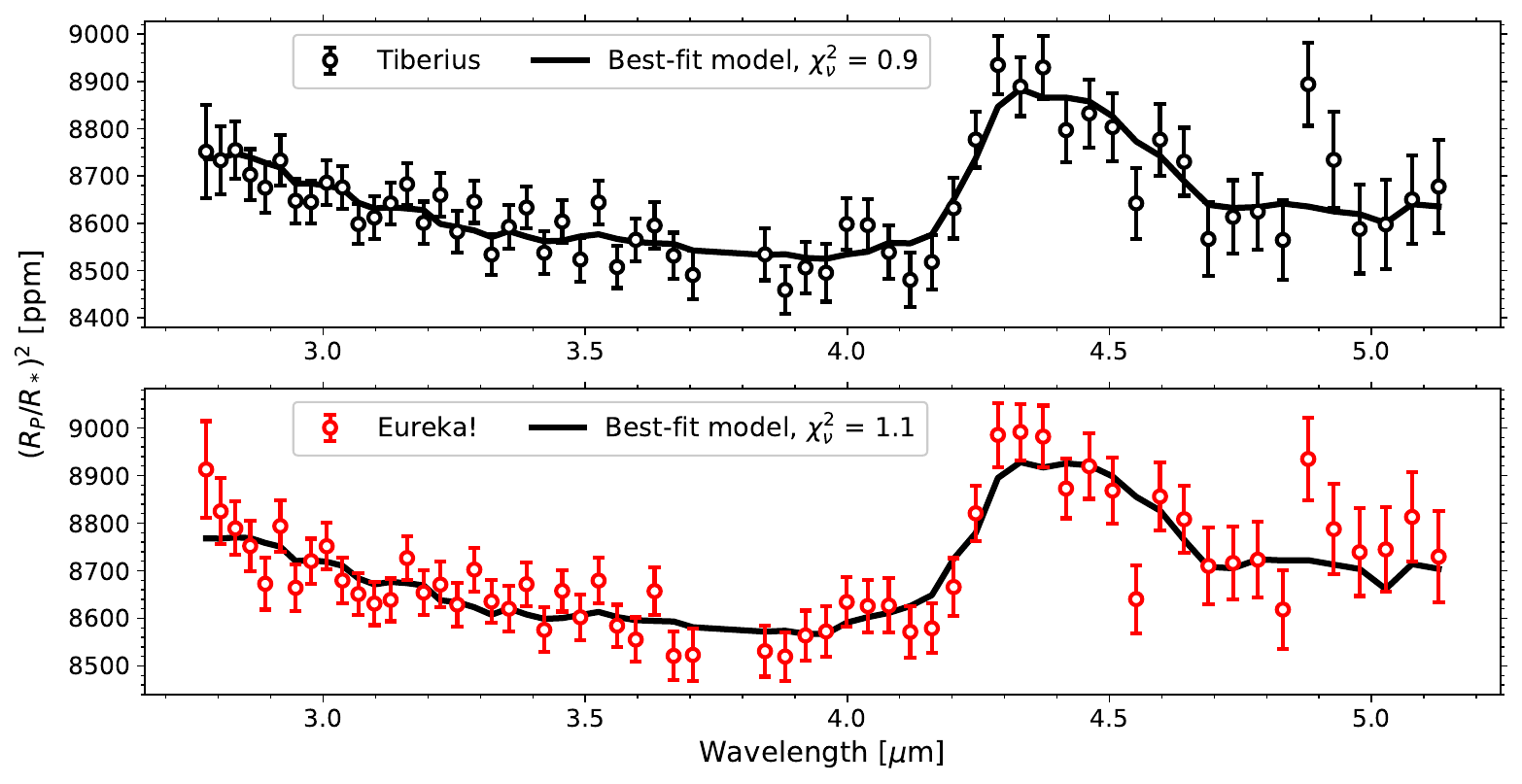}%
    }}
    \caption{The corner plot and best-fit models from our 1D chemical equilibrium atmosphere retrievals with \texttt{PLATON} run on the $R=100$ spectra, after excluding the unphysically high metallicity solutions. The black contours correspond to the \texttt{Tiberius} retrieval and the red contours to the \texttt{Eureka!} retrieval. The vertical dashed lines indicate the 16th, 50th (median) and 84th percentiles, which are also given in the axes titles.}
    \label{fig:PLATON_corner_lowZ}
\end{figure*}

In our \texttt{PLATON} retrievals we have five free parameters: the planet's radius ($\mathrm{R_P}$), the temperature of the isothermal atmosphere ($\mathrm{T_{iso}}$), the atmospheric metallicity ($\log Z$), the atmospheric C/O and the pressure of a grey cloud deck ($\log P_{\mathrm{cloud}}$). The metallicity is defined relative to solar \citep{Asplund2009} and \texttt{PLATON}'s default C/O ratio is 0.53. We place wide flat priors on each parameter of $1.20 < \mathrm{R_P} < 1.47$\,\Rjup, $300 < \mathrm{T_{iso}} < 2500$\,K, $-1 < \log Z < 3$, $0.05 < \mathrm{C/O} < 2.0$, and $-1 < \log P_{\mathrm{cloud}} < 5$\,Pa. We fixed the stellar radius to 1.477\,\Rsun\ and planet mass to 0.542\,\Mjup\ \citep{Bonomo2017}. We explored the parameter space using nested sampling, implemented through \texttt{DYNESTY} \citep{Speagle2020}, with 1000 live points. We ran these retrievals for both the \texttt{Tiberius} and \texttt{Eureka!} spectra at $R=100$ and $R=400$. The posterior medians, 16th and 84th percentiles are given for each fit parameter in Table \ref{tab:all_retrievals} with the best-fitting models plotted in Figure \ref{fig:PLATON_corner_lowZ}. 

As shown in Table \ref{tab:all_retrievals}, our \texttt{PLATON} analysis favours super-solar metallicities of $>29\times$ solar and $>38 \times$ solar to $1\sigma$ for \texttt{Tiberius} and \texttt{Eureka!}, respectively. The median retrieved C/O for both reductions of the data are consistent with solar albeit with relatively large uncertainties (C/O = $0.45^{+0.18}_{-0.20}$ and $0.54^{+0.15}_{-0.21}$ for \texttt{Tiberius} and \texttt{Eureka!}, respectively). However, as we showed in Section \ref{sec:interior_model}, our interior structure models place a $2\sigma$ upper limit on WASP-15b's metal mass fraction of 0.36. If we exclude the retrieval samples with metal mass fractions greater than this and recalculate our posteriors, we find a revised metallicity constraint of $35^{+23}_{-19}$ and $41^{+21}_{-22} \times$ solar for \texttt{Tiberius} and \texttt{Eureka!}, respectively, while C/O does not change significantly (Table \ref{tab:all_retrievals}). We consider these to be a more accurate inference of WASP-15b's true atmospheric metallicity. Figure \ref{fig:PLATON_corner_lowZ} shows the corner plot after excluding the unphysically high metallicity solutions. Figure \ref{fig:PLATON_corner} shows the corner plot over the full metalliticy range.

Table \ref{tab:all_retrievals} demonstrates that our inferences with \texttt{PLATON} are insensitive to whether we run our retrievals on the $R=100$ or $R=400$ spectra. For this reason, we only plot the results from our $R=100$ retrievals (Figure \ref{fig:PLATON_corner_lowZ}).

In summary, all of our retrieval analyses converge to a super-solar metallicity with a C/O consistent with solar. For \texttt{PLATON} and \texttt{BeAR}, the loose metallicity prior allows for unphysically high metallicities which we exclude based on interior structure models of WASP-15b (Section \ref{sec:interior_model}). We discuss the comparison between the retrievals in more detail in Section \ref{sec:discussion_differences_retrieval}.

\section{Interpreting the spectrum with a 3D general circulation model}
\label{sec:GCM}

To investigate the potential impact of spatial inhomogeneities in WASP-15b's atmosphere on the observed transmission spectrum, we performed simulations of WASP-15b's atmosphere using the \texttt{Met Office Unified Model} (\texttt{UM}), which is a 3D climate model of a planetary atmosphere (also known as a general circulation model, or a GCM). We used the same basic \texttt{UM} setup as in \citet{Drummond2020}, \citet{Zamyatina2023} and \citet{Zamyatina24_quenchingdriven} that provides the coupling between the dynamics, radiative transfer and chemistry. In brief, \texttt{UM}'s dynamical core \citep[\texttt{ENDGame},][]{Wood2014,Mayne2014a,Mayne2014} solves the full, deep-atmosphere, non-hydrostatic equations of motion on a constant angular grid using a semi-implicit semi-Lagrangian scheme. The \texttt{UM}'s radiative transfer component \citep[\texttt{SOCRATES,}][]{Edwards1996,Edwards1996a,Amundsen2014,Amundsen2016,Amundsen2017} solves the two-stream equations and treats opacities using the correlated-k and equivalent extinction methods. Opacity sources considered in the radiative transfer include the absorption due to \ce{H2O}, \ce{CO}, \ce{CO2}, \ce{CH4}, \ce{NH3}, \ce{HCN}, \ce{Li}, \ce{Na}, \ce{K}, \ce{Rb}, \ce{Cs} and collision-induced absorption due to \ce{H2}-\ce{H2} and \ce{H2}-\ce{He} as well as Rayleigh scattering due to \ce{H2} and \ce{He} \citep[for the line list information, see][]{Goyal2020}. Lastly, the \texttt{UM}'s chemistry component provides several chemical schemes for simulating the evolution of hot-Jupiter gas-phase chemistry, with the ``equilibrium'' and ``kinetics'' chemical schemes used in this study and described below. We chose a model grid resolution of \ang{2.5} in longitude by \ang{2} in latitude and 86 vertical levels equally spaced in height (covering pressures from $\sim$\SI{200}{\bar} to $\sim$\SI{e-5}{\bar}). This grid resolution is too coarse to resolve convection; however, we do not find that a convective adjustment or a similar correction is required. Even so, a longitudinal filter is used to maintain numerical stability, with the filtering constant of the horizontal wind, $K$, equal to $0.04$ \citep[see][]{Mayne2014a,Mayne2014,Christie2024}. For the stellar spectrum we used a \texttt{PHOENIX} BT-Settl model \citep{Rajpurohit2013} with parameters from Table~\ref{tab:wasp15_parameters_um}, and for the planet --- parameters from Table~\ref{tab:wasp15b_parameters_um}.

We performed two simulations. One simulation (which we refer to as the \texttt{UM} ``equilibrium'' simulation), used a chemical equilibrium scheme to compute the change in chemical species abundances induced by the changes in atmospheric pressure and temperature only. The other simulation (which we refer to as the \texttt{UM} ``kinetics'' simulation), allowed for departures from chemical equilibrium due to disequilibrium thermochemistry (but omitting photolysis) and used a kinetics chemical scheme to compute the change in chemical species abundances caused not only by the changes in atmospheric pressure and temperature, but also by the changes in the production and loss of these species during their atmospheric transport. Chemical species included in both \texttt{UM} simulations were those present in the \citet{Venot2019} \ce{C}-\ce{O}-\ce{N}-\ce{H} reduced chemical network. Abundances of alkali metals not included in the \citet{Venot2019} network --- \ce{Li}, \ce{Na}, \ce{K}, \ce{Rb} and \ce{Cs} --- were calculated using a threshold method outlined in \citet{Amundsen2016}. Both \texttt{UM} simulations assumed an aerosol-free atmosphere with 10$\times$ solar metallicity and \ce{C}/\ce{O} of 0.55 \citep{Asplund2009}. Additional details about the simulations, i.e. initialisation, runtime and the calculation of transmission spectra, are given in Appendix~\ref{appendix:um}. 

Figure~\ref{fig:um_vs_tiberius_vs_eureka1} shows the comparison of WASP-15b's limb-average transmission spectra observed with JWST NIRSpec/G395H and predicted by the \texttt{UM}. For the purpose of this comparison, a vertical offset was applied to the \texttt{UM} spectra, the value of which was determined from the \texttt{Tiberius}  reduction at $R=100$ using a least squares fit. The resulting offsets applied were $-1523$\,ppm and $-1515$\,ppm for the equilibrium and kinetics simulations, respectively. The results from the \texttt{Eureka!} reduction are presented alongside those from \texttt{Tiberius} but were not used to obtain the applied offset values. Overall, this comparison demonstrates that the \texttt{UM} simulations predict the general shape of WASP-15b's limb-average transmission spectrum rather well. Both simulations suggest that \ce{H2O}, \ce{CO} and \ce{CO2} are the major contributors to WASP-15b's opacity at the limbs. The enhancements in the observed transit depths at wavelengths where \ce{SO2} and \ce{OCS} absorb are not captured by the \texttt{UM} due to the lack of sulphur chemistry in the model.

\begin{figure}
    \centering
    \includegraphics[width=1\columnwidth]{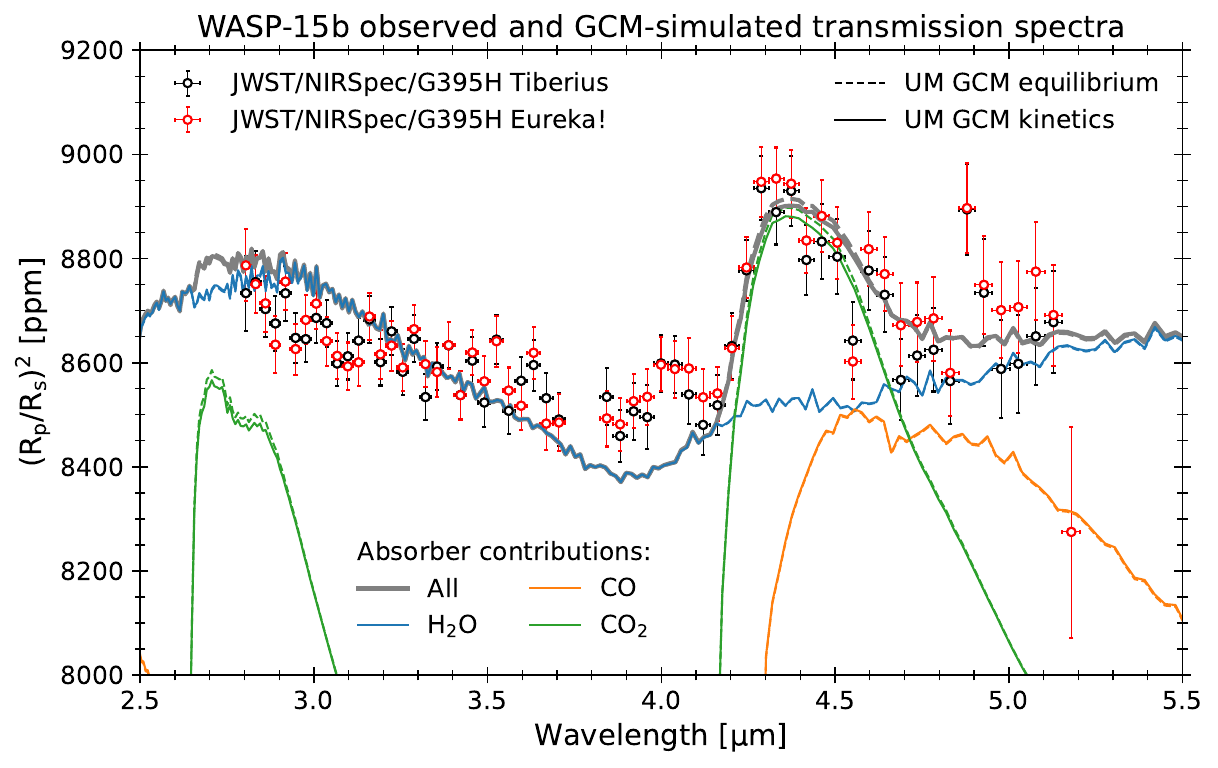}
    \caption{WASP-15b's $R=100$ limb-averaged transmission spectra obtained using the \texttt{Tiberius} (black points) and \texttt{Eureka!} (red points) reductions compared to the spectra predicted by the \texttt{UM} equilibrium (dashed lines) and kinetics (solid lines) simulations. \texttt{UM} simulations suggest that \ce{H2O} (blue lines), \ce{CO} (orange lines) and \ce{CO2} (green lines) are the major contributors to the observed limb-averaged transmission spectrum.}
    \label{fig:um_vs_tiberius_vs_eureka1}
\end{figure}

\begin{figure}
    \centering
    \includegraphics[scale=0.416]{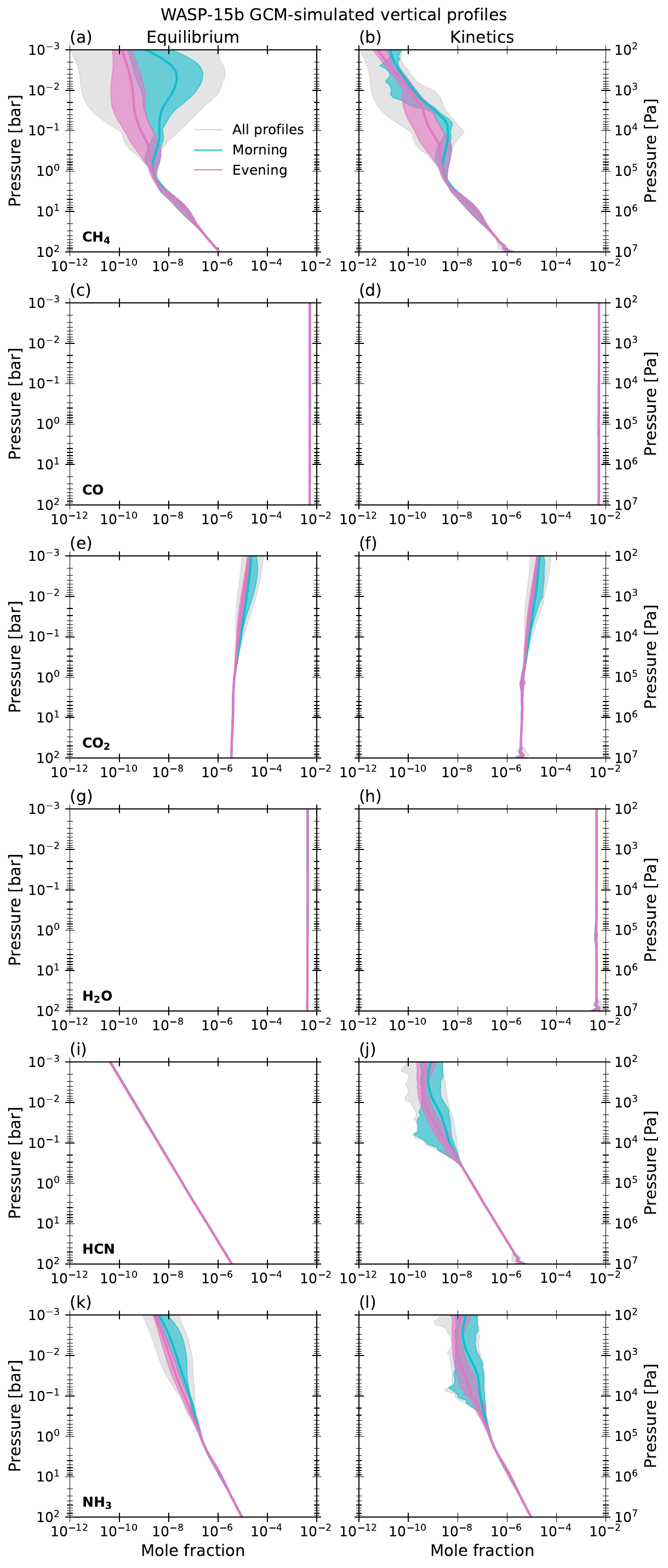}
    \caption{\ce{CH4}, \ce{CO}, \ce{CO2}, \ce{H2O}, \ce{HCN} and \ce{NH3} vertical profiles predicted by the \texttt{UM} equilibrium (left column) and kinetics (right column) simulations of WASP-15b's atmosphere. Grey shading shows the range of abundances for the entire atmosphere, cyan shading --- for the morning terminator only, pink shading  --- for the evening terminator only. Solid cyan and pink lines indicate the meridional mean for the morning and evening terminator, respectively.}
    \label{fig:um_chem_profiles}
\end{figure}

Our GCM simulations suggest that the constraints on \ce{H2O} and \ce{CO2} abundances, and by extension, on \ce{C}/\ce{O}, derived from JWST NIRSpec/G395H transit observations of planets like WASP-15b, could be informative not only in the context of the limbs of such planets but also in the context of their entire photospheres. To corroborate that we show the vertical profiles of major \ce{C}- and \ce{O}-bearing radiatively active species, \ce{CH4}, \ce{CO}, \ce{CO2}, \ce{H2O} and \ce{HCN} (and \ce{NH3} for completeness) predicted by the equilibrium and kinetics simulations in Figure \ref{fig:um_chem_profiles}. Firstly, we see that transport-induced quenching, a process capable of altering \ce{C} and \ce{O} budget and distribution at observable pressures in transmission \citep[see][for review]{Moses2014}, causes \ce{CH4}, \ce{HCN} and \ce{NH3} profiles in the kinetics simulation to diverge from those at chemical equilibrium: \ce{CH4} is depleted while \ce{HCN} and \ce{NH3} abundances are enhanced at pressures lower than $\sim$\SI{e-1}{\bar}. However, these disequilibrium changes in \ce{CH4} and \ce{NH3} would not cause the contributions of \ce{CH4} and \ce{HCN} to the overall absorption by \ce{C}-bearing species in WASP-15b's photosphere to be much different from their contributions at chemical equilibrium, because \textit{both} simulations predict that abundances of \ce{CH4} and \ce{HCN} are low (lower than 1 ppm) throughout the entire GCM model domain. Meanwhile, \ce{H2O}, \ce{CO} and \ce{CO2} are more abundant than \ce{CH4} and \ce{HCN} in \textit{both} simulations: \ce{H2O} and \ce{CO} reach $\sim$10,000 ppm and \ce{CO2} $\sim$10 ppm. \ce{H2O}, \ce{CO} and \ce{CO2} are also rather uniformly distributed throughout the GCM model domain. Together, low \ce{CH4} and \ce{HCN} abundances but high \ce{H2O}, \ce{CO} and \ce{CO2} abundances and their uniform spatial distribution, imply that \ce{H2O}, \ce{CO} and \ce{CO2} would (1) account for the majority of \ce{C} and \ce{O} in WASP-15b's photosphere and (2) their contribution to the opacity at the limbs of the planet would be representative of their contribution to the opacity across the entire planet's photosphere.

As shown in \cite{Zamyatina24_quenchingdriven}, for WASP-96b, differences in metallicity and the assumption or relaxation of chemical equilibrium can produce differences in the spectra within the 3--5\,$\mu$m range. However, these will largely manifest as differences in the applied vertical offset, which is not known and need not be the same across the two simulations, with differences as a function of wavelength being smaller than observational uncertainties. In light of our results in this work, we plan to revisit WASP-15b to explore, in more detail, the impacts of metallicity on the spectra within a 3D context, but this is beyond the scope of this current work.

\section{Photochemical modelling}
\label{sec:photochem}

The evidence for absorption by \ce{SO2} indicates that photochemistry is an active process in the atmosphere of WASP-15b. We model the photochemical processes in this atmosphere using VULCAN, a 1D kinetics model that treats photochemical \citep{Tsai2021} and thermochemical \citep{Tsai2017} reactions. The VULCAN setup used in this work solves the Eulerian continuity equations, including chemical sources/sinks and diffusive transport. We use the updated C-H-N-O-S network\footnote{\url{https://github.com/exoclime/VULCAN/blob/master/thermo/SNCHO_photo_network_2024.txt}} for hydrogen-dominated atmospheres that importantly considers S-bearing species.

We consider a grid of models based on the temperature profiles at the terminators of WASP-15b as calculated in the equilibrium \texttt{UM} GCM described in Section \ref{sec:GCM} and shown in Section \ref{appendix:photochem_inputs}. We use a K$_{zz}$ profile based on the scaling relations in \citet{moses2022} using an internal temperature of $T_\mathrm{int} = 100$\,K. We use a host star stellar spectrum for a 6500 K star from the stellar spectral grid in \citet{rugheimer2013}, which combines synthetic ATLAS spectra \citep{kurucz1979} with observed spectra from the International Ultraviolet Explorer for wavelengths less than 300\,nm. We do not consider the impact of aerosol opacity in our models. The grid spans a range of planet metallicities from 10 ${\times}$ solar, consistent with the input \texttt{UM} GCM, to 100 ${\times}$ solar in intervals of 10. By extending to metallicities higher than those considered by our GCM model, we made the modelling trade-off of considering the differences in photochemistry on the terminators of WASP-15b based on the GCM modelling in-lieu of more self-consistent atmospheric modelling. To create spectra from the photochemical models, we use the \texttt{PICASO} radiative transfer code \citep{Batalha2019} with a resolution of R = 100. The molecular opacities used to generate the transmission spectra in this work are taken from the references in Appendix \ref{appendix:picaso_inputs}. 

\begin{figure}
    \centering
    \includegraphics[width=1\columnwidth]{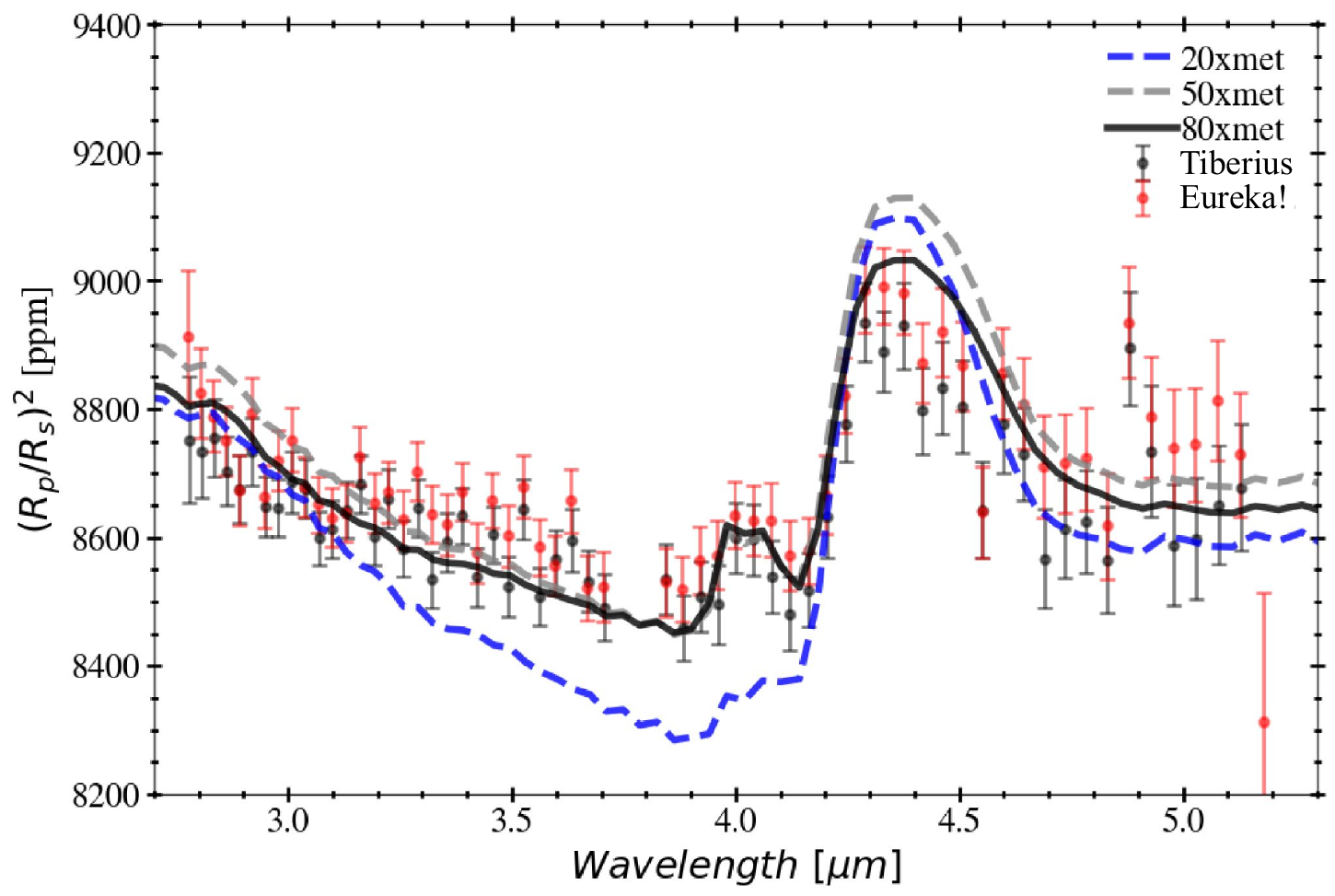}
    \caption{Photochemical model spectra for WASP-15b at $80\times$ (solid, black), $50\times$ (dashed, gray), and $20\times$ (dashed, blue) solar metallicites as compared to the Tiberius (black points) and Eureka! (red points) reductions. Photochemical models suggest that the observed \ce{SO2} feature is indicative of a high metallicity atmosphere. }
    \label{fig:photo_spec}
\end{figure}

\begin{figure}
    \centering
    \includegraphics[width=0.9\columnwidth]{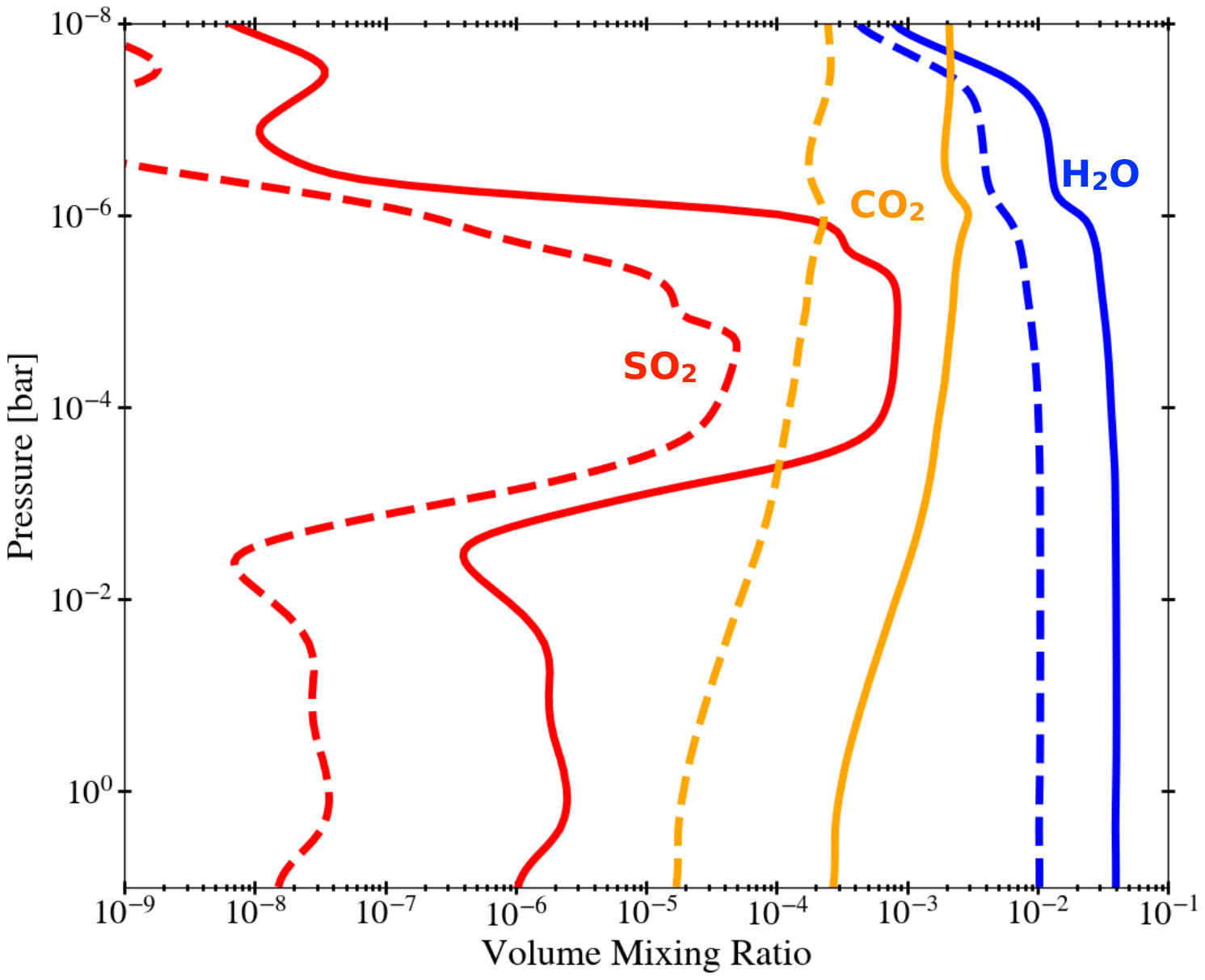}
    \includegraphics[width=0.9\columnwidth]{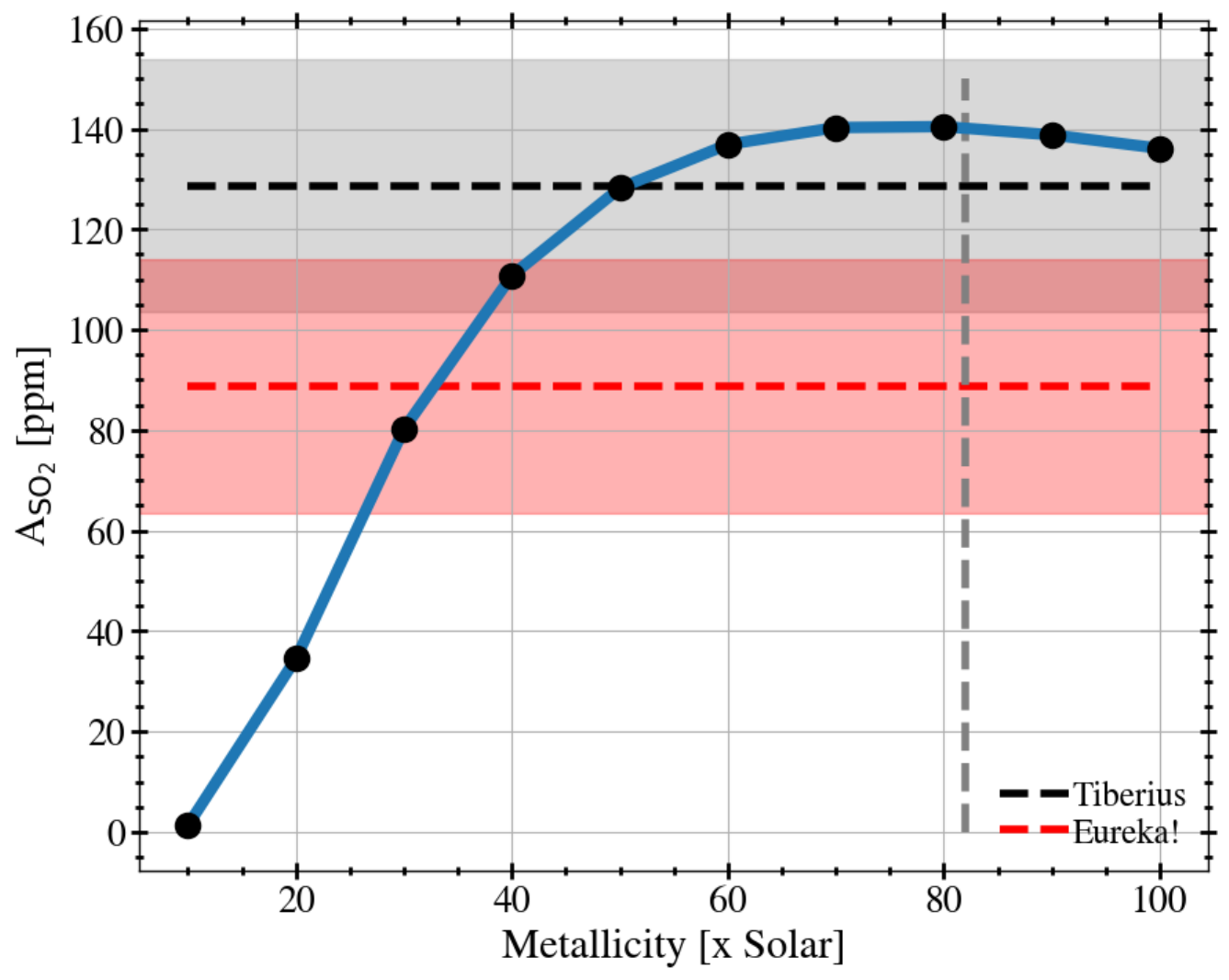}
    \caption{Top: The vertical distribution of \ce{SO2}, \ce{CO2}, and \ce{H2O} as predicted by the VULCAN photochemical model for an atmosphere with $80\times$ solar metallicity (solid lines) and $20\times$ solar metallicity (dashed lines). All three key species have abundances that change substantially with increased metallicity. Bottom: The predicted amplitude of the 4.05 $\mu$m feature ($\mathrm{A_{SO_2}}$) varies by more than 100 ppm as a function of atmospheric metallicity. The observed amplitude and uncertainty of the feature are shown by the horizontal dashed lines and shaded regions for \texttt{Tiberius} (grey) and \texttt{Eureka!} (red). This indicates a substantially metal-rich atmosphere based on our photochemical modelling. The vertical dashed grey line indicates the upper limit on the atmospheric metallicity derived from the interior structure model (Section \ref{sec:interior_model}).}
    \label{fig:photo_spag_amp}
\end{figure}

Our photochemical models indicate that WASP-15b is likely to be substantially enhanced in metallicity compared to solar to reproduce the observed \ce{SO2} feature, as shown in Figure \ref{fig:photo_spec}. To reproduce the amplitudes of all of the observed features, namely \ce{SO2}, \ce{H2O}, and \ce{CO2}, a metallicity of either $\sim80 \times$ solar or greater is preferred. We emphasise that the high metallicity required to reproduce the spectrum with photochemical modelling is driven by both the \ce{SO2} and \ce{CO2} feature amplitude (and to a lesser extent the \ce{H2O} feature amplitude as demonstrated in Section \ref{sec:1D_models}). Figure \ref{fig:photo_spag_amp} (bottom panel) shows how the amplitude of the \ce{SO2} feature changes with metallicity as compared to the amplitude of the observed \ce{SO2} feature. The abundance of \ce{SO2} is highly sensitive to metallicity due to the net chemical reaction network that produces \ce{SO2} which requires two \ce{H2O} molecules to interact with every \ce{H2S} molecule \citep{Tsai2023, Powell2024}. This effect can be seen in our model grid in Figure \ref{fig:photo_spag_amp} (top panel) where we find that the peak \ce{SO2} abundance increases by more than an order of magnitude for substantially metal-rich atmospheres (e.g., $80\times$ solar) as compared to more moderately enriched atmospheres (e.g., $20\times$ solar). 

We note that while both reductions are consistent with \ce{SO2} amplitudes produced from models with lower metallicities than our $80\times$ solar metallicity case, the \ce{CO2} amplitude is systematically higher than our best-fit $80\times$ solar metallicity case for those lower metallicity models (e.g., Figure \ref{fig:photo_spec}). However, there are two caveats to this high-metallicity interpretation. Firstly, the GCM pressure-temperature profile that we use for our photochemical models (Appendix \ref{appendix:photochem_inputs}) may overly broaden the spectral features, pushing us to higher metallicities to fit the observed CO$_2$ feature. Secondly, these photochemical models do not include clouds. As we saw in our \texttt{petitRADTRANS} forward models, the gradient of the H$_2$O feature can be fitted with cloud opacity, without the need for such high metallicities (Figure \ref{fig:prt_forward_model1}).

The photochemical modelling results thus point towards a very high metallicity atmosphere for WASP-15b if the elements vary according to a solar abundance. Indeed, under the assumption of a cloud-free atmosphere, our photochemical models indicate that WASP-15b may have a metallicity near the maximum metallicity inferred from interior structure models ($82\times$ solar, Section \ref{sec:interior_model}). However, based on the \ch{SO2} feature alone, our photochemical models are able to fit the amplitude of this feature with metallicities $\lesssim 40\times$ solar (Figure \ref{fig:photo_spag_amp}, bottom panel). This is consistent with the metallicities inferred from our 1D retrievals which include clouds and a freely-fitted, yet isothermal, temperature profile (Section \ref{sec:1D_models}). Future work that varies the individual abundance ratios of the different atomic species will be useful to better understand the chemistry of WASP-15b's atmosphere.

\section{Discussion}
\label{sec:discussion}

\subsection{Differences between reductions and retrievals}
\label{sec:discussion_differences}

Above, we presented inferences regarding WASP-15b's atmosphere from two different data reduction pipelines (Section \ref{sec:data_reduction}) and three different retrieval models (Section \ref{sec:1D_models}). While all approaches are consistent with super-solar metallicity atmospheres and solar C/O, the spectra and retrieval results do have some differences. We discuss these differences in greater detail in the following sub-sections.

\subsubsection{Reduction differences}
\label{sec:discussion_differences_reduction}

\begin{figure}
    \centering
    \includegraphics[width=1.0\linewidth]{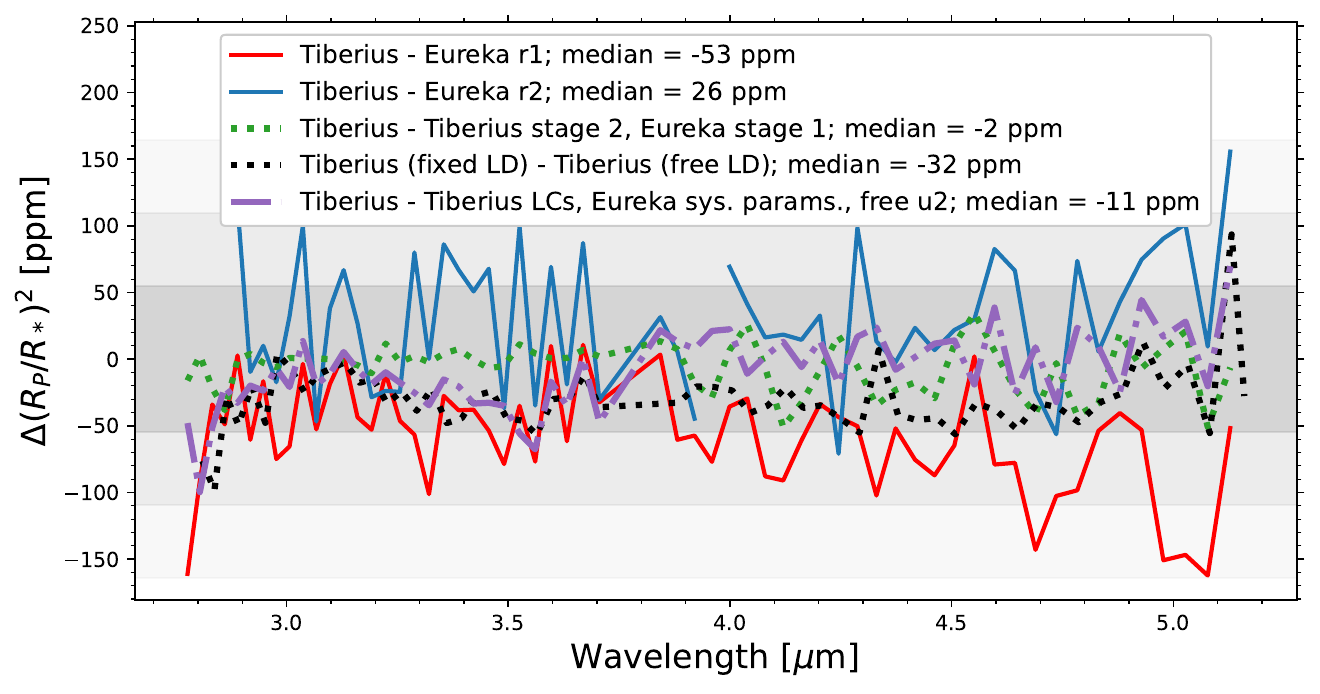}
    \caption{Figure showing the differences between the $R=100$ spectra resulting from different reduction setups. In red is the difference between the nominal \texttt{Tiberius} (Section \ref{sec:tiberius_reduction}) and \texttt{Eureka} (`r1', Section \ref{sec:eureka_reduction}) reductions. In blue is the difference between the nominal \texttt{Tiberius} reduction and the second \texttt{Eureka!} reduction (`r2', Appendix \ref{sec:eureka_VP}) which used the \texttt{Tiberius} stage 1 outputs. In dotted green is the difference between the nominal \texttt{Tiberius} reduction and a test \texttt{Tiberius} reduction run on the \texttt{Eureka!} r1 stage 1 outputs. In dotted black is a comparison between two \texttt{Tiberius} reductions with fixed vs completely free quadratic limb darkening, using an old binning scheme. In dot-dashed purple is a comparison between the nominal \texttt{Tiberius} reduction and a spectrum resulting from fitting the \texttt{Tiberius} light curves with the system parameters fixed to \texttt{Eureka!} r1 and the second limb darkening coefficient as a free parameter, to match the approach of \texttt{Eureka!} r1. No offsets have been applied to any of these residuals. The grey shaded regions indicate $1\times$, $2\times$ and $3\times$ the median transit depth uncertainty of the nominal \texttt{Tiberius} reduction.}
    \label{fig:reduction_comparisons}
\end{figure}

As stated in Section \ref{sec:trans_spec} there is an offset between the median transit depths of the $R=100$ \texttt{Tiberius} and \texttt{Eureka!} spectra. In Figure \ref{fig:reduction_comparisons}, we present a comparison between the $R=100$ spectra resulting from the two nominal reductions presented in Section \ref{sec:data_reduction} and additional test reductions. Considering first the difference between the nominal \texttt{Tiberius} (Section \ref{sec:tiberius_reduction}) and \texttt{Eureka!} reductions (Section \ref{sec:eureka_reduction}), we see there is both an offset and a slope in the NRS2 residuals (red line, Figure \ref{fig:reduction_comparisons}).

A possible cause for this difference is the differing treatment of limb darkening. While both \texttt{Tiberius} and \texttt{Eureka!} used quadratic limb darkening with coefficients informed by the same stellar parameters, 3D model and interpolation code, \texttt{Tiberius} fixed both $u1$ and $u2$ while \texttt{Eureka!} fixed $u1$ and fitted for $u2$. In a separate test with \texttt{Tiberius}, we performed a fit where both limb darkening coefficients were free parameters. This led to a transmission spectrum with a median depth 32\,ppm deeper than the spectrum with fixed coefficients, which lessens the offset between \texttt{Eureka!} and \texttt{Tiberius}, but  did not lead to a slope difference in NRS2 (black dotted line, Figure \ref{fig:reduction_comparisons}). Therefore, limb darkening could be partly, but not solely, responsible for the differences between \texttt{Tiberius} and \texttt{Eureka!}.

An additional consideration is the role of the stage 1 extraction, which includes the 1/f correction. The primary differences between the 1/f corrections of \texttt{Tiberius} and \texttt{Eureka!} as implemented in this study are that \texttt{Tiberius}' is a row-by-row median correction while \texttt{Eureka!} applies the same Stage 3 background subtraction algorithm to the group-level files (after masking the trace). This means that \texttt{Eureka!}'s correction has more free parameters and options to correct for 1/f noise, but can also introduce systematics, e.g., if outliers are not masked properly. In addition, \texttt{Eureka!} also allows for a bias correction in the form of a multiplication factor that could explain some of the offsets between reductions.

We tested the impacts of the differing stage 1 approaches in two different ways. Firstly, we ran the \texttt{Tiberius} stage 2 spectral extraction and light curve fitting on the \texttt{Eureka!} stage 1 outputs. This led to a transmission spectrum within $1\sigma$ of the nominal \texttt{Tiberius} reduction and a baseline offset of just 2\,ppm (green dotted line, Figure \ref{fig:reduction_comparisons}). Secondly, we performed an additional \texttt{Eureka!} reduction (presented in Appendix \ref{sec:eureka_VP}) which used \texttt{Tiberius}'s stage 1 output but \texttt{Eureka!} from stage 2 onwards. For this second \texttt{Eureka!} reduction, we fixed both quadratic limb darkening coefficients to the same values as \texttt{Tiberius}. This revealed a higher scatter in the residuals between this second \texttt{Eureka} reduction and the nominal \texttt{Tiberius} reduction (blue solid line, Figure \ref{fig:reduction_comparisons}), and a median depth offset of $+26$\,ppm but no slope in the NRS2 residuals. This suggests that the differences may be emerging at the stage 2 (spectral extraction and light curve fitting) stage, including how limb darkening is treated. 

Differences in the system parameters, namely $a/R_*$ and $i$, can also lead to depth offsets and slopes between reductions \citep[e.g.,][]{Alexoudi2018,Alexoudi2020}. As a final test, we ran a set of fits to the \texttt{Tiberius} light curves but fixing the NRS1/NRS2 system parameters to the NRS1/NRS2 best fit values from the nominal \texttt{Eureka!} reduction (Table \ref{tab:system_params}). In this test we also left the second limb darkening coefficient, $u2$, as a free parameter. This approach meant that we adopted the same fitting setup as the nominal \texttt{Eureka!} reduction but with the \texttt{Tiberius} light curves. Again, no slope difference was produced in NRS2's transmission spectrum (purple dash-dotted line, Figure \ref{fig:reduction_comparisons}). Therefore, we do not believe differing system parameters are the cause. Instead, this suggests differences in the light curves, which could be related to spectral extraction and/or background subtraction.

Although, we cannot conclusively determine the origins of the reduction differences, we are encouraged by the fact that despite these differences, our broad conclusions about the planet's super-solar metallicity and sub-solar-to-solar C/O are not impacted (Section \ref{sec:1D_models}). However, we discuss specific retrieval differences in Section \ref{sec:discussion_differences_retrieval}. Understanding the origins of reduction differences, such as slopes and offsets, is a broader problem faced by the community as a whole \citep[e.g.,][]{CarterMay2024}. Thus far there is no consensus on best practices, in particular, because these effects seem to vary between stellar types and brightnesses, planet types, instruments used, etc.\ Further investigation is needed to understand the origins of these differences and if and how they may impact population-level atmospheric inferences. In the meantime, the BOWIE-ALIGN approach is to have at least two independent reductions and retrievals per planet and one uniform reduction setup threading through all planets' analyses. The independent approaches will allow us to assess the robustness and uncertainties of our inferences while the uniform approach should allow us to mitigate biases arising from different approaches when we combine our results at the end of our survey \citep{Kirk2024_align}.

\subsubsection{Retrieval differences}
\label{sec:discussion_differences_retrieval}

When comparing the results between retrieval codes, it is important to consider how each code defines its parameters. For example, \texttt{petitRADTRANS}'s chemical equilibrium models parameterise metallicity as [Fe/H] with chemical equilibrium grids calculated over a range of [Fe/H]. However, when adjusting the C and O abundances, \texttt{petitRADTRANS} scales C/H with metallicity and then sets O/H according to C/O. \texttt{PLATON} takes the opposite approach, adjusting O/H based on metallicity, Z ([M/H]), and then setting C/H based on C/O. The approach of both \texttt{petitRADTRANS} and \texttt{PLATON} means that (C+O)/H does not respect scaled metallicity abundances. 

Furthermore, in our \texttt{petitRADTRANS} retrievals, we included the planet's surface gravity as a free parameter with a Gaussian prior. \texttt{PLATON} does not include the surface gravity as a fit parameter, instead allowing the planet mass to be a fit parameter (in addition to the planet radius). In our \texttt{PLATON} retrievals, we fix the planet's mass and only fit for the radius. In a further test, we ran a retrieval with mass as a fit parameter with a Gaussian prior from \cite{Bonomo2017} but found this made no difference to our resulting posteriors. With regard to the cloud-top pressure, \texttt{PLATON} recovers a lower pressure cloud deck than \texttt{petitRADTRANS}, albeit with large uncertainties. However, this is misleading as the \texttt{PLATON} corner plot (Figure \ref{fig:PLATON_corner_lowZ}) shows that the cloud-top pressure is actually unconstrained with a lower limit of $\gtrsim 10$\,Pa (0.1\,mbar). It is important to note that the reference pressure is fixed to 1\,mbar in our \texttt{petitRADTRANS} retrievals and to 1\,bar in our \texttt{PLATON} retrievals.

Of the different \texttt{petitRADTRANS} setups, we favour the `hybrid chemistry' retrievals owing to their better fits to the data (Figure \ref{fig:prt_best_fit_models}). These follow equilibrium chemistry, but with the addition of free abundances for \ch{SO2} and OCS. However, all \texttt{petitRADTRANS} approaches produce consistent results (Table \ref{tab:all_retrievals}).

\begin{figure}
    \centering
    \includegraphics[width=0.95\linewidth]{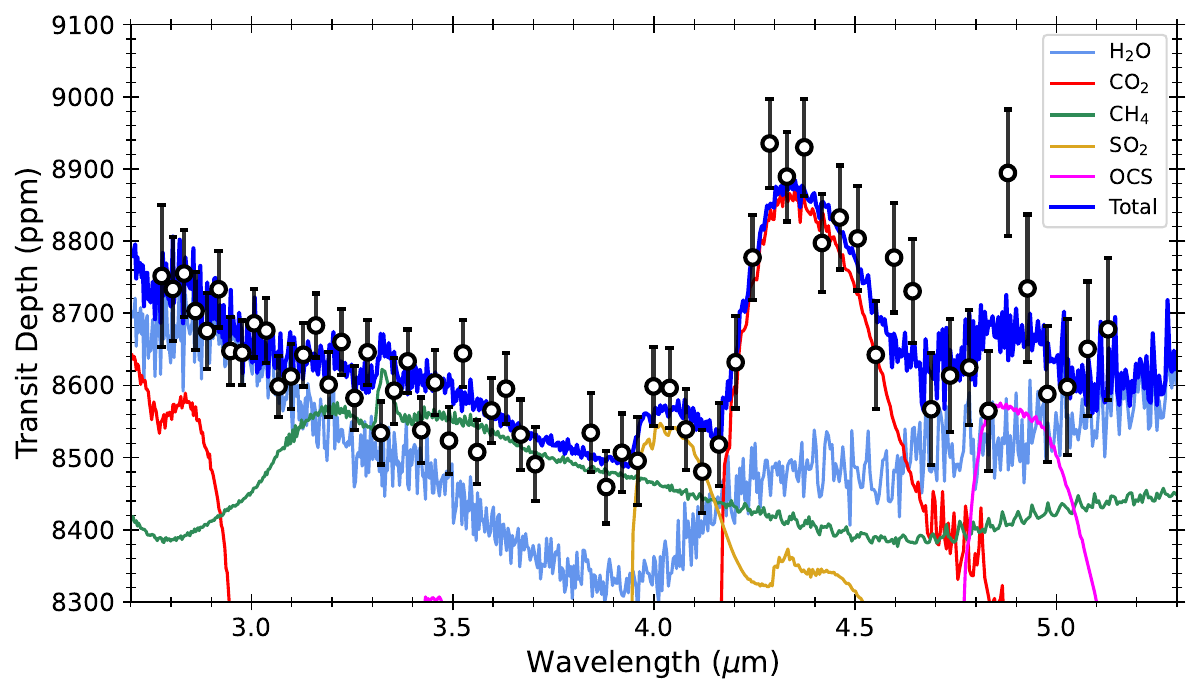}
    \includegraphics[width=0.95\linewidth]{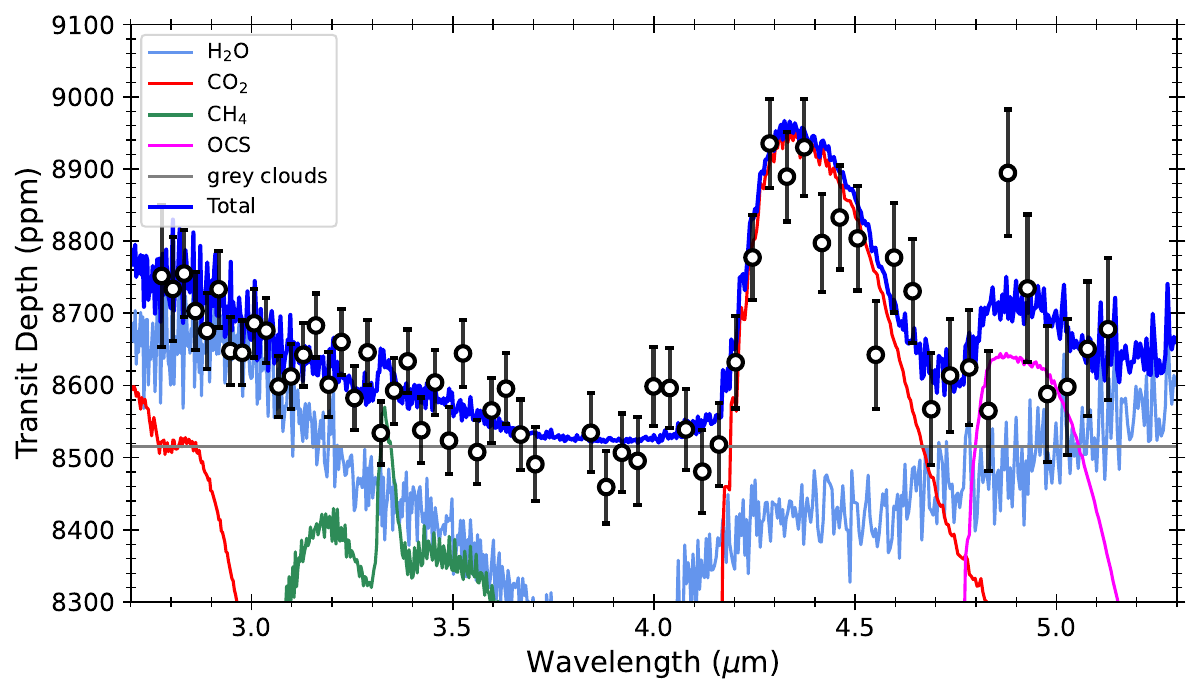}
    \caption{Figure showing the contributions of opacities from the best fitting \texttt{petitRADTRANS} hybrid chemistry (equilibrium plus free S) retrieval (top panel) and free chemistry retrieval (bottom panel) to the \texttt{Tiberius} $R=100$ spectrum. The difference in Bayesian evidences, $\Delta\ln \mathrm{Z}$, is 2.7 in favour of the hybrid model (top panel).}
    \label{fig:gas_opac}
\end{figure}

Figure \ref{fig:gas_opac} shows the contributions of gas and cloud opacities from the best-fitting hybrid and free chemistry retrievals to the \texttt{Tiberius} $R=100$ spectrum. This figure demonstrates how the relatively shallow gradient of the \ch{H2O} feature drives the hybrid model to include \ch{CH4} and the free chemistry model to include grey cloud opacity. This same shallow NRS1 slope is likely responsible for driving the \texttt{PLATON} and \texttt{BeAR} retrievals to high metallicities. As we saw in Section \ref{sec:retrievals_pRT}, the Bayesian evidence favours the hybrid and equilibrium models over free chemistry by $\sim3\sigma$ for \texttt{Tiberius} and the hybrid model over free chemistry by $\sim 4\sigma$ for \texttt{Eureka!}.

From our retrievals, the isothermal temperatures converge to temperatures that are 600--700\,K cooler than the equilibrium temperature of the planet ($1676 \pm 29$\,K), albeit with large uncertainties. Considering only the \texttt{petitRADTRANS} retrievals, we see that the hybrid retrievals favour isothermal temperatures of around 900\,K while the free retrievals find temperatures around 1100\,K (Table \ref{tab:all_retrievals}). This could be why these two retrievals disagree at to whether \ch{CH4} or clouds are needed to fit the NRS1 data (Figure \ref{fig:gas_opac}).

Retrieved temperatures are often cooler than the equilibrium temperatures. This can be caused both by the assumption of 1D atmospheres that neglect limb asymmetries \citep{MacDonald2020} and modelling choices, such as how the pressure-temperature profile is parameterised \citep{Welbanks2022}. Our three retrieval setups with \texttt{petitRADTRANS} (Section \ref{sec:retrievals_pRT}), \texttt{BeAR} (Section \ref{sec:retrievals_bear}), and \texttt{PLATON} (Section \ref{sec:retrievals_platon}) all used isothermal pressure-temperature profiles. Our GCM model predicts maximal differences of $\sim100$\,ppm in the amplitude of \ch{CO2} between the morning and evening limbs (Figure \ref{fig:um_chem_profiles}). Therefore, the cool temperatures we retrieve may be a product of both our pressure-temperature parameterisation \citep{Welbanks2022} and real thermal limb asymmetries \citep{MacDonald2020}.

So far, our retrieval approaches have neglected the possibility of a transit depth offset between NRS1 and NRS2 (Section \ref{sec:discussion_differences_reduction}). To correct this, we ran a further two \texttt{PLATON} retrievals on the nominal \texttt{Tiberius} (Section \ref{sec:tiberius_reduction}) and \texttt{Eureka!} (Section \ref{sec:eureka_reduction}) reductions but this time with an additional free parameter that fits for a transit depth offset between NRS1 and NRS2. This offset parameter had a uniform prior bounded between $-200$ and $+200$\,ppm. For \texttt{Tiberius}, the resulting offset was consistent with zero ($14^{+21}_{-20}$\,ppm) and hence there was no significant difference in the posteriors of the other parameters ($Z = 35^{+23}_{-19}\times$ solar, no offset; $Z = 29^{+33}_{-18}\times$ solar, with offset; $\mathrm{C/O} = 0.49^{+0.15}_{-0.20}$, no offset; $\mathrm{C/O} = 0.43^{+0.18}_{-0.21}$, with offset). The Bayesian evidence difference between the \texttt{Tiberius} retrievals with and without an offset was less than 1, indicating no statistical preference for a depth offset between NRS1 and NRS2. For \texttt{Eureka!}, the resulting offset was $52^{+22}_{-21}$\,ppm. This led to a slightly larger change in the retrieved Z and C/O but neither of which was statistically significant ($Z = 41^{+21}_{-22}\times$ solar, no offset; $Z = 32^{+31}_{-19}\times$ solar, with offset; $\mathrm{C/O} = 0.57^{+0.11}_{-0.18}$, no offset; $\mathrm{C/O} = 0.38^{+0.19}_{-0.21}$, with offset). For \texttt{Eureka!}, the Bayesian evidence difference was 5 in favour of the retrieval with an offset, indicating a $>3\sigma$ preference \citep[e.g.,][]{benneke2013distinguish}.

Despite these differences, we are encouraged by the fact that our conclusion regarding the planet's super-solar metallicity and sub-solar-to-solar C/O is not impacted. To help the community consolidate our suite of reductions and retrievals, we recommend the use of the results from the \texttt{petitRADTRANS} hybrid chemistry retrievals run on the \texttt{Tiberius} spectra. This is because the \texttt{Tiberius} spectrum does not require a depth offset and our \texttt{petitRADTRANS} hybrid chemistry retrievals allow for disequilibrium sulphur chemistry (which \texttt{PLATON} does not) and do not suffer the high metallicity problems of \texttt{BeAR} and \texttt{PLATON}.

\subsection{WASP-15b's atmospheric composition and predictions from migration scenarios}

Despite the differences in abundances retrieved by each code (Table \ref{tab:all_retrievals}), they consistently point towards a super solar metallicity and an approximately solar C/O ratio. Our GCM simulations (Section \ref{sec:GCM}), that included C-O-N-H gas-phase chemistry and assumed aerosol-free conditions, suggest that CO, \ch{H2O} and \ch{CO2} are the most abundant C- and O-bearing chemical species in WASP-15b's atmosphere. This supports our use of NIRSpec/G395H to measure the planet's C/O from its \ch{H2O} and \ch{CO2} features. Given that these GCM simulations also predict that the spatial distribution of CO, \ch{H2O} and \ch{CO2} is rather uniform throughout the entire GCM model domain (with \ch{CO2} variations in the vertical being less than an order of magnitude, Figure \ref{fig:um_chem_profiles}), this suggests that CO, \ch{H2O} and \ch{CO2}'s contribution to the opacity at the limbs of the planet is representative of their contribution to the opacity across the entire planet's photosphere. 

As explained in the introduction, WASP-15b is part of our ongoing survey \citep{Kirk2024_align} to determine whether aligned and misaligned hot Jupiters around F stars have different C/O ratios and metallicities based on their likely different migration mechanisms (disc vs.\ disc-free/high-eccentricity). As shown in \citetalias{PenzlinBooth2024} \citeyear{PenzlinBooth2024}, without the full BOWIE-ALIGN sample we cannot make conclusive statements about where the planet formed due to lack of knowledge of where C and O are in the solids; however, the misaligned nature and high metallicity allow us to speculate in general terms.

WASP-15b is a misaligned hot Jupiter, which suggests it formed exterior to $\sim0.6\,{\rm au}$ and underwent disc-free (high-eccentricity) migration \citep{2016Munoz}. The super-solar metallicity we infer for WASP-15b indicates the late accretion of solids, which would serve to drive up the O/H of its atmosphere \citep{Booth2017,Schneider2021}. The fact that we also see evidence for sulphur content (Section \ref{subsec:OCS}), additionally points to the planet acquiring its high metallicity via the accretion of solids rather than high metallicity gas \citep{Chachan2023}. This is because sulphur in the disc is bound in solids and thus cannot be delivered by metal-rich gas accretion \citep{2023Danti}, except at temperatures too high to be compatible with high-eccentricity migration (above $\sim700$~K, \citealt{Lodders2003,2023Timmermann}). Given the mass of WASP-15b is much larger than the pebble isolation mass ($\sim 20\,M_{\earth}$, \citealt{2018Bitsch}), the planet likely only had minor pebble accretion late in its formation, implying planetesimal accretion as the main driver of the solid enrichment.

Alternatively, the planet's composition may be explained without the accretion of planetesimals if WASP-15b accreted its envelope more-or-less in situ. Here, pebbles might be small enough to be accreted alongside the gas \citep{2023Morbidelli}, possibly enabling a high metallicity. Further, metal-rich gas accretion could explain the planet's composition if the accretion happened close enough to the star that refractory sulphur sublimates. While in situ formation of hot Jupiters is often disfavoured \citep[e.g.,][]{Dawson2018}, composition alone cannot rule out this scenario if an alternative explanation for the misaligned orbit of the planet can be found.

We find a C/O ratio in the range 0.4--0.6, which is consistent with the solar C/O. However, there is no clear interpretation of where the planet formed because the amount of carbon and oxygen contained in solids is unknown. As shown in \citetalias{PenzlinBooth2024} \citeyear{PenzlinBooth2024}, the prospects for such analyses may be better once the full sample of planets is available. This is one of the main goals of our observational programme \citep{Kirk2024_align}.

\subsection{Sulphur chemistry in WASP-15b's atmosphere}
\label{subsec:OCS}

As described in Sections \ref{sec:retrievals_pRT} and \ref{sec:photochem}, the absorption features we see at 4.0 and 4.9\,\micron\ can be fit with \ch{SO2} and \ch{OCS}, respectively. We discuss the plausibility of this sulphur chemistry below. 

Firstly, we compared WASP-15b to the ERS observations of the Saturn-mass planet WASP-39b which resulted in the first detection of \ch{SO2} in an exoplanet's atmosphere \citep{Alderson2023,Rustamkulov2023,Tsai2023}. Since this detection in WASP-39b, there have been additional detections of \ch{SO2} in the atmospheres of the Neptune-mass planet WASP-107b \citep{Dyrek2024,Sing2024,Welbanks2024}, the sub-Neptune GJ\,3470b \citep{Beatty2024} and hints of \ch{SO2} in the sub-Neptune TOI-270d \citep{Benneke2024,Holmberg2024}.

The abundance of \ch{SO2} seen in WASP-39b's spectrum was several orders of magnitude higher than expectations based on equilibrium chemistry. This implies that photochemistry is responsible for the observed abundance of \ch{SO2}, with the reaction chain beginning with the photodissociation of water in the planet's atmosphere \citep{Tsai2023}. This is why our equilibrium chemistry models, that do not include photochemistry, do not attempt to fit the 4.0\,\micron\ feature in WASP-15b's spectrum (Figures \ref{fig:prt_forward_model1}, \ref{fig:prt_best_fit_models}, \ref{fig:PLATON_corner_lowZ} and \ref{fig:um_vs_tiberius_vs_eureka1}). Given that WASP-15b orbits an earlier spectral type star than WASP-39b (F7, \citealt{Triaud2010}, vs.\ G8, \citealt{Faedi2011}), and is more irradiated ($T_{\mathrm{eq}} = 1676 \pm 29$\,K, \citealt{Southworth2013}, vs.\ $T_{\mathrm{eq}} = 1116^{+33}_{-32}$\,K, \citealt{Faedi2011}), it is likely that photochemistry is also important for its atmosphere. 

\begin{figure}
    \centering
    \includegraphics[width=1\columnwidth]{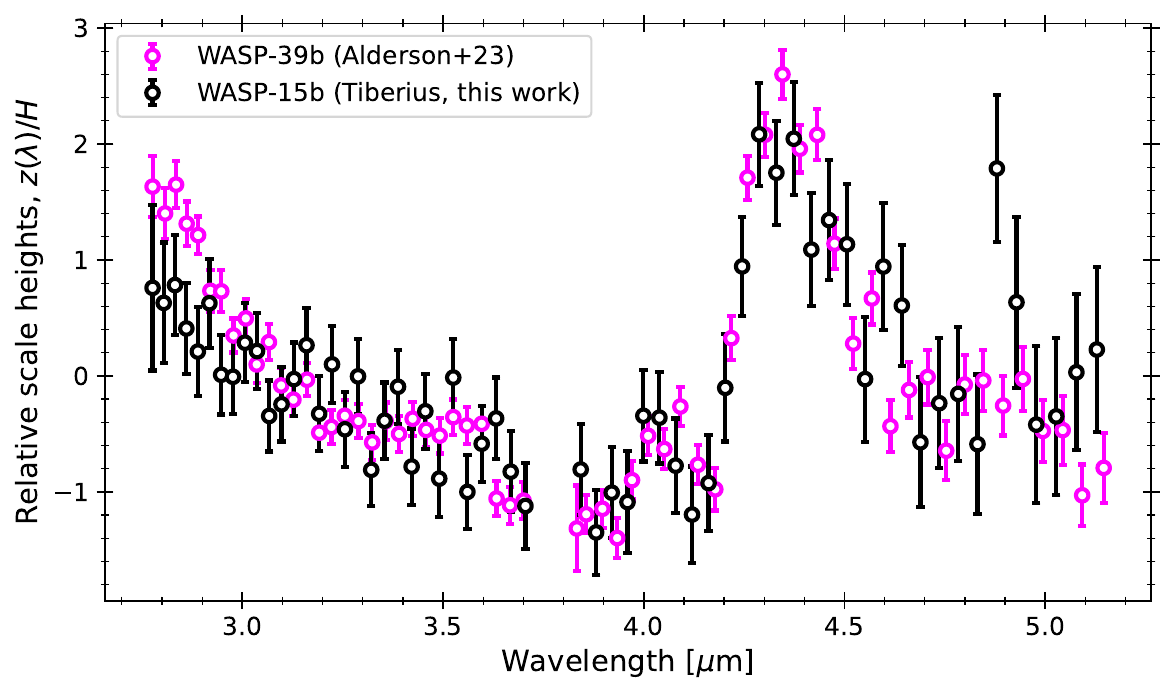}
    \caption{The transmission spectrum of WASP-15b (black, this work) as compared with that of WASP-39b (magenta, \protect\citealt{Alderson2023}) binned to $R=100$. The transmission spectra have been scaled with respect to the planets' atmospheric scale heights.}
    \label{fig:W39_comparison}
\end{figure}

To determine the amplitude of \ch{SO2} absorption in WASP-15b's atmosphere relative to WASP-39b, we plot the NIRSpec/G395H transmission spectrum of WASP-39b from \cite{Alderson2023} along with WASP-15b's transmission spectrum in Figure \ref{fig:W39_comparison}. For this comparison, we binned the WASP-39b spectrum to $R=100$ and then scaled the transit depths of both WASP-39b and WASP-15b by the transit depth corresponding to one scale height for both planets (421\,ppm for WASP-39b and 139\,ppm for WASP-15b). This figure indicates that the amplitude of the \ch{SO2} absorption in both planets is approximately one atmospheric scale height. For WASP-39b, the smaller spectral uncertainties, resulting from the planet's larger scale height, led to a $4.8\sigma$ detection of \ch{SO2} while our larger uncertainties for WASP-15b prevent a statistically significant detection. Follow-up observations of WASP-15b would improve the spectral precision and could search for additional \ch{SO2} absorption features in the mid-infrared \citep{Powell2024}.

Our hybrid \texttt{petitRADTRANS} retrievals place upper limits on the abundance of \ch{SO2} in WASP-15b's atmosphere of $\lesssim 100$\,ppm (Figure \ref{fig:prt_Tiberius_hybrid_corner}), while the forward models favour an abundance of $\sim5$\,ppm (Section \ref{sec:forward_model_pRT}). This abundance of \ch{SO2} would be consistent with the abundances seen in the modestly super-solar metallicity WASP-39b \citep[0.5--25\,ppm,][]{Powell2024} and WASP-107b \citep[6--9\,ppm,][]{Dyrek2024,Sing2024,Welbanks2024} but somewhat smaller than the $270^{+220}_{-120}$\,ppm seen in the $125\pm40\times$ solar metalliticy GJ\,3470b. This is consistent with expectations that \ch{SO2} abundance is correlated with metallicity \citep{Tsai2023}.

The most notable difference between the spectra of WASP-39b and WASP-15b is the feature at 4.9\,\micron\ which is only present in WASP-15b's spectrum. This absorption feature appears in both the \texttt{Tiberius} and \texttt{Eureka!} spectra at $R=100$ and $R=400$ (Figure \ref{fig:trans_spec_R100_R400_comparison}). The feature is confined to a single bin (= 61 pixels) at $R=100$ and a handful of bins in the $R=400$ spectrum. By fitting a Gaussian to the $R=400$ transmission spectrum at these wavelengths, we determine that this feature has a central wavelength of 4.90\,\micron\ and a full width at half maximum of 0.05\,\micron.

We investigated whether this feature is correlated to any unusual behaviour in limb darkening, systematics coefficients, excess red and white noise, and excess bad pixels at these wavelengths. We found no correlation between any of these parameters and the outliers at these wavelengths. Given these tests, and the fact that the absorption feature is independently seen in both the \texttt{Tiberius} and \texttt{Eureka!} reductions, we conclude that this is a real absorption feature.

To interpret what may be the cause of this feature, we explored a number of different molecules that would be present in both chemical equilibrium and disequilibrium, including \ch{CH4}, CO, \ch{H2S}, \ch{HCN}, \ch{O3}, \ch{PH3}, \ch{NH3}, \ch{C2H2}, \ch{SO2}, \ch{NO}, \ch{HC3N}, \ch{H2CO}, \ch{CS2}, \ch{SO}, \ch{SH}, \ch{OCS}, \ch{OH}, \ch{AlH}, \ch{AlO}, \ch{CaH}, \ch{CrH}, \ch{FeH}, MgH, MgO, NaH, SiO, and SiO2.
Of these molecules, the only ones that matched the wavelength of the feature we see are OCS (carbonyl sulphide) and \ch{O3} (ozone). We ruled out \ch{O3} based on the feature width (\ch{O3} would result in a much broader feature) and the implausibility of finding the large abundances of \ch{O3} necessary to match the feature strength in a \ch{H2}-dominated atmosphere.

In Figure \ref{fig:prt_best_fit_models} we showed how the inclusion of OCS in our free and hybrid retreivals with \texttt{petitRADTRANS} leads to a better fit of the 4.9\,\micron\ absorption feature. However, this figure also showed that the width of the feature at 4.9\,\micron\ is narrower than expected for OCS. We investigated if the narrowness of the feature could be caused by OCS at lower pressures or temperatures than the bulk atmospheric composition responsible for the other spectral features. To do this we ran the \texttt{petitRADTRANS} forward model with an abundance of OCS localised between 1 and 100 $\mu$bar, and attempted to vary the temperature and OCS abundance in this part of the atmosphere to fit the feature. A temperature of 300 K with a $2\%$ OCS mixing ratio in this part of the atmosphere gave a feature that was still slightly broader and weaker than the feature observed in the $R=400$ spectrum. Such a cold temperature and high abundance of OCS localised to the upper part of the atmosphere of WASP-15b is highly unphysical.

Aside from the quality of the fits, there are caveats regarding the physics of the OCS interpretation, namely that OCS is not expected to be abundant at the low pressures ($\lesssim$ mbar) probed by transmission \citep{Tsai2021,Tsai2023}. Similar to H$_2$S, OCS is destroyed by photodissociation as well as by photochemically produced atomic H and atomic S. To have OCS at the high altitudes we are observing, either the photochemical sinks must be suppressed, or there exist unidentified production mechanisms. \cite{Jordan2021} found that OCS can survive at high altitudes in a Venus-like atmosphere around M stars with significantly lower NUV flux. However, OCS is expected to be depleted around F-type stars like WASP-15 or even stars with solar-like UV radiation. Alternatively, it is conceivable that OCS might be produced through the oxidation of CS or CS$_2$ after their formation in the upper atmosphere, although the abundances of CS and CS$_2$ remain low in our models. The identification of plausible OCS production is beyond the scope of this study.

In summary, we believe that the absorption we see at 4.9\,\micron\ is astrophysical and not an instrumental artefact. While OCS is the leading candidate, there are several caveats to this interpretation. Alternatively, the feature observed at 4.9\,\micron\ could be produced by a molecule not included in currently available line lists, necessitating additional laboratory work \citep{Fortney2019}.

\section{Conclusions}
\label{sec:conclusions}

We present the 2.8--5.2\,\micron\ transmission spectrum of the misaligned hot Jupiter WASP-15b obtained from a single transit observation with JWST/NIRSpec/G395H. We reduce our data with three independent approaches and find minimal red noise in our data, likely due to the quiet and relatively faint star which allows for a high number groups per integration (44). This allows us to measure a precise transmission spectrum (median uncertainty of 55\,ppm at $R=100$ and 106\,ppm at $R=400$). 

We interpret WASP-15b's spectrum using three independent retrieval codes and GCM simulations. Our spectrum reveals significant absorption from H$_2$O ($4.9\sigma$) and CO$_2$ ($8.9\sigma$), with evidence for SO$_2$ and absorption at 4.9\,\micron\ for which the current best candidate is OCS, albeit with several caveats. If further observations of this planet are able to confirm if the feature at 4.9\,\micron\ is indeed OCS this would be the first detection of this molecule in an exoplanet atmosphere and would allow for new tests of sulphur chemistry in exoplanet atmospheres.

Despite some differences between the absolute abundances inferred from which reduction and retrieval code is adopted, all methods converge on a super-solar metallicity atmosphere ($\gtrsim 15\times$ solar) and a C/O that is consistent with solar but with relatively large uncertainties. Our GCM simulations for WASP-15b suggest that the C/O we measure at the limb is likely representative of the entire photosphere due to the mostly uniform spatial distribution of \ch{H2O}, \ch{CO2} and CO.

The super-solar metallicity we infer for WASP-15b indicates the late accretion of planetesimals. The fact that we also see evidence for sulphur content, may additionally point to planetesimal accretion as sulphur cannot be delivered by metal-rich gas accretion in the inner disc \citep{Bitsch2022}. Given the mass of WASP-15b is much larger than the pebble isolation mass, it likely only had minor pebble accretion late in its formation, instead implying planetesimal accretion as the main driver of solid enrichment. However, we refrain from making comparisons between the planet's C/O and formation models, such as those of \citetalias{PenzlinBooth2024} \citeyear{PenzlinBooth2024}, until we have analysed the rest of the planets in our programme \citep{Kirk2024_align}, as these comparisons need to be performed for a sample of planets rather than individual objects.

This is the first planet to be observed as part of our BOWIE-ALIGN programme that seeks to determine whether a hot Jupiter's atmospheric composition depends on its method of migration, as indicated by its obliquity around an F star (GO 3838, PIs: Kirk \& Ahrer, \citealt{Kirk2024_align}). By combining WASP-15b with the results from the rest of our programme, we will test models of planet formation and demonstrate whether atmospheric composition can be reliably traced to formation history.

\section*{Acknowledgements}

The authors thank Bertram Bitsch for insightful discussion and the anonymous reviewer whose suggestions made our comparative analyses more impactful and our conclusions more robust. This work is based on observations made with the NASA/ESA/CSA James Webb Space Telescope. The data were obtained from the Mikulski Archive for Space Telescopes at the Space Telescope Science Institute, which is operated by the Association of Universities for Research in Astronomy, Inc., under NASA contract NAS 5-03127 for JWST. These observations are associated with program \#3838. This work was inspired by collaboration through the UK-led BOWIE+ collaboration. JK acknowledges financial support from Imperial College London through an Imperial College Research Fellowship grant. N.J.M., D.E.S. and M.Z. acknowledge support from a UKRI Future Leaders Fellowship [Grant MR/T040866/1], a Science and Technology Facilities Funding Council Nucleus Award [Grant ST/T000082/1], and the Leverhulme Trust through a research project grant [RPG-2020-82]. RAB thanks the Royal Society for their support through a University Research Fellowship. PJW acknowledges support from STFC through consolidated grant ST/X001121/1. VP acknowledges support from the UKRI Future Leaders Fellowship grant MR/S035214/1 and UKRI Science and Technology Facilities Council (STFC) through the consolidated grant ST/X001121/1.

\section*{Data Availability}

The data products associated with this manuscript can be found online at Zenodo at \url{https://doi.org/10.5281/zenodo.14779026}. We describe the data products resulting from our survey in \cite{Kirk2024_align}.



\bibliographystyle{mnras}
\bibliography{main} 




\appendix

\section{Allan variance plots for spectroscopic light curves}
\label{sec:Allan_variance_spec}

\begin{figure}
    \centering
    \includegraphics[width=1\columnwidth]{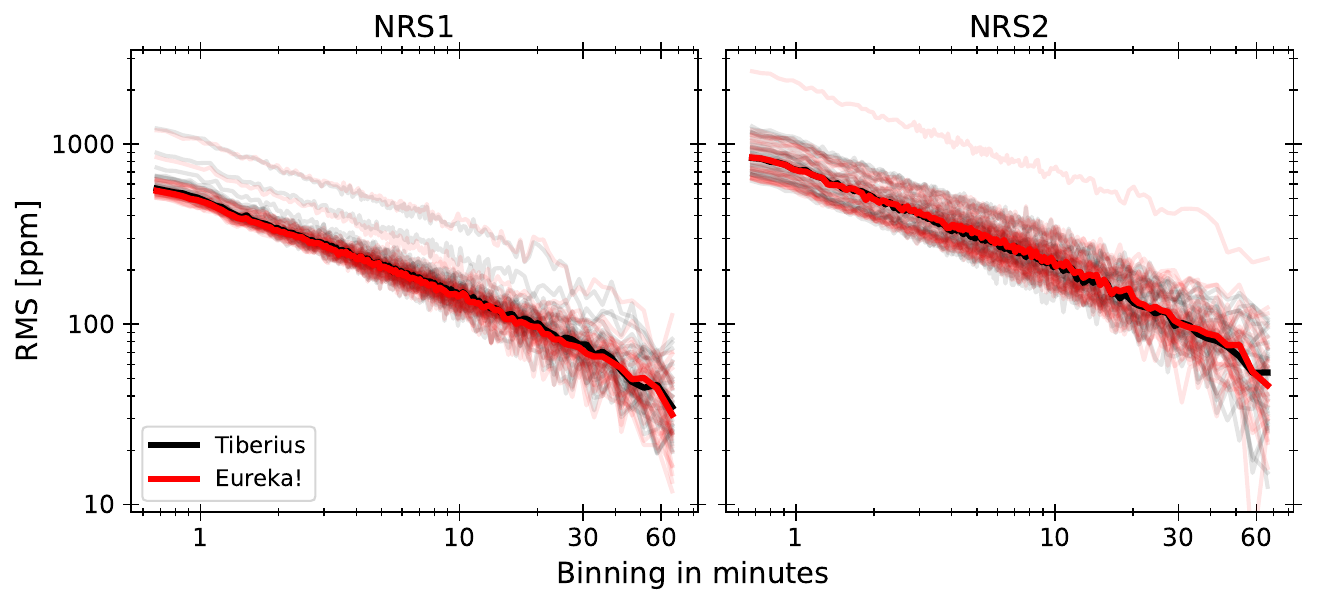}
    \caption{The Allan variance plot from the \texttt{Tiberius} (black) and \texttt{Eureka!} (red) $R=100$ spectroscopic light curve fits for NRS1 (left panel) and NRS2 (right panel). The thick lines show the median Allan variances.}
    \label{fig:Allan_variance_spec}
\end{figure}

This appendix includes the Allan variance plots for the spectroscopic light curves from both the \texttt{Tiberius} and \texttt{Eureka!} reductions in Figure \ref{fig:Allan_variance_spec}.

\section{Second Eureka! reduction}
\label{sec:eureka_VP}

As mentioned in Section \ref{sec:data_reduction}, we performed a second, independent reduction with \texttt{Eureka!} to determine how robust our spectrum is to choices made during the spectral extraction process. Aside from a different choice of wavelength binning to generate the spectroscopic light curves, the key difference between this second reduction and the \texttt{Eureka!} reduction presented in Section \ref{sec:data_reduction} is that we start with the Stage 1 output of \texttt{Tiberius} as described in Section \ref{sec:data_reduction} and use them as inputs to the Stage 2 of \texttt{Eureka!}. We perform the spectral extraction in \texttt{Eureka!}'s Stage 3 the same way as done in Section \ref{sec:data_reduction} except we use a $>$10$\sigma$ threshold for performing the double-iterative masking of outliers along the time axis, extract the background from the area $>$10 pixels away from central pixel of the trace, and use a full width of 8 pixels for optimal spectral extraction. For the light curve fitting, we followed the same steps in the first reduction as described in Section \ref{sec:data_reduction}, except we fixed both the quadratic limb-darkening parameters u1 and u2 to the \texttt{ExoTiC-LD} values.      

\begin{figure}
    \centering
    \includegraphics[width=1\columnwidth]{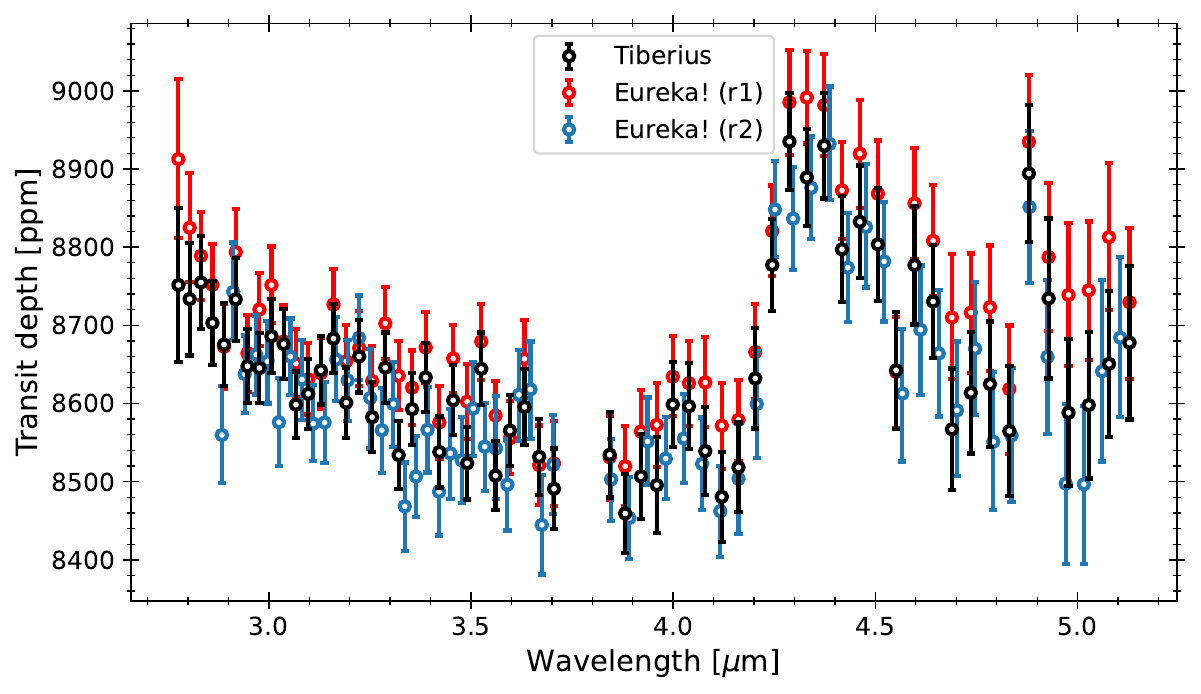}
    \caption{The comparison between the transmission spectra obtained with \texttt{Tiberius} (black), \texttt{Eureka!} reduction 1 (red, Section \ref{sec:data_reduction}), and \texttt{Eureka!} reduction 2 (blue).}
    \label{fig:trans_spec_R100_comparison_all}
\end{figure}

We present the comparison between the spectra from both \texttt{Eureka!} reductions and the \texttt{Tiberius} reduction in Figure \ref{fig:trans_spec_R100_comparison_all}. This figure shows that the spectra from each reduction are consistent with one another. Unlike in Figure \ref{fig:trans_spec_R100_R400_comparison}, no transit depth offset has been applied between the spectra in this plot. The differences in the median transit depths are: \texttt{Eureka!} r1 -- \texttt{Tiberius} $=38$\,ppm, \texttt{Eureka!} r2 -- \texttt{Tiberius} $=-17$\,ppm (in the overlapping wavelength range, $\geq2.9$\,\micron). These differences are both less than the median transit depth uncertainties of each spectrum: \texttt{Eureka!} r1 $= 54$\,ppm, \texttt{Tiberius} $=54$\,ppm, \texttt{Eureka!} r2 $=61$\,ppm. 

\section{TESS light curve fitting}
\label{sec:TESS}

Given the lack of optical wavelength coverage from our G395H transmission spectrum, we decided to fit the TESS light curves. Our goal was to place a constraint on the transit depth at visible wavelengths which could help constrain cloud and haze parameters \citep[e.g.,][]{Wakeford2018,Fairman2024}. Rather than taking the literature TESS transit depth for WASP-15b \citep{Patel2022}, we re-fitted the TESS light curve with the values for the system parameters ($T_0$, $a/R_*$, $i$) that we derived from our JWST light curves (\texttt{Tiberius} values, Table \ref{tab:system_params}). This was done to avoid a bias in the transit depth arising from inconsistent system parameters. 

We used \texttt{lightkurve} \citep{lightkurve} to extract the short cadence (SPOC) TESS light curve and phase-folded this using the period from \cite{Patel2022} and the $T_0$ from our own JWST light curve fits. We trimmed the phase-folded TESS light curve to have the same out-of-transit baseline as our JWST data and fitted the trimmed light curve using a \texttt{batman} model multiplied by a linear polynomial to be consistent with our JWST light curve fits. The resulting  $R_P/R_*$ we derive from the TESS light curve is $0.092155 \pm 0.000470$ ($(R_P/R_*)^2 = 8493 \pm 87$\,ppm), which is within $1\sigma$ of \cite{Patel2022} but is more precise owing to the fewer free parameters in our model. However, despite this improved precision, the inclusion of the TESS data did not improve the precision of our retrievals and did not substantially change the fit of the GCM spectra to our JWST data, so we opted against using the TESS data in our final analysis. 

\section{Nightside dilution calculation}

To determine the amplitude of dilution in the transmission spectrum caused by thermal emission from the planet's nightside, we used both the equations from \cite{Kipping2010} and the \texttt{ExoTETHyS} package \citep{Morello2021}. \texttt{ExoTETHyS} additionally accounts for dilution from the planet's phase curve variations over the course of a transit observation. Both approaches give consistent results, with the dilution amplitude ranging from 5\,ppm at the bluest wavelengths we consider to 14\,ppm at the reddest wavelengths. These variations are significantly smaller than the uncertainties in our $R=100$ spectra and so we do not apply a dilution correction to our final transmission spectrum. 

\section{Parameters and additional plots from atmospheric retrievals}

\begin{figure*}
    \centering
    \settototalheight{\dimen0}{\includegraphics[width=\textwidth]{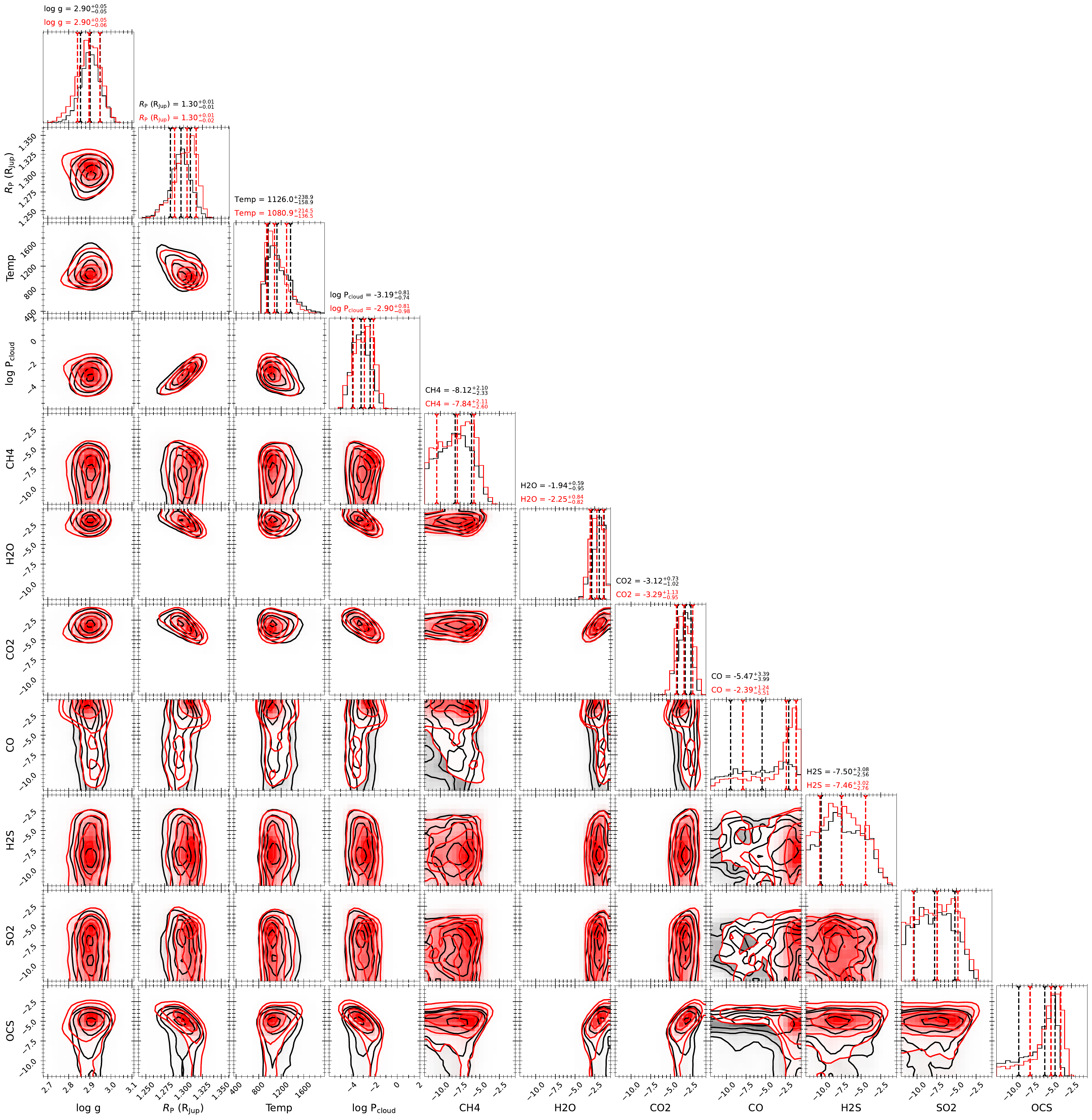}}
    \includegraphics[width=\textwidth]{figures/prt_free_both_corner.pdf}%
    \llap{\raisebox{\dimen0-5cm}{%
    \includegraphics[height=5cm]{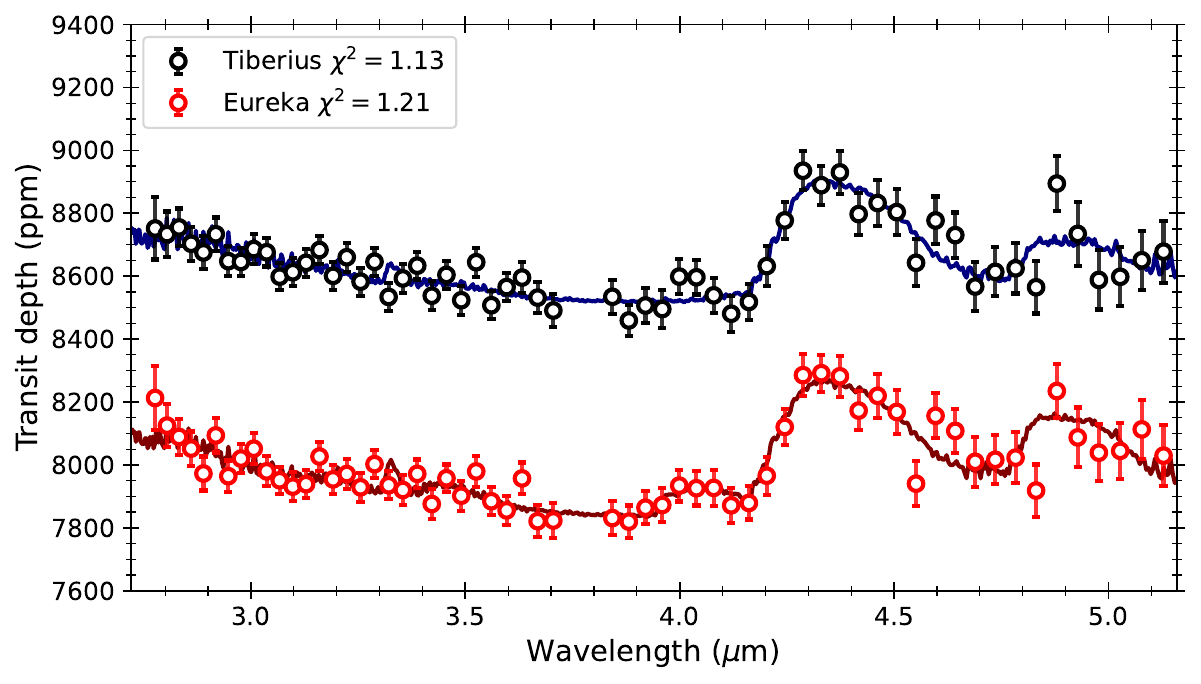}%
    }}
    \caption{Corner plot showing the posterior PDFs from the free \texttt{petitRADTRANS} retrieval performed on the \texttt{Tiberius} $R=100$ spectrum (black) and \texttt{Eureka!} $R=100$ spectrum (red). The best fitting model and residuals are displayed in the top right.}
    \label{fig:prt_free_both_corner}
\end{figure*}

\begin{figure*}
    \centering
    \settototalheight{\dimen0}{\includegraphics[width=\textwidth]{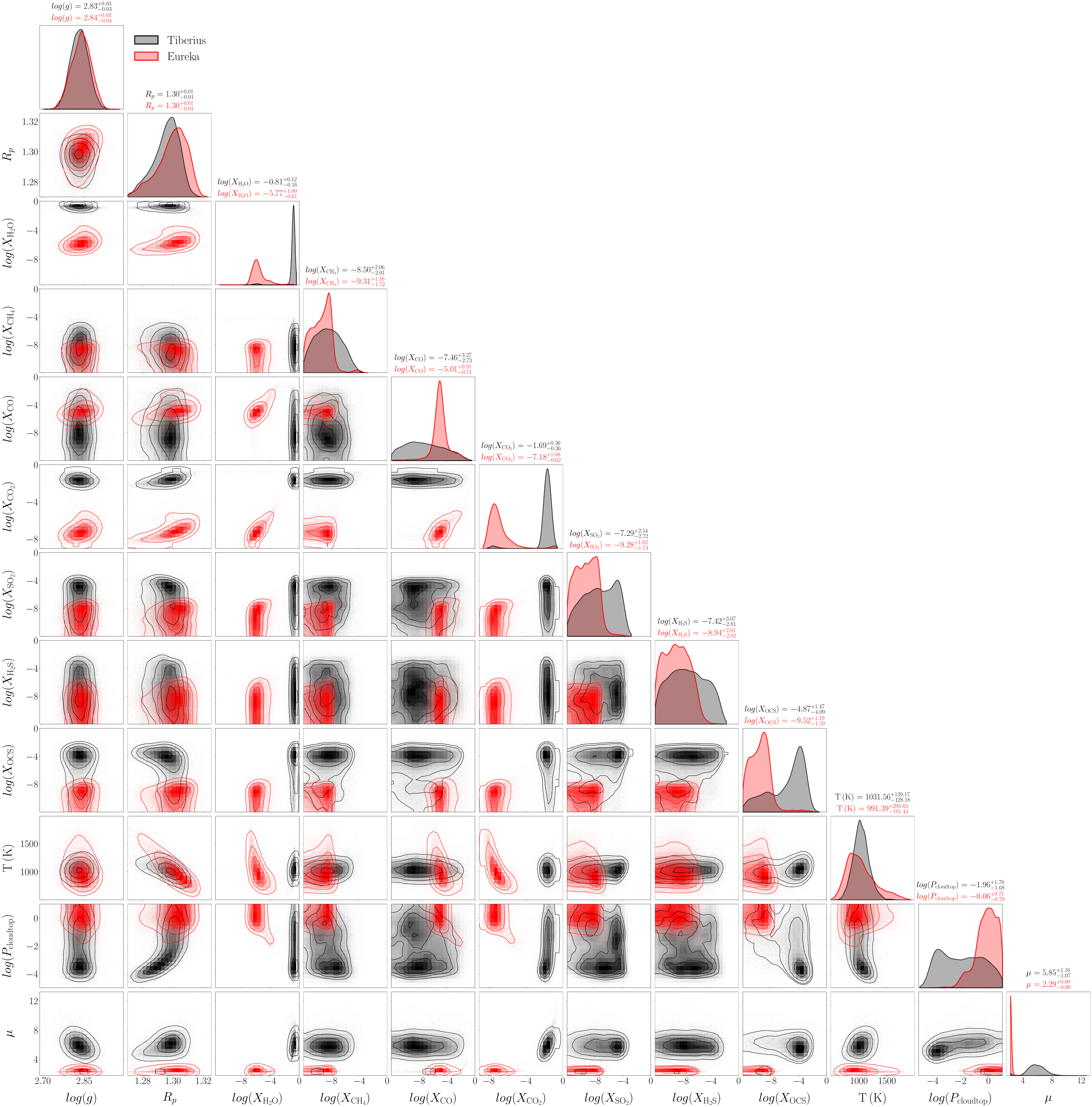}}
    \includegraphics[width=\textwidth]{figures/BeAR_retrieval_comparison_R100_rb.pdf}%
    \llap{\raisebox{\dimen0-6cm}{%
    \includegraphics[height=6cm]{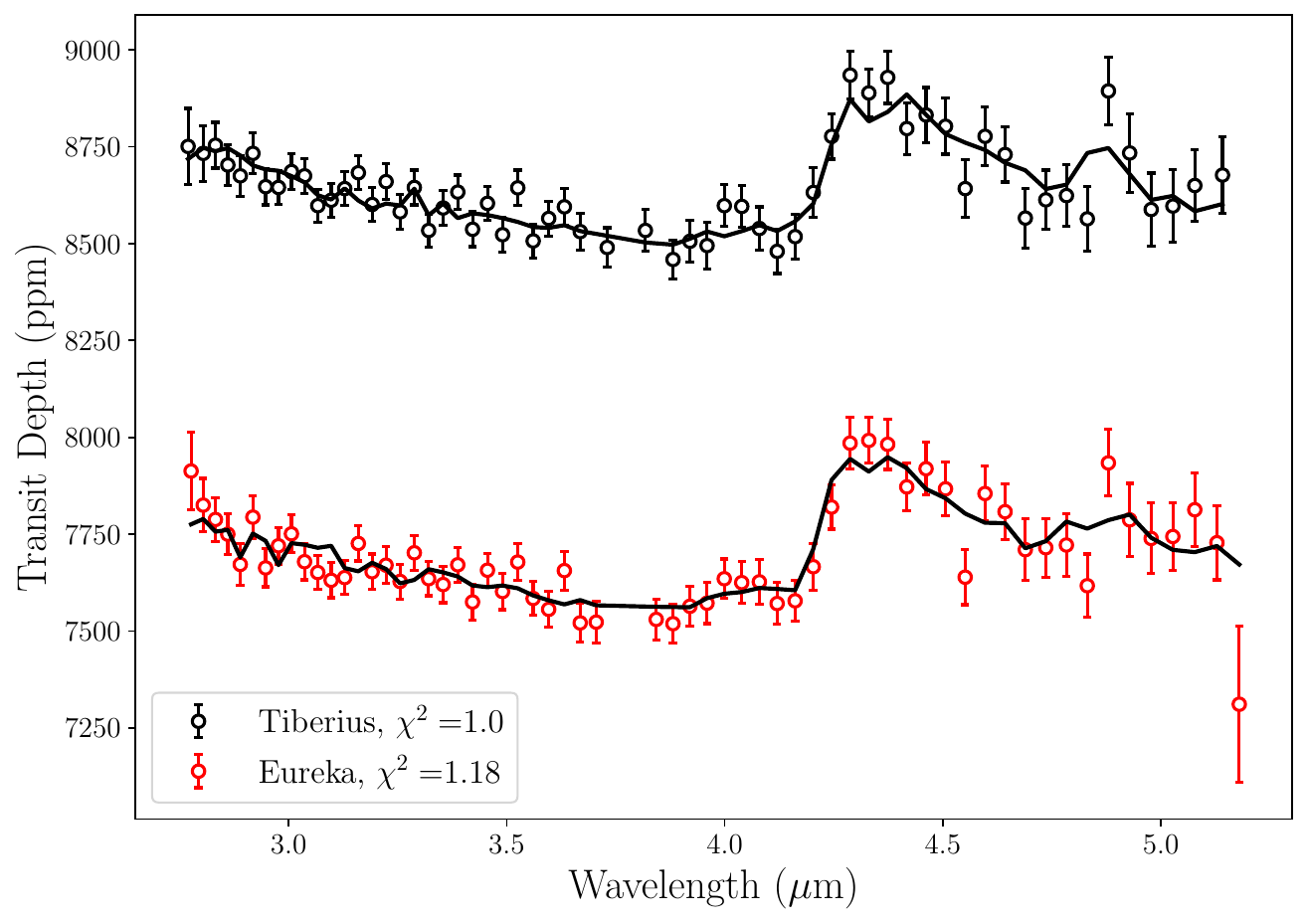}%
    }}
    \caption{Cornerplot showing the posteriors from the BeAR retrievals on the \texttt{Tiberius} (black) and \texttt{Eureka!} (red) reductions of WASP-15 b at $R=100$ prior to excluding the high mean molecular weight solutions. Note that the mean molecular weight, $\mu$, is not a free parameter in the retrievals, but is derived from the retrieved abundances. The top right insert shows the best-fit models for the \texttt{Tiberius} (black) and \texttt{Eureka!} (red) reductions. The \texttt{Eureka!} spectrum is offset by 1000 ppm for visualisation purposes. The legend in the bottom left indicates the reduced $\chi^2$ values for each of the fits.}
    \label{fig:BeAR_cornerplot_R100_unrestricted}
\end{figure*}

\begin{figure*}
    \centering
    \includegraphics[width=\textwidth]{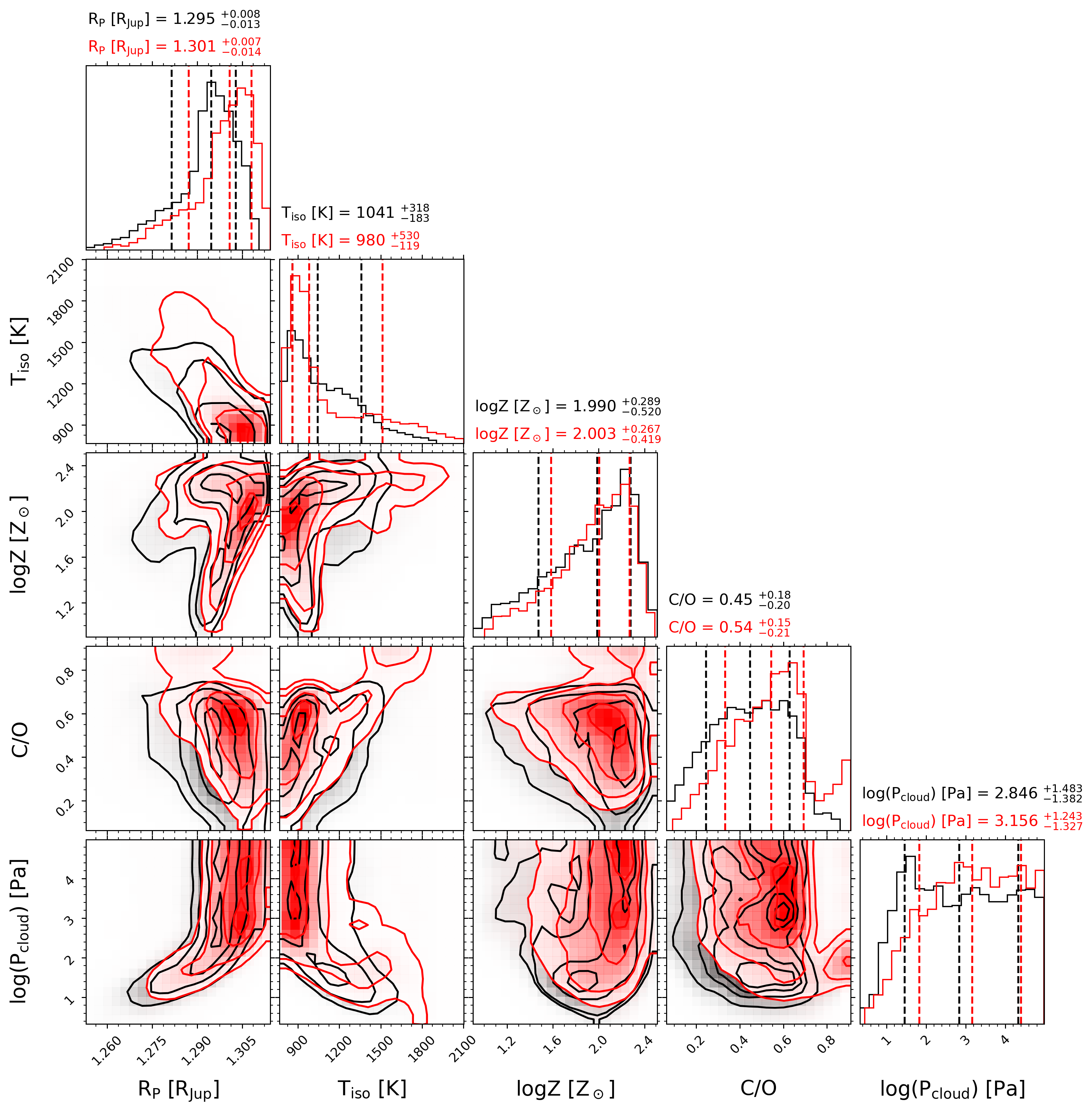}
    \caption{The corner plot from our 1D chemical equilibrium atmosphere retrievals with \texttt{PLATON} run on the $R=100$ spectra over the full, unrestricted metallicity prior range [-1,3]. The black contours correspond to the \texttt{Tiberius} retrieval and the red contours to the \texttt{Eureka!} retrieval. The vertical dashed lines indicate the 16th, 50th (median) and 84th percentiles, which are also given in the axes titles.}
    \label{fig:PLATON_corner}
\end{figure*}

This appendix includes the tabulated parameters from our 1D atmospheric retrievals (Section \ref{sec:1D_models}). We also include a \texttt{petitRADTRANS} corner plot for the free retrievals (Figure \ref{fig:prt_free_both_corner}), the \texttt{BeAR} $R=100$ best fit and corner plot without the restricted metallicity (Figure \ref{fig:BeAR_cornerplot_R100_unrestricted}) and the \texttt{PLATON} corner plot without the restricted metallicity (Figure \ref{fig:PLATON_corner}).

\begin{landscape}
\begin{table}
    \centering
    \caption{The results from our retrievals using \texttt{petitRADTANS} (free and equilibrium chemistry, Section \ref{sec:retrievals_pRT}), \texttt{BeAR} (free chemistry, Section \ref{sec:retrievals_bear}) and \texttt{PLATON} (equilibrium chemistry, Section \ref{sec:retrievals_platon}). In this table, the \texttt{petitRADTRANS} abundances have been converted from the mass fractions presented in the \texttt{petitRADTRANS} corner plots.  Both \texttt{BeAR} and \texttt{PLATON} include additional rows where the posteriors have been recalculated after excluding solutions with $Z > 82\times$ solar, motivated by our interior structure model (Section \ref{sec:interior_model}). For the \texttt{BeAR} retrievals this is limited to the $R=100$ case, as removing the high metallicity solutions from the $R=400$ retrievals does not leave sufficient posterior samples. Parameters that are not fit parameters for each code are marked as "--".}
    \label{tab:all_retrievals}
    \begin{tabular}{l|c|c|c|c|c|c|c|c|c|c|c} \hline
         Input spectrum & $\mathrm{R_P}$ (\Rjup) & $\log g$ (cgs) & $\mathrm{T_{iso}}$ (K) & $\log P_{\mathrm{cloud}}$ (bar) & $Z$ ($\times$ solar) & C/O & $\log(X_{\mathrm{H_2O}})$ & $\log(X_{\mathrm{CO_2}})$ & $\log(X_{\mathrm{CO}})$ & $\log(X_{\mathrm{SO_2}})$ & $\log(X_{\mathrm{OCS}})$ \\
         \hline\texttt{petitRADTRANS} & & & & & & & & & & &   \\
         \textit{Equilibrium Chemistry:} & & & & & & & & & & &   \\
         \texttt{Tiberius}, $R=100$ & $1.302\pm0.003$ & $2.88\pm0.05$ & $910${\raisebox{0.5ex}{\tiny$\substack{+110 \\ -73}$}} & $-0.50${\raisebox{0.5ex}{\tiny$\substack{+1.61 \\ -2.04}$}} & $18${\raisebox{0.5ex}{\tiny$\substack{+22 \\ -8}$}} & $0.48${\raisebox{0.5ex}{\tiny$\substack{+0.11 \\ -0.16}$}} & -- & -- & -- & -- & --   \\
         \texttt{Eureka!}, $R=100$ & $1.305\pm0.002$ & $2.87${\raisebox{0.5ex}{\tiny$\substack{+0.05 \\ -0.04}$}} & $914${\raisebox{0.5ex}{\tiny$\substack{+86 \\ -62}$}} & $-0.45${\raisebox{0.5ex}{\tiny$\substack{+1.62 \\ -1.76}$}} & $22${\raisebox{0.5ex}{\tiny$\substack{+27 \\ -9}$}} & $0.53${\raisebox{0.5ex}{\tiny$\substack{+0.09 \\ -0.16}$}} & -- & -- & -- & -- & --   \\
         \textit{Free Chemistry:} & & & & & & & & & & &   \\
         \texttt{Tiberius}, $R=100$ & $1.296\pm0.013$ & $2.90\pm0.05$ & $1126${\raisebox{0.5ex}{\tiny$\substack{+239 \\ -159}$}} & $-3.19${\raisebox{0.5ex}{\tiny$\substack{+0.81 \\ -0.74}$}} & -- & -- & $-2.83${\raisebox{0.5ex}{\tiny$\substack{+0.59 \\ -0.95}$}} & $-4.40${\raisebox{0.5ex}{\tiny$\substack{+0.73 \\ -1.02}$}} & $-6.55${\raisebox{0.5ex}{\tiny$\substack{+3.39 \\ -3.99}$}} & $-9.20${\raisebox{0.5ex}{\tiny$\substack{+2.56 \\ -2.61}$}} & $-7.31${\raisebox{0.5ex}{\tiny$\substack{+1.31 \\ -3.29}$}}  \\
         \texttt{Eureka!}, $R=100$ & $1.304\pm0.002$ & $2.90${\raisebox{0.5ex}{\tiny$\substack{+0.05 \\ -0.06}$}} & $1081${\raisebox{0.5ex}{\tiny$\substack{+215 \\ -137}$}} & $-2.90${\raisebox{0.5ex}{\tiny$\substack{+0.81 \\ -0.98}$}} & -- & -- & $-3.14${\raisebox{0.5ex}{\tiny$\substack{+0.84 \\ -0.82}$}} & $-4.57${\raisebox{0.5ex}{\tiny$\substack{+1.13 \\ -0.95}$}} & $-3.48${\raisebox{0.5ex}{\tiny$\substack{+1.24 \\ -5.51}$}} & $-8.88${\raisebox{0.5ex}{\tiny$\substack{+2.64 \\ -3.02}$}} & $-6.52${\raisebox{0.5ex}{\tiny$\substack{+1.22 \\ -2.64}$}}   \\
         \textit{Hybrid Chemistry:} & & & & & & & & & & &   \\
         \texttt{Tiberius}, $R=100$$^\dagger$ & $1.306\pm0.002$ & $2.87${\raisebox{0.5ex}{\tiny$\substack{+0.04 \\ -0.05}$}}  & $887${\raisebox{0.5ex}{\tiny$\substack{+65 \\ -56}$}} & $-0.66${\raisebox{0.5ex}{\tiny$\substack{+1.78 \\ -1.83}$}} & $22${\raisebox{0.5ex}{\tiny$\substack{+20 \\ -10}$}} & $0.53${\raisebox{0.5ex}{\tiny$\substack{+0.09 \\ -0.15}$}} & -- & -- & -- & $-6.56${\raisebox{0.5ex}{\tiny$\substack{+1.11 \\ -4.46}$}} & $-8.84${\raisebox{0.5ex}{\tiny$\substack{+2.04 \\ -3.10}$}}  \\
         \texttt{Tiberius}, $R=400$ & $1.306\pm0.002$ & $2.87\pm0.05$  & $882\pm58$ & $-0.65${\raisebox{0.5ex}{\tiny$\substack{+1.76 \\ -1.90}$}} & $17${\raisebox{0.5ex}{\tiny$\substack{+14 \\ -8}$}} & $0.51${\raisebox{0.5ex}{\tiny$\substack{+0.10 \\ -0.16}$}} & -- & -- & -- & $-7.13${\raisebox{0.5ex}{\tiny$\substack{+1.54 \\ -4.14}$}} & $-9.14${\raisebox{0.5ex}{\tiny$\substack{+2.09 \\ -2.72}$}}  \\
         \texttt{Eureka!}, $R=100$ & $1.303\pm0.002$ & $2.84\pm0.05$ & $891${\raisebox{0.5ex}{\tiny$\substack{+47 \\ -49}$}} & $-0.48${\raisebox{0.5ex}{\tiny$\substack{+1.68 \\ -1.78}$}} & $24${\raisebox{0.5ex}{\tiny$\substack{+19 \\ -11}$}} & $0.56${\raisebox{0.5ex}{\tiny$\substack{+0.07 \\ -0.13}$}} & -- & -- & -- & $-5.49${\raisebox{0.5ex}{\tiny$\substack{+0.34 \\ -1.63}$}} & $-6.52${\raisebox{0.5ex}{\tiny$\substack{+0.65 \\ -2.15}$}}  \\
         \texttt{Eureka!}, $R=400$ & $1.303\pm0.002$ & $2.85${\raisebox{0.5ex}{\tiny$\substack{+0.04 \\ -0.05}$}} & $882${\raisebox{0.5ex}{\tiny$\substack{+43 \\ -46}$}} & $-0.63${\raisebox{0.5ex}{\tiny$\substack{+1.75 \\ -1.70}$}} & $22${\raisebox{0.5ex}{\tiny$\substack{+16 \\ -9}$}} & $0.57${\raisebox{0.5ex}{\tiny$\substack{+0.07 \\ -0.12}$}} & -- & -- & -- & $-5.53${\raisebox{0.5ex}{\tiny$\substack{+0.35 \\ -1.87}$}} & $-6.90${\raisebox{0.5ex}{\tiny$\substack{+0.76 \\ -3.25}$}}  \\
         \hline\texttt{BeAR} & & & & & & & & & & &   \\
         \textit{Full prior range:} & & & & & & & & & & &   \\
         \texttt{Tiberius}, $R=100$ & $1.297${\raisebox{0.5ex}{\tiny$\substack{+0.007 \\ -0.01}$}} & $2.83\pm0.03$ & $1031${\raisebox{0.5ex}{\tiny$\substack{+139 \\ -128}$}} & $-1.96${\raisebox{0.5ex}{\tiny$\substack{+1.78 \\ -1.68}$}} & -- & -- & $-0.81${\raisebox{0.5ex}{\tiny$\substack{+0.12 \\ -0.16}$}} & $-1.69\pm0.36$ & $-7.46${\raisebox{0.5ex}{\tiny$\substack{+3.27 \\ -2.73}$}} & $-7.29${\raisebox{0.5ex}{\tiny$\substack{+2.54 \\ -2.72}$}} & $-4.87${\raisebox{0.5ex}{\tiny$\substack{+1.47 \\ -4.09}$}} \\
         \texttt{Tiberius}, $R=400$ & $1.302\pm0.007$ & $2.84\pm0.03$ & $1163${\raisebox{0.5ex}{\tiny$\substack{+212 \\ -192}$}} & $-1.46${\raisebox{0.5ex}{\tiny$\substack{+1.39 \\ -1.38}$}} & -- & -- & $-0.82${\raisebox{0.5ex}{\tiny$\substack{+0.15 \\ -0.29}$}} & $-1.13${\raisebox{0.5ex}{\tiny$\substack{+0.23 \\ -0.2}$}} & $-7.59${\raisebox{0.5ex}{\tiny$\substack{+3.2 \\ -2.67}$}} & $-7.99${\raisebox{0.5ex}{\tiny$\substack{+2.28 \\ -2.21}$}} & $-7.69${\raisebox{0.5ex}{\tiny$\substack{+2.3 \\ -2.44}$}} \\
         \texttt{Eureka!}, $R=100$ & $1.301${\raisebox{0.5ex}{\tiny$\substack{+0.008 \\ -0.011}$}} & $2.84${\raisebox{0.5ex}{\tiny$\substack{+0.03 \\ -0.04}$}} & $991${\raisebox{0.5ex}{\tiny$\substack{+294 \\ -191}$}} & $-0.06${\raisebox{0.5ex}{\tiny$\substack{+0.71 \\ -0.78}$}} & -- & -- & $-5.77${\raisebox{0.5ex}{\tiny$\substack{+1.0 \\ -0.61}$}} & $-7.18${\raisebox{0.5ex}{\tiny$\substack{+1.08 \\ -0.62}$}} & $-5.01${\raisebox{0.5ex}{\tiny$\substack{+0.91 \\ -0.71}$}} & $-9.28${\raisebox{0.5ex}{\tiny$\substack{+1.62 \\ -1.74}$}} & $-9.52${\raisebox{0.5ex}{\tiny$\substack{+1.19 \\ -1.59}$}} \\
         \texttt{Eureka!}, $R=400$ & $1.296${\raisebox{0.5ex}{\tiny$\substack{+0.01 \\ -0.011}$}} & $2.84\pm0.03$ & $1498${\raisebox{0.5ex}{\tiny$\substack{+274 \\ -292}$}} & $-2.09${\raisebox{0.5ex}{\tiny$\substack{+1.58 \\ -1.37}$}} & -- & -- & $-0.91${\raisebox{0.5ex}{\tiny$\substack{+0.16 \\ -0.32}$}} & $-0.98${\raisebox{0.5ex}{\tiny$\substack{+0.16 \\ -0.14}$}} & $-6.5${\raisebox{0.5ex}{\tiny$\substack{+3.27 \\ -3.36}$}} & $-7.77${\raisebox{0.5ex}{\tiny$\substack{+2.62 \\ -2.57}$}} & $-7.1${\raisebox{0.5ex}{\tiny$\substack{+3.17 \\ -2.94}$}} \\
         \textit{High Z solutions excluded} & & & & & & & & & & &   \\
         \texttt{Tiberius}, $R=100$ & $1.291\pm0.007$ & $2.84\pm0.03$ & $1169${\raisebox{0.5ex}{\tiny$\substack{+190 \\ -198}$}} & $0.11${\raisebox{0.5ex}{\tiny$\substack{+0.53 \\ -0.81}$}} & -- & -- & $-5.79${\raisebox{0.5ex}{\tiny$\substack{+0.4 \\ -0.47}$}} & $-7.44${\raisebox{0.5ex}{\tiny$\substack{+0.5 \\ -0.32}$}} & $-7.13${\raisebox{0.5ex}{\tiny$\substack{+1.27 \\ -2.58}$}} & $-9.79${\raisebox{0.5ex}{\tiny$\substack{+1.42 \\ -1.29}$}} & $-9.06${\raisebox{0.5ex}{\tiny$\substack{+0.68 \\ -1.6}$}} \\
         \texttt{Eureka!}, $R=100$ & $1.301${\raisebox{0.5ex}{\tiny$\substack{+0.008 \\ -0.011}$}} & $2.84${\raisebox{0.5ex}{\tiny$\substack{+0.03 \\ -0.04}$}} & $992${\raisebox{0.5ex}{\tiny$\substack{+298 \\ -190}$}} & $-0.03${\raisebox{0.5ex}{\tiny$\substack{+0.69 \\ -0.74}$}} & -- & -- & $-5.79${\raisebox{0.5ex}{\tiny$\substack{+0.85 \\ -0.59}$}} & $-7.22${\raisebox{0.5ex}{\tiny$\substack{+0.9 \\ -0.6}$}} & $-5.02${\raisebox{0.5ex}{\tiny$\substack{+0.83 \\ -0.67}$}} & $-9.32${\raisebox{0.5ex}{\tiny$\substack{+1.6 \\ -1.71}$}} & $-9.59${\raisebox{0.5ex}{\tiny$\substack{+1.15 \\ -1.54}$}} \\

         \hline\texttt{PLATON} & & & & & & & & & & &   \\
         \textit{Full prior range:} & & & & & & & & & & &   \\
         \texttt{Tiberius}, $R=100$ & $1.295${\raisebox{0.5ex}{\tiny$\substack{+0.008 \\ -0.013}$}} & -- & $1041${\raisebox{0.5ex}{\tiny$\substack{+318 \\ -183}$}} & $-3.846${\raisebox{0.5ex}{\tiny$\substack{+1.483 \\ -1.382}$}} & $97${\raisebox{0.5ex}{\tiny$\substack{+92 \\ -68}$}} & $0.45${\raisebox{0.5ex}{\tiny$\substack{+0.18 \\ -0.20}$}} & -- & -- & -- & -- & -- \\
         \texttt{Tiberius}, $R=400$ & $1.300${\raisebox{0.5ex}{\tiny$\substack{+0.006 \\ -0.008}$}} & -- & $1016${\raisebox{0.5ex}{\tiny$\substack{+310 \\ -175}$}} & $-2.280${\raisebox{0.5ex}{\tiny$\substack{+1.173 \\ -1.391}$}} & $162${\raisebox{0.5ex}{\tiny$\substack{+77 \\ -92}$}} & $0.45${\raisebox{0.5ex}{\tiny$\substack{+0.17 \\ -0.20}$}} & -- & -- & -- & -- & -- \\
         \texttt{Eureka!}, $R=100$ & $1.301${\raisebox{0.5ex}{\tiny$\substack{+0.007 \\ -0.014}$}} & -- & $980${\raisebox{0.5ex}{\tiny$\substack{+530 \\ -119}$}} & $-2.156${\raisebox{0.5ex}{\tiny$\substack{+1.243 \\ -1.327}$}} & $100${\raisebox{0.5ex}{\tiny$\substack{+85 \\ -62}$}} & $0.54${\raisebox{0.5ex}{\tiny$\substack{+0.15 \\ -0.21}$}} & -- & -- & -- & -- & -- \\
         \texttt{Eureka!}, $R=400$ & $1.304${\raisebox{0.5ex}{\tiny$\substack{+0.005 \\ -0.008}$}} & -- & $960${\raisebox{0.5ex}{\tiny$\substack{+285 \\ -93}$}} & $-2.440${\raisebox{0.5ex}{\tiny$\substack{+1.067 \\ -1.242}$}} & $123${\raisebox{0.5ex}{\tiny$\substack{+78 \\ -71}$}} & $0.57${\raisebox{0.5ex}{\tiny$\substack{+0.10 \\ -0.18}$}} & -- & -- & -- & -- & -- \\ 
        \textit{High Z solutions excluded} & & & & & & & & & & &   \\
         \texttt{Tiberius}, $R=100$ & $1.293${\raisebox{0.5ex}{\tiny$\substack{+0.005 \\ -0.017}$}} & -- & $933${\raisebox{0.5ex}{\tiny$\substack{+293 \\ -97}$}} & $-3.772${\raisebox{0.5ex}{\tiny$\substack{+1.511 \\ -1.319}$}} & $35${\raisebox{0.5ex}{\tiny$\substack{+23 \\ -19}$}} & $0.49${\raisebox{0.5ex}{\tiny$\substack{+0.15 \\ -0.20}$}} & -- & -- & -- & -- & -- \\
         \texttt{Tiberius}, $R=400$ & $1.299${\raisebox{0.5ex}{\tiny$\substack{+0.003 \\ -0.005}$}} & -- & $852${\raisebox{0.5ex}{\tiny$\substack{+84 \\ -54}$}} & $-2.539${\raisebox{0.5ex}{\tiny$\substack{+0.915 \\ -1.216}$}} & $44${\raisebox{0.5ex}{\tiny$\substack{+18 \\ -21}$}} & $0.51${\raisebox{0.5ex}{\tiny$\substack{+0.12 \\ -0.19}$}} & -- & -- & -- & -- & -- \\
         \texttt{Eureka!}, $R=100$ & $1.299${\raisebox{0.5ex}{\tiny$\substack{+0.005 \\ -0.011}$}} & -- & $923${\raisebox{0.5ex}{\tiny$\substack{+195 \\ -78}$}} & $-2.141${\raisebox{0.5ex}{\tiny$\substack{+1.213 \\ -1.180}$}} & $41${\raisebox{0.5ex}{\tiny$\substack{+21 \\ -22}$}} & $0.57${\raisebox{0.5ex}{\tiny$\substack{+0.11 \\ -0.18}$}} & -- & -- & -- & -- & -- \\
         \texttt{Eureka!}, $R=400$ & $1.300${\raisebox{0.5ex}{\tiny$\substack{+0.003 \\ -0.005}$}} & -- & $915${\raisebox{0.5ex}{\tiny$\substack{+77 \\ -61}$}} & $-2.469${\raisebox{0.5ex}{\tiny$\substack{+1.010 \\ -1.018}$}} & $42${\raisebox{0.5ex}{\tiny$\substack{+19 \\ -21}$}} & $0.61${\raisebox{0.5ex}{\tiny$\substack{+0.07 \\ -0.15}$}} & -- & -- & -- & -- & -- \\ 
          \hline
          \multicolumn{12}{l}{$^\dagger$Our favoured interpretation for the reasons discussed in Section \protect\ref{sec:discussion_differences_retrieval}.}
    \end{tabular}
\end{table}
\end{landscape}


\section{Additional details about the \texttt{UM}}
\label{appendix:um}

\subsection{WASP-15 system parameters used in the \texttt{UM} simulations}
\label{appendix:um_wasp15_system_parameters}

Tables~\ref{tab:wasp15_parameters_um} and \ref{tab:wasp15b_parameters_um} show the stellar and planetary parameters, respectively, used in the \texttt{UM} simulations presented in this study.

\begin{table}
\caption{WASP-15 parameters used in the \texttt{UM} simulations.}
\label{tab:wasp15_parameters_um}
\centering
\begin{tabular}{lcc}
\hline
Parameter & Value & Unit\\
\hline
Type & F7 & \\
Radius & \num{10.03e8} $^a$ & \unit{\metre}\\
Effective temperature & 6300 $^b$ & \unit{\kelvin}\\
Stellar constant at 1 au & 4235.10 & \unit{\watt\per\square\metre}\\
$\log_{10}$(surface gravity) & 4.00 $^c$ & Gal (cgs)\\
$[\ce{Fe}/\ce{H}]$ & 0.00 $^d$ & dex\\
\hline
\multicolumn{3}{l}{$^a$ 10.03$\times$\num{e8} = $1.48$\,\Rsun $\approx$ $1.477 \pm 0.072$\,\Rsun \citep{Bonomo2017}} \\
\multicolumn{3}{l}{$^b$ \SI{6300}{\kelvin} $\approx$ $6372 \pm 13$\,K  \citep{GaiaDR3}} \\
\multicolumn{3}{l}{$^c$ 4.00 (cgs) $\approx$ $4.17$ (cgs) \citep{Bonomo2017}} \\
\multicolumn{3}{l}{$^d$ 0.00 $\approx$ $-0.17$ \citep{Bonomo2017}} \\
\end{tabular}
\end{table}

\begin{table}
\centering
\caption{WASP-15b parameters used in the \texttt{UM} simulations.}
\label{tab:wasp15b_parameters_um}
\begin{tabular}[t]{lcc}
\hline
Parameter & Value & Unit\\
\hline
Inner radius & \num{9.07e7} $^a$ & \unit{\metre}\\
Domain height & \num{1.50e7} $^a$ & \unit{\metre}\\
Semi-major axis & 0.0520 $^b$ & au \\
Orbital period & 3.7521 $^c$ & Earth day\\
Rotation rate & \num{1.94e-05} & \unit{\radian\per\second}\\
Surface gravity at inner radius & 8.30 $^d$ & \unit{\metre\per\square\second}\\ 
Intrinsic temperature & 300 & \unit{\kelvin}\\
Metallicity [M/H] & 10$\times$solar & \citet{Asplund2009}\\
\ce{C}/\ce{O} & 0.55 & \citet{Asplund2009}\\
Specific gas constant & 3256.02 & \unit{\joule\per\kelvin\per\kilogram}\\ 
Specific heat capacity & \num{1.25e+04} & \unit{\joule\per\kelvin\per\kilogram}\\ 
Stellar irradiance & \num{1.57e+06} & \unit{\watt\per\square\metre}\\ 
Effective temperature & 1555 $^e$ & \unit{\kelvin}\\
\hline
\multicolumn{3}{l}{$^a$ (9.07+1.50)$\times$\num{e7} \unit{\metre} = 1.06$\times$\num{e8} \unit{\metre} = $1.48$\,\Rjup}\\
\multicolumn{3}{l}{$^b$ 0.0520 au $\approx$ 0.05165 au \citep{Southworth2013}} \\
\multicolumn{3}{l}{$^c$ 3.7521 Earth day $\approx$ 3.75209748 Earth day \citep{Southworth2013}} \\
\multicolumn{3}{l}{$^d$ With the inner radius boundary initially placed at \SI{200}{\bar}.} \\
\multicolumn{3}{l}{$^e$ Calculated at pseudo-steady state as $\mathrm{(OLR/\sigma)^{1/4}}$, where OLR is} \\
\multicolumn{3}{l}{the global mean top-of-the-atmosphere outgoing longwave radiation} \\
\multicolumn{3}{l}{and $\sigma$ is the Stefan–Boltzmann constant.} \\
\end{tabular}
\end{table}

\subsection{\texttt{UM} initialisation and runtime}
\label{sec:um_initialisation_runtime}

The \texttt{UM} equilibrium simulation was performed first. We initialised this simulation at rest with a piecewise power-law  pressure-temperature profile crudely approximating the results from initial tests with analytic chemistry. This simulation was then ran for 1500 Earth days to let the upper atmosphere (from \SI{e-3}{\bar} to \SI{1}{\bar}) reach a pseudo-steady state dynamically, radiatively and chemically. The \texttt{UM} kinetics simulation was initialised from day 1000 of the \texttt{UM} equilibrium simulation, and ran for another 1000 Earth days.

\subsection{Constructing \texttt{UM} high resolution transmission spectrum}
\label{appendix:um_high_res_spectra}

During normal \texttt{UM} runtime, radiative transfer is computed for 32 spectral bands covering \SIrange{0.2}{322}{\micro\metre}. During diagnostic \texttt{UM} runs required to obtain a planet's transmission spectrum \citep{Lines2018}, radiative transfer was computed at a higher spectral resolution for two sets of spectral bands, (1) 500 spectral bands covering \SIrange{0.2}{10000}{\micro\metre} and (2) 500 spectral bands covering \SIrange{2}{10000}{\micro\metre}. The resulting two high resolution transmission spectra were combined into one high resolution spectrum via post-processing.

\subsection{\texttt{UM} pressure-temperature and chemical species vertical profiles}
\label{appendix:um_pt_chem_profiles}

\begin{figure}
    \centering
    \includegraphics[scale=0.416]{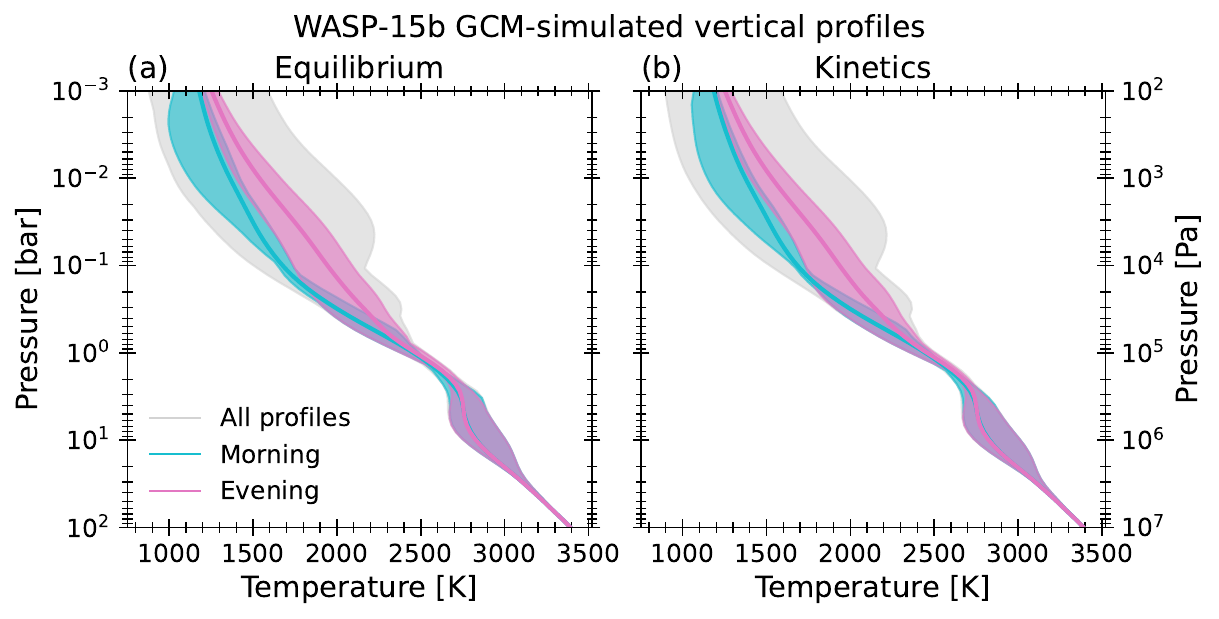}
    \caption{Pressure-temperature vertical profiles predicted by the \texttt{UM} equilibrium (left column) and kinetics (right column) simulations of WASP-15b's atmosphere. Grey shading shows the range of abundances for the entire atmosphere, cyan shading --- for the morning terminator only, pink shading  --- for the evening terminator only. Solid cyan and pink lines indicate the meridional mean for the morning and evening terminator, respectively.}
    \label{fig:um_pt_profiles_appendix}
\end{figure}

Figure \ref{fig:um_pt_profiles_appendix} shows pressure-temperature profiles predicted by the \texttt{UM} equilibrium and kinetics simulations for WASP-15b's entire atmosphere and its terminator region (separately for the morning and evening terminators, i.e., exactly \ang{90}E and \ang{270}E, respectively, without averaging over the opening angle). The data were averaged over the last 200 simulation days.

\section{Additional details regarding photochemical models}
\label{appendix:photochem}

\subsection{Terminator Pressure-Temperature Profiles, K\MakeLowercase{zz} Profile, and limb separated transmission spectra}
\label{appendix:photochem_inputs}
\begin{figure}
     \centering
     \includegraphics[width=0.9\columnwidth]{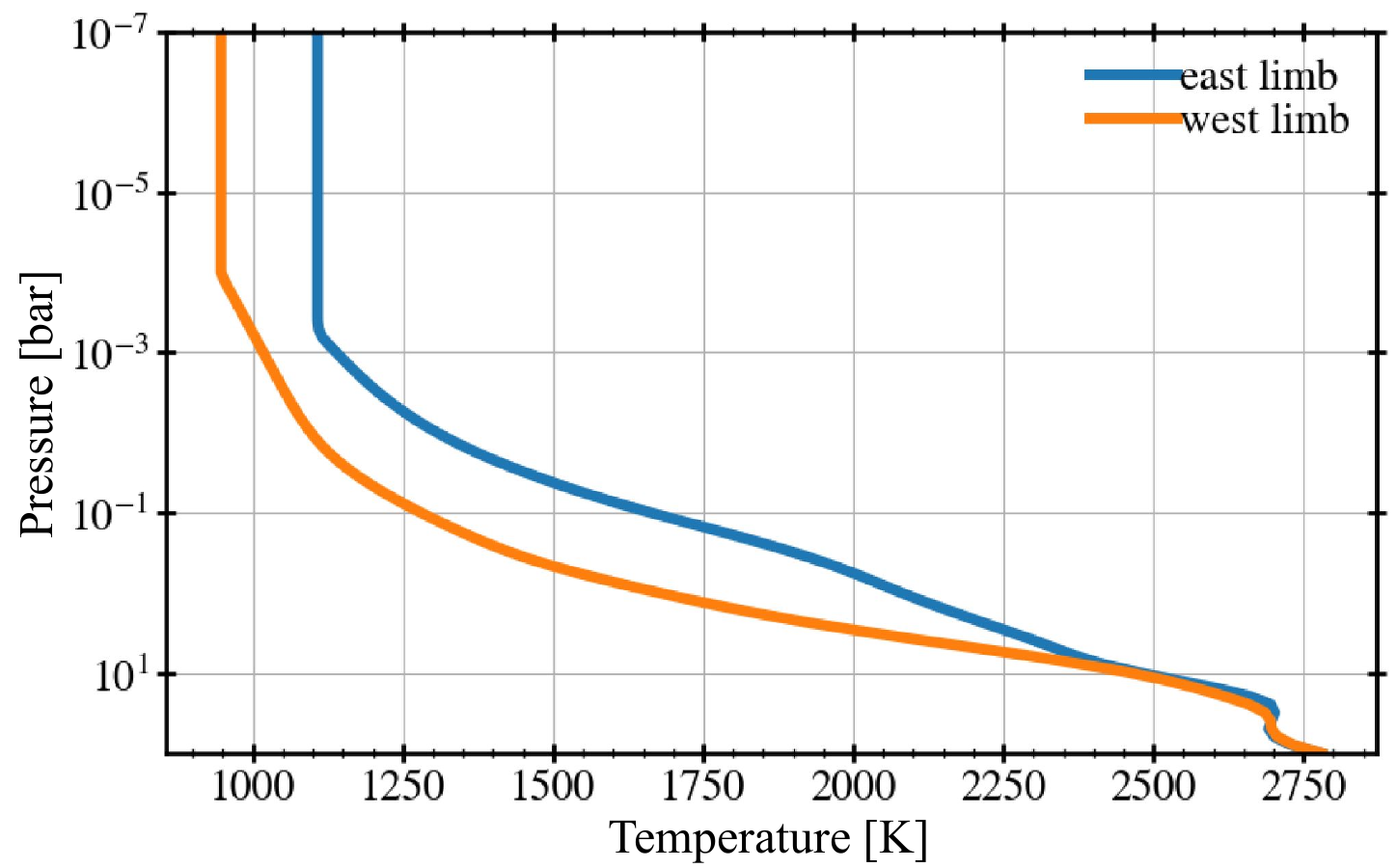}
     \includegraphics[width=0.9\columnwidth]{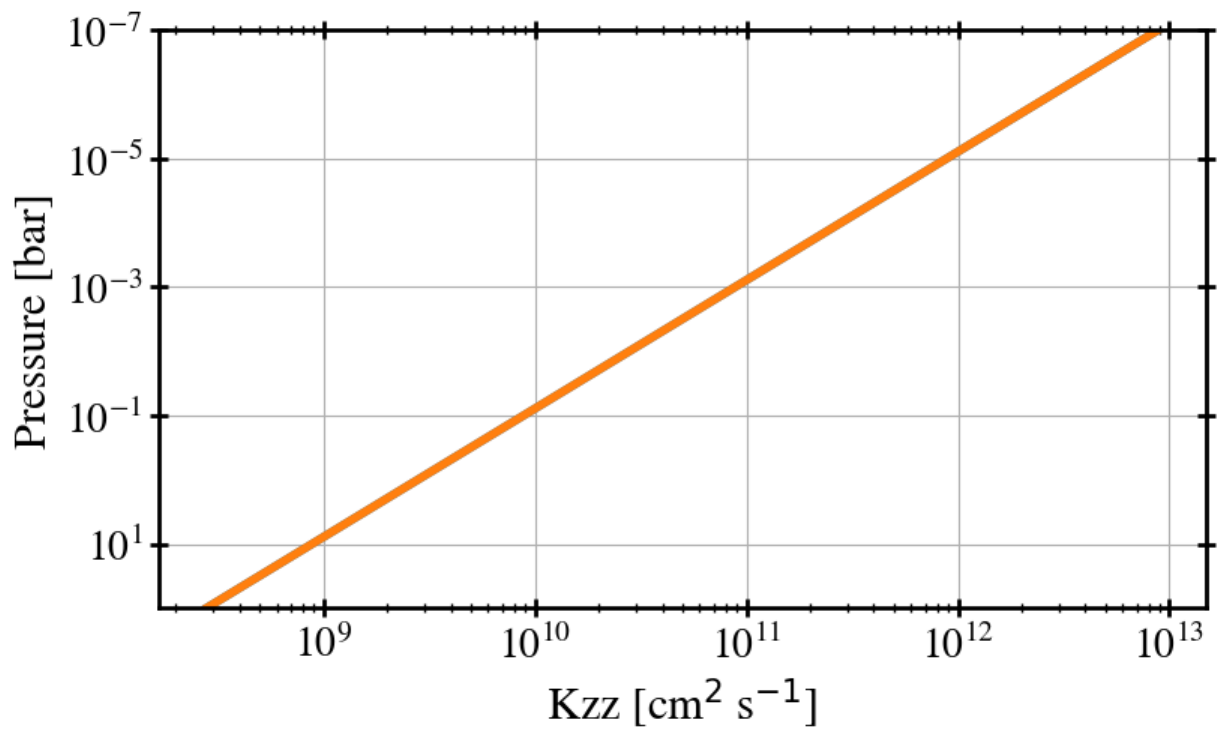}
     \caption{Top: the pressure-temperature profiles at the east and west terminators as calculated from the equilibrium UM GCM, Bottom: The K$_{zz}$ profile used in the photochemical models.}
     \label{fig:pt_kzz}
 \end{figure}

The pressure/temperature profiles at the east and west limbs from the equilibrium UM GCM used for the photochemical modeling are shown in Figure \ref{fig:pt_kzz} where we isothermally extend the atmospheric structure to high pressures where photochemistry is the most active. These profiles were averaged over $\pm$ 20$^\circ$ of each terminator. The K$_{zz}$ profile used in our modeling is also shown in Figure \ref{fig:pt_kzz}. 

\begin{figure}
     \centering
     \includegraphics[width=0.9\columnwidth]{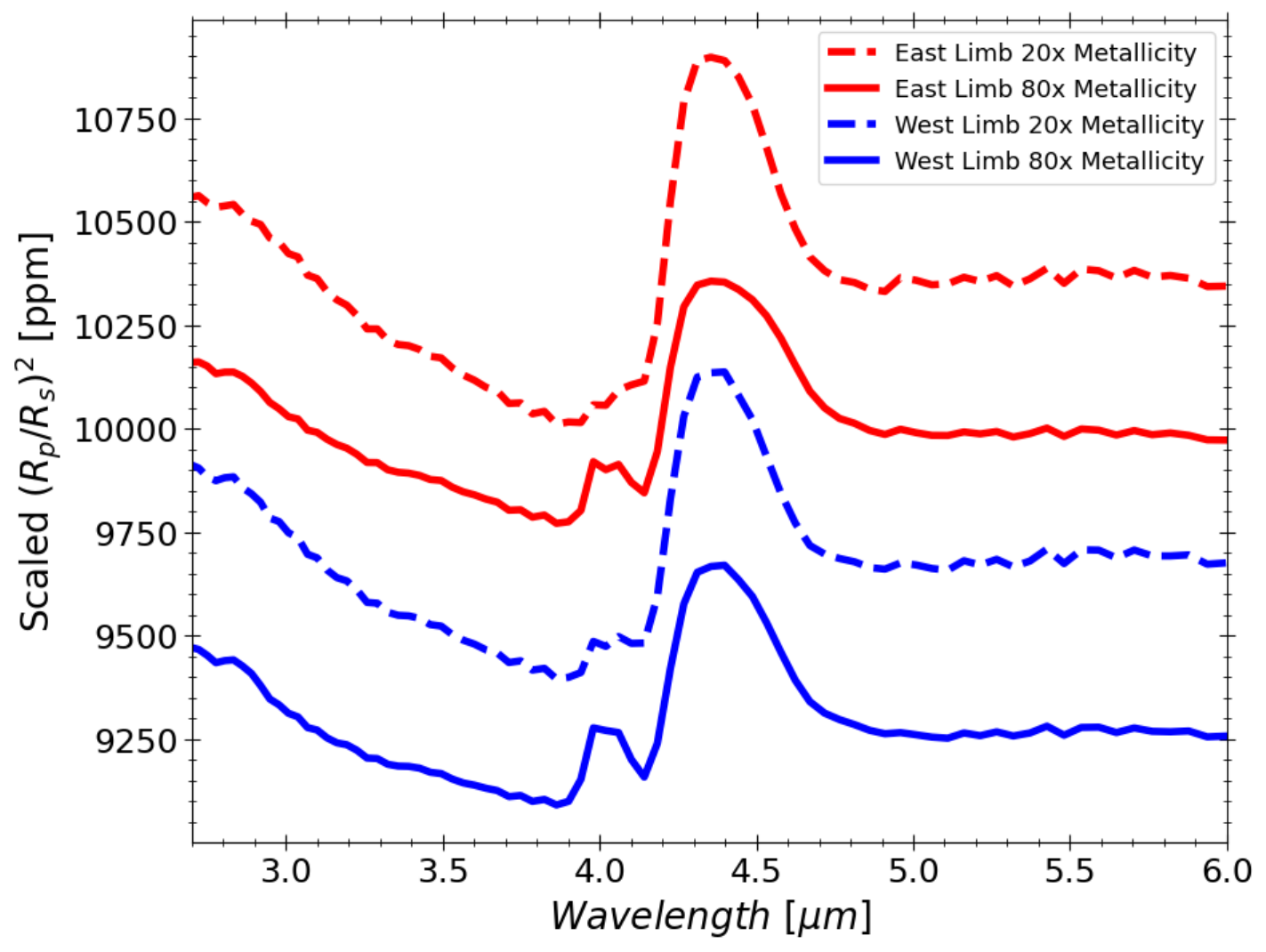}
     \caption{The limb separated transmission spectra based on the photochemical models for WASP-15b for both the $20\times$ and $80\times$ solar models.}
     \label{fig:photo_limbs}
 \end{figure}

In the main text we show a limb-averaged transmission spectrum for WASP-15b. In Figure \ref{fig:photo_limbs} we show the transmission spectra separated for the east/west limbs for our best-fit $80\times$ solar metallicity case. The \ce{SO2} feature is visible on both limbs with an enhanced amplitude on the cooler west limb as was seen in \citet{Tsai2023}.

\subsection{Molecular line lists used for PICASO opacities}
\label{appendix:picaso_inputs}

The molecular line lists used to create the opacities in the PICASO radiative transfer modelling are given in Table \ref{tab:wasp15b_linelists}. 

\begin{table}
\centering
\caption{Line lists used to make PICASO Opacities.}
\label{tab:wasp15b_linelists}
\begin{tabular}[t]{lc}
\hline
Species & Reference\\
\hline
        CO2 &  \citet{HUANG2014reliable} \\ 
        CH4 &  \citet{Yurchenko2013,yurchenko_2014} \\ 
        CO &  \citet{Rothman2010,Li2015,Gordon2017} \\ 
        H2 &  \citet{Gordon2017} \\ 
        H2O &  \citet{Polyansky2018} \\ 
        H2S &  \citet{azzam16exomol} \\ 
        H2--H2 &  \citet{Saumon12,Lenzuni1991h2h2} \\ 
        H2--He &  \citet{Saumon12} \\ 
        H2--H &  \citet{Saumon12} \\ 
        H2--CH4 &  \citet{Saumon12} \\ \hline
\end{tabular}
\end{table}


\bsp	
\label{lastpage}
\end{document}